\documentclass[a4paper,11pt]{article}

\usepackage{jheppub} 

\usepackage{amsmath}
\usepackage{algorithm}
\usepackage{algpseudocode}

\arxivnumber{2611.19306}

\title{\boldmath Spatiotemporal K-Shape: Reconstructing inhomogeneous condensate profiles on the lattice}

\author[a]{Keita Horie,}
\author[b]{Rin Kagami,}
\author[c,d,e]{Chiho Nonaka}
\affiliation[a]{ChudenCTI Co., Ltd., Nagoya 461-0005, Japan}
\affiliation[b]{Department of Physics, Kyoto University, Kyoto 606-8502, Japan}
\affiliation[c]{Physics Program, Graduate School of Advanced Science and Engineering, Hiroshima University, Higashi-Hiroshima 739-8511, Japan}
\affiliation[d]{International Institute for Sustainability with Knotted Chiral Meta Matter, Hiroshima University, Higashi-Hiroshima 739-8511, Japan}
\affiliation[e]{Kobayashi Maskawa Institute, Nagoya University, Nagoya 464-8602, Japan}

\emailAdd{keita-horie@hiroshima-u.ac.jp}

\abstract{
Spatially modulated chiral condensates can affect the phase structure of dense QCD effective models. Understanding such order requires characterizing its real-space structure as well as detecting a modulation. Spatial two-point correlators retain the ensemble-averaged power spectrum but discard relative Fourier phases, leaving the configuration shapes undetermined in general. We propose the Spatiotemporal K-Shape method to extract representative shapes from lattice Monte Carlo ensembles without assuming a functional form. The method extends K-Shape clustering to periodic lattice fields, aligning configurations modulo translations and, optionally, a global sign or phase before extracting cluster centroids. Comparisons across clusterings assess whether one representative shape suffices. Controlled tests recover distinct profiles with identical correlators and separate mixtures of different shapes. Applied to the $(1+1)$-dimensional Gross--Neveu model with $N_f=8$, the method extracts non-sinusoidal profiles with two sign changes at two chemical potentials and separates multiple shape families at a third, where a single centroid is insufficient. Including the discrete chiral sign symmetry in the alignment can fold domain-length fluctuations into an asymmetric centroid without implying a preferred ensemble asymmetry. The method thus complements correlator diagnostics with representative real-space shapes and information on their diversity. The application to the Gross--Neveu model provides a testing ground for configuration-based studies of inhomogeneous order relevant to the QCD phase diagram.
}


\begin{document}
\maketitle
\flushbottom

\section{Introduction}
\label{sec:intro}

The phase structure of strongly interacting matter at finite temperature and baryon density is a central open problem~\cite{Jacak:2012dx, Baym:2017whm}. At finite baryon chemical potential lattice QCD is severely limited by the sign problem, and the low-temperature, high-density region has been explored mainly with effective models and with theories that are tractable in particular limits.

In this region chiral symmetry may be broken by a spatially modulated condensate rather than by a uniform one. A Fermi surface can favour particle--hole pairing at non-zero total momentum, and modulated chiral condensates have been argued to arise in QCD at a large number of colours~\cite{Deryagin:1992rw}---although at finite $N_c$ only for an extremely large number of colours~\cite{Shuster:1999tn}---and have been found in Nambu--Jona-Lasinio (NJL) and quark--meson models, as chiral density waves and more general one-dimensional modulations~\cite{Nakano:2004cd, Nickel:2009wj, Buballa:2014tba}. Allowing spatially modulated condensates can change the phase structure inferred from homogeneous ansatzes. In the NJL model, for example, an inhomogeneous region can replace the first-order transition line obtained when only homogeneous condensates are considered~\cite{Nickel:2009wj}. A wave number, however, does not determine the real-space profile: minimizations of the free energy over general one-dimensional profiles find that strongly non-sinusoidal, kink-like modulations can be favoured~\cite{Nickel:2009wj, Buballa:2014tba}. A localized domain wall is a superposition of several Fourier components with definite relative phases, and whether a modulation is close to a single harmonic or consists of well-separated walls is a physical property of the state. Understanding such spatial order therefore requires characterizing the real-space structure of the condensate in addition to detecting its presence and characteristic wavelength.

The Gross--Neveu model in $1+1$ dimensions (GN$_2$)~\cite{Gross:1974jv} provides a tractable setting for investigating this spatial order. It shares with QCD asymptotic freedom and the dynamical breaking of a chiral symmetry, here the discrete symmetry $\sigma\to-\sigma$, and its phase diagram is known analytically at large flavour number $N_f$. Its inhomogeneous phase is described by a self-consistent kink--antikink crystal, ranging from a nearly harmonic modulation to well-separated kinks and antikinks as temperature and chemical potential vary~\cite{Thies:2003kk, Schnetz:2004vr, Thies:2006ti}. There is also a concrete connection to QCD effective models: known GN-model solutions can be used to construct one-dimensional modulations in the $(3+1)$-dimensional NJL model within mean-field theory~\cite{Nickel:2009wj}. In the GN$_2$ model the spatial structure of the condensate is furthermore related to the baryon distribution~\cite{Lenz:2020cuv}. These connections make the model a useful testing ground for methods aimed at characterizing inhomogeneous order in theories relevant to the QCD phase diagram.

The chiral variant also has crystalline condensates~\cite{Schon:2000he, Basar:2008im}, and a chiral-spiral ground state has been argued to persist at finite $N$ and zero temperature~\cite{Ciccone:2022zkg}. At finite $N_f$, both models have been studied with lattice Monte Carlo simulations~\cite{Lenz:2020bxk, Lenz:2020cuv, Nonaka:2021pwm, Lenz:2021kzo}. For GN$_2$, an inhomogeneous regime was identified from the negative minimum of the spatial correlation function~\cite{Lenz:2020bxk}. The question addressed here is how to extract representative real-space shapes from such fluctuating configurations, and how to assess whether one shape suffices to describe an ensemble, without assuming a functional form for the condensate. The GN$_2$ application tests this analysis; applications to higher-dimensional theories remain a subsequent step.

The question is not trivial, because in a finite periodic volume no position is preferred. Schematically, the $i$-th configuration of an ensemble may be written as
\begin{equation}
    \sigma_i(m) \simeq \bar{\sigma}_i + A_i \, f_i(m - X_i) + \eta_i(m) \,,
    \label{eq:intro_decomposition}
\end{equation}
where $m$ labels the spatial sites, $\bar{\sigma}_i$ is a uniform component, $A_i$ the amplitude of the modulation, $f_i$ its shape, normalized to zero spatial mean and unit variance, $X_i$ its position (a translational collective coordinate), and $\eta_i$ the fluctuations. Even if all configurations share one shape, averaging over its position in a translationally invariant ensemble makes the one-point function spatially uniform (section~\ref{subsec:ensemble_loss}).

Two-point correlation functions avoid this cancellation, being invariant under a common translation of the field. For a single harmonic, the correlator $C(n)$ retains the wave number and amplitude. For general profiles, however, detecting a modulation and reconstructing its shape are different tasks. Up to a Fourier transform, $C(n)$ is the ensemble-averaged power spectrum: it retains the averaged squared moduli of the Fourier modes but not their relative phases. Different profiles containing several modes can therefore share the same correlator (section~\ref{subsec:kink_difficulty}). Ensemble averaging also combines contributions from different shapes without uniquely determining those shapes or their relative populations (section~\ref{subsec:ensemble_loss}).

A complementary strategy is to treat the collective coordinate explicitly: rather than integrating over $X_i$, one estimates the relative translations of the configurations, aligns them, and only then forms a representative profile, or $K$ of them if the ensemble does not carry a single shape.

In this paper we develop such an analysis, the Spatiotemporal K-Shape method, which extends the K-Shape algorithm for time-series clustering~\cite{paparrizos2015k} to multi-dimensional lattice fields with periodic boundary conditions. Each configuration is Z-normalized, so that the spatial mean, the amplitude and the normalized profile, the three pieces of information in eq.~\eqref{eq:intro_decomposition}, are kept separately, up to the contribution of the fluctuations $\eta_i$, and only the last is clustered. Configurations are compared by a dissimilarity built from the maximum of their normalized circular cross-correlation over all lattice translations, optionally extended to a global sign or phase. The centroid of a cluster is the leading eigenvector of a Hermitian matrix built from its aligned members, restored to the units of the field; no functional form of the profile is imposed. A delete-block jackknife gives an error band for the normalized shape. Shape-correlation criteria, calibrated against synthetic ensembles built from one and from two shapes, compare the results at different $K$ to assess whether one representative shape suffices.

Alignment and classification of noisy signals have established counterparts in single-particle cryo-electron microscopy~\cite{Penczek:1992, Bendory:2020}, generalized Procrustes analysis~\cite{Gower:1975}, and multi-reference alignment, which estimates a signal from noisy copies with unknown cyclic shifts~\cite{Bandeira:2014, Bendory:2020}. The leading-eigenvector construction for the centroid also has a precedent in statistical shape analysis~\cite{Kent:1994}; its relation to the present definition is discussed in section~\ref{subsec:centroid}. In multi-reference alignment, a signal is generically not determined by its power spectrum, whereas under suitable conditions its bispectrum determines it up to a cyclic shift~\cite{Bendory:2018}. The contribution here is to combine periodic lattice translations, optional global sign or phase identification, a correlation-based dissimilarity, a Rayleigh-quotient centroid and clustering with a block-jackknife treatment of Markov-chain data.

The method is not meant to replace correlation functions. The correlator detects spatial order and gives its characteristic length; the centroids characterize representative real-space shapes and help assess whether one shape describes the ensemble. A centroid describes an ensemble within a stated analysis, including the choice of the symmetry group and of $K$, and is not an order parameter (section~\ref{subsec:centroid_descriptor}).

We first test the method on controlled mock ensembles, including profiles with identical correlators and mixtures of distinct shapes. We then apply it to GN$_2$ ensembles with $N_f=8$ at $T/\sigma_0=0.0985$ and six chemical potentials. At $\mu/\sigma_0=0.485$ and $0.582$, the centroids show non-sinusoidal profiles with two sign changes. At $\mu/\sigma_0=0.630$, the clustering separates families of different shape, for which a single centroid is insufficient. The choice of symmetry group affects how variations in the domain lengths appear in the centroid: including $\sigma\to-\sigma$ folds symmetry-related orientations and can produce an asymmetric representative without implying a preferred asymmetry of the ensemble. The profiles are described as non-sinusoidal; a quantitative comparison with the large-$N_f$ kink--antikink crystal remains future work. Long autocorrelations limit the statistical precision, and cluster populations are not interpreted as equilibrium mixing ratios.

The paper is organized as follows. Section~\ref{sec:limitations} examines what correlators can and cannot tell about the real-space shape, section~\ref{sec:formulation} formulates the Spatiotemporal K-Shape method, section~\ref{sec:benchmarks} presents the tests on mock ensembles, and section~\ref{sec:application} the application to the GN$_2$ model. Section~\ref{sec:discussion} discusses the interpretation of the centroid, practical aspects, extensions and limitations, and section~\ref{sec:conclusion} concludes. Appendices~\ref{app:math}, \ref{app:impl} and~\ref{app:gn2_supplement} collect mathematical details, implementation details and autocorrelation measurements, and supplementary figures and numbers for the GN$_2$ analysis.

\section{What Correlators Determine and What They Leave Open}
\label{sec:limitations}

Spatial correlation functions are a natural and powerful tool for identifying inhomogeneous ordering in lattice field theories. Being insensitive to the arbitrary position of a modulated condensate, they have been used successfully to locate the inhomogeneous region of the GN$_2$ model at a finite number of flavours~\cite{Lenz:2020bxk, Lenz:2020cuv} and in related lattice studies of the chiral GN model~\cite{Nonaka:2021pwm, Lenz:2021kzo}, and further criteria continue to be proposed~\cite{Motta:2024rvk}. Beyond the spatial correlator, the second-order boundary between the symmetric and the inhomogeneous phase of the GN$_2$ model has been located, in the mean-field approximation, from an instability of the homogeneous phase towards finite-momentum fluctuations in the bosonic two-point function~\cite{Koenigstein:2021llr}; like $C(n)$, this criterion detects the onset of a modulation but not the real-space profile that the condensate develops. We address a complementary question: how much of the real-space structure of an individual configuration can be reconstructed from correlator-based observables?

Following ref.~\cite{Lenz:2020bxk}, we take the equal-time spatial correlator of a real scalar field $\sigma(t, m)$ to be
\begin{equation}
    C(n) = \left\langle \frac{1}{N_t N_s} \sum_{t, m} \sigma(t, m+n) \, \sigma(t, m) \right\rangle \,,
    \label{eq:correlator_def}
\end{equation}
where $m$ labels the spatial sites and $n$ the separation, $N_t$ and $N_s$ are the numbers of sites in the Euclidean time and spatial directions, $\langle \cdots \rangle$ is the ensemble average over Monte Carlo configurations, periodic boundary conditions are understood, and the sums over $t$ and $m$ reduce statistical errors. Phase identification is then based on the single number $C_{\text{min}} = \min_{n} C(n)$, for which eq.~(60) of ref.~\cite{Lenz:2020bxk} gives
\begin{equation}
    C_{\text{min}} 
    \begin{cases}
        \gg 0 & \text{inside a homogeneously broken phase,} \\
        \approx 0 & \text{inside a symmetric phase,} \\
        \ll 0 & \text{inside an inhomogeneous phase.}
    \end{cases}
    \label{eq:conventional_criterion}
\end{equation}
Equation~\eqref{eq:conventional_criterion} should be read as its authors present it: an empirical observation, with no numerical thresholds separating the three cases, used to draw what ref.~\cite{Lenz:2020bxk} itself calls ``a crude phase diagram''. That work further cautions that $C_{\text{min}}$ ``is not the expectation value of a product of local operators'' and that ``one must be cautious, when using non-local quantities like $C_{\text{min}}$, since they can fake non-existing phase transitions''. We take up this caution along a different axis. The question is sharp in the GN$_2$ model, whose inhomogeneous phase at large $N_f$ is a self-consistent kink--antikink crystal rather than a single harmonic~\cite{Schnetz:2004vr, Thies:2003kk}. Three aspects are discussed in turn.

\subsection{Loss of Real-Space Shape Information}
\label{subsec:kink_difficulty}

The correlator of eq.~\eqref{eq:correlator_def} is an autocorrelation of the field rather than the field profile itself. It is instructive to see first when this is harmless: for a purely sinusoidal condensate $\sigma(t, m) = A \cos(2\pi q_0 m / N_s + \varphi)$, with integer $q_0$ satisfying $0 < q_0 < N_s/2$,
\begin{equation}
    C(n) = \frac{A^2}{2} \cos\!\left( \frac{2\pi q_0 n}{N_s} \right) \,,
    \label{eq:sine_correlator}
\end{equation}
independently of the translational phase $\varphi$. The arbitrary position drops out while the wave number $q_0$ and the amplitude $A$ remain directly visible. If a separation corresponding to half a period is available on the lattice, $C_{\text{min}} = -A^2/2 < 0$. For nearly harmonic modulations the correlator thus retains essentially all the information needed to characterize the state.

The situation changes once several Fourier modes contribute. With the spatial Fourier coefficients of a configuration at fixed Euclidean time,
\begin{equation}
    \hat{\sigma}(q) = \frac{1}{\sqrt{N_s}} \sum_{m} \sigma(t, m) \, e^{-2\pi i q m / N_s} \,, \qquad q = 0, 1, \dots, N_s - 1 \,,
    \label{eq:fourier_coefficients}
\end{equation}
the spatial sum in eq.~\eqref{eq:correlator_def} becomes diagonal in $q$ by periodicity, and
\begin{equation}
    C(n) = \frac{1}{N_s} \sum_{q} \left\langle \left| \hat{\sigma}(q) \right|^2 \right\rangle e^{2\pi i q n / N_s} \,,
    \label{eq:power_spectrum}
\end{equation}
where $\langle \cdots \rangle$ now includes the average over Euclidean time. Up to a Fourier transform, the correlator is the averaged power spectrum $\langle |\hat{\sigma}(q)|^2 \rangle$ and nothing more. Consistently with this, ref.~\cite{Lenz:2020bxk} analyses the Fourier transform of the correlator, denoted $\tilde{C}(k)$ there, alongside $C$ itself, and states that it ``also provides information about the absolute values of the Fourier coefficients of the field''. The information carried by the conventional observables is therefore exhausted by the moduli $\{ |\hat{\sigma}(q)| \}$, while the relative phases $\arg \hat{\sigma}(q)$ enter nowhere. The ensemble average in eq.~\eqref{eq:power_spectrum} makes the loss strictly larger, since not even the moduli of a single configuration are recovered; the phase freedom discussed below therefore exists already at the level of one configuration.

Those phases determine the real-space shape. A lattice translation $\sigma(m) \to \sigma(m+\tau)$ acts on the Fourier phases $\theta_q = \arg \hat{\sigma}(q)$ as
\begin{equation}
    \theta_q \longrightarrow
    \theta_q + \frac{2\pi q\tau}{N_s} \,,
    \label{eq:phase_translation}
\end{equation}
so that translations change the phases linearly in $q$. A general change of the relative phases at fixed Fourier moduli cannot be represented by such a translation and therefore gives a different real-space shape with the same correlator. The counting of the remaining phase degrees of freedom is given in appendix~\ref{app:phase_freedom}. For a single harmonic with non-zero modes only at $q=\pm q_0$, the remaining phase instead specifies the position of the same sinusoidal profile, allowing for offsets by a fraction of a lattice spacing. The correlator therefore determines a single-harmonic profile up to its position, but does not in general determine a profile containing several harmonics.

This is a statement about shape reconstruction, not a claim that eq.~\eqref{eq:conventional_criterion} fails to register kink-like modulations; the opposite is the case. Consider an idealized kink--antikink lattice of vanishing spatial mean, modelled by a square wave of amplitude $A$ and period $\ell$ taking the values $\pm A$ on alternating half periods. A shift by half a period maps it onto minus itself, and a direct evaluation of eq.~\eqref{eq:correlator_def} gives the triangular wave
\begin{equation}
    C(n) = A^2 \left( 1 - \frac{4 n}{\ell} \right) \,, \qquad 0 \le n \le \frac{\ell}{2} \,,
    \label{eq:square_correlator}
\end{equation}
continued as an even, $\ell$-periodic function of $n$, so that $C(0) = A^2$ and $C_{\text{min}} = -A^2$. This is deeper by a factor of two than the $-A^2/2$ of a sine wave of the same peak amplitude, although that comparison is a matter of normalization, the two minima coinciding at equal variance. What is convention independent is that any zero-mean profile antisymmetric under a shift by half a period --- the sine wave of eq.~\eqref{eq:sine_correlator} and the idealized kink--antikink lattice alike --- obeys $C(n + \ell/2) = -C(n)$ and hence $C_{\text{min}}/C(0) = -1$ exactly. Sharp domain walls therefore make eq.~\eqref{eq:conventional_criterion} no less decisive than a single harmonic does. What the criterion cannot do is tell the two profiles apart: they share the same normalized minimum while having entirely different morphologies. This is a weaker form of the degeneracy established above, holding at the level of the criterion alone: the correlators themselves, a triangular and a cosine wave, differ. The distinction between detecting a modulation and reconstructing its morphology is the central motivation for the present work.

\subsection{Coexistence of Homogeneous and Modulated Components}
\label{subsec:biased_states}

A second aspect concerns a modulation coexisting with a non-vanishing homogeneous component. Writing $\sigma(t, m) = \sigma_{\text{const}} + \delta\sigma(m)$ with constant $\sigma_{\text{const}}$ and $\frac{1}{N_s}\sum_m \delta\sigma(m) = 0$ by construction, the cross terms drop out of the spatial average exactly and
\begin{equation}
    C(n) = \sigma_{\text{const}}^2 + \left\langle \frac{1}{N_s} \sum_{m} \delta\sigma(m+n) \, \delta\sigma(m) \right\rangle \,.
    \label{eq:offset_decomposition}
\end{equation}
The homogeneous component produces an $n$-independent positive offset. For $\delta\sigma(m) = A \cos(2\pi q m / N_s)$, and if the shift $n = N_s/(2q)$ is available on the lattice,
\begin{equation}
    C_{\text{min}} = \sigma_{\text{const}}^2 - \frac{A^2}{2} \,,
    \label{eq:cmin_offset}
\end{equation}
so a criterion based on the sign of $C_{\text{min}}$ alone stops registering the modulation once $|\sigma_{\text{const}}| > A/\sqrt{2}$, even though the finite-$q$ component is unchanged.

For the fixed background considered above, the connected correlator, $C_{\text{conn}}(n)=C(n)-\langle\sigma\rangle^2$, removes the offset. If the spatial mean varies between configurations, however, this subtraction leaves its variance as an $n$-independent contribution. In a finite volume, the exact ensemble average of the GN$_2$ model obeys $\langle\sigma\rangle=0$ by the discrete chiral symmetry $\sigma\to-\sigma$, so that the connected and disconnected correlators coincide. A finite Monte Carlo history need not sample the two sectors equally, as illustrated by the homogeneous ensembles of section~\ref{subsec:homo_inhomo}. Subtracting the spatial mean of each configuration removes this contribution, and that is what the Z-normalization of section~\ref{subsec:exsbd} performs. Our point is a different one: removing the spatial mean of each configuration discards the relative size of the zero-momentum and finite-momentum components, although that ratio is itself a physical property of the state. Compressing the spatial information into the single scalar $C_{\text{min}}$ forces such a choice, whereas a shape-based analysis can in principle keep the spatial mean, the fluctuation amplitude and the normalized fluctuation profile as separate quantities. Equation~\eqref{eq:cmin_offset} is not a prediction that the GN$_2$ condensate develops such an offset; it is a controlled illustration of a situation relevant wherever uniform and modulated order coexist~\cite{Buballa:2014tba}.

\subsection{Ensemble Averaging and Configuration-Level Information}
\label{subsec:ensemble_loss}

The third aspect concerns the ensemble average itself. In a finite periodic volume no spatial origin is preferred. If the configurations carry a common profile at fluctuating positions, $\sigma_i(m) \simeq f(m - X_i) + \eta_i(m)$, with the collective coordinate $X_i$ giving the position of the structure in the $i$-th configuration and $\eta_i$ denoting fluctuations with $\langle \eta \rangle = 0$, the one-point function reads
\begin{equation}
    \langle \sigma(m) \rangle = \sum_{X} P(X) \, f(m - X) \,,
    \label{eq:onepoint_average}
\end{equation}
which is independent of $m$ whenever $P(X)$ samples the translational orbit uniformly, as it must in a translationally invariant ensemble. The profile $f$ is not absent from the ensemble; it is invisible to the one-point function.

The two-point function evades this cancellation, which is one reason why it is useful. It does so, however, by integrating over the collective coordinate rather than by measuring it, so that configuration-level information is not recoverable from the averaged correlator. A related difficulty arises when the Markov chain visits configurations of genuinely different shape, such as competing structures near a phase boundary: the averaged correlator then combines contributions from different shapes and does not uniquely determine those shapes or their relative populations.

Both difficulties suggest the same change of strategy: rather than averaging over the collective coordinate, estimate it for each configuration and align the configurations before forming a representative profile; and rather than describing the ensemble by a single average, describe it by $K$ representative shapes. The first step turns translational invariance from a source of information loss into an alignment problem, implemented by the circular cross-correlation of section~\ref{subsec:correlation} and the centroid extraction of section~\ref{subsec:centroid}; the second is the role of the number of clusters $K$, used in section~\ref{subsec:multimodality} to separate configurations of different shape within one GN$_2$ ensemble. Data-driven analyses of configurations along these lines have been explored in other contexts~\cite{Carrasquilla:2016oun, vanNieuwenburg:2016zsd, Hu:2017wey, Boyda:2022nmh}, and machine-learning approaches to inverse problems in quantum chromodynamics have been reviewed recently~\cite{Aarts:2025gyp}. Each aspect of this section is tested on controlled configurations in section~\ref{sec:benchmarks}.

\section{Formulation of the Spatiotemporal K-Shape Method}
\label{sec:formulation}

In this section, we formulate the Spatiotemporal K-Shape method, which extends the K-Shape algorithm~\cite{paparrizos2015k}, designed for one-dimensional time series with open endpoints, to multi-dimensional lattice fields with periodic boundary conditions that may be complex-valued and have several internal components. Its two ingredients are a dissimilarity that is invariant under lattice translations (sections~\ref{subsec:correlation} and \ref{subsec:exsbd}), so that configurations related by a translation count as the same shape, and a centroid given by the leading eigenvector of a Hermitian matrix built from the aligned configurations (section~\ref{subsec:centroid}). The clustering iteration that combines the two is specified in section~\ref{subsec:algorithm}, and the statistical uncertainty of the resulting centroids is defined in section~\ref{subsec:error_estimation}.

\subsection{Lattice Notations and Conventions}
\label{subsec:notations}

We consider field configurations generated by Monte Carlo simulations on a $D$-dimensional periodic lattice $\Lambda$. The field is allowed to be complex-valued and to have several internal components (e.g.\ a complex scalar field $\Delta(x) = \sigma(x) + i\pi(x)$). Let $\Phi_{\alpha, \mathbf{n}}$ denote the field value, where $\alpha$ is the index for the internal field components and $\mathbf{n} \in \Lambda$ is the $D$-dimensional coordinate on the lattice. An ensemble consists of $N_{\text{conf}}$ configurations $\Phi_i$, $i = 1, \dots, N_{\text{conf}}$.

The lattice extent is $N_i$ in the $i$-th spatial direction ($i = 1, \dots, D-1$) and $N_t$ in the Euclidean time direction; for the isotropic spatial volumes considered in this paper we write $N_i = N_s$. The total number of lattice sites $N_{\text{lat}}$ and the dimension $N_v$ of the vectorized configuration space introduced in section~\ref{subsec:centroid} are then
\begin{equation}
    N_{\text{lat}} = N_t \prod_{i=1}^{D-1} N_i \,, \qquad
    N_v = N_{\text{comp}} \, N_{\text{lat}} \,,
    \label{eq:site_counting}
\end{equation}
where $N_{\text{comp}}$ is the number of internal field components, i.e.\ the number of values taken by the index $\alpha$. If the configurations are averaged over some lattice directions before the analysis, as they are over Euclidean time in section~\ref{subsec:setup}, everything below applies with $\Lambda$ replaced by the lattice of the remaining directions and $N_{\text{lat}}$ by its number of sites.

The formulation is given for a general complex, multi-component field, since none of its steps requires the field to be real; in this paper it is applied to real scalar fields with $N_{\text{comp}} = 1$ only. Our implementation handles real fields and single-component complex fields, the latter tested so far on synthetic configurations only (appendix~\ref{app:tolerance}); fields with $N_{\text{comp}} > 1$ are not implemented.

A tilde on a lattice field or centroid always denotes a Z-normalized quantity, so that $\tilde{\Phi}$, $\boldsymbol{\tilde{\phi}}$ and $\tilde{\Phi}_c$ all have zero mean in every internal component and a mean squared modulus of one over all components, $\| \tilde{\Phi} \|^2 = N_v$. Concerning terminology, we reserve the term \textit{order parameter} for quantities associated with spontaneous symmetry breaking, and refer to the spatial pattern encoded in a cluster centroid as its \textit{morphology} or \textit{shape}. Likewise we call the exSBD of section~\ref{subsec:exsbd} a \textit{dissimilarity} rather than a distance, keeping the word ``distance'' only in the proper name Shape-Based Distance inherited from the original K-Shape algorithm.

\subsection{Circular Cross-Correlation and Boundary Conditions}
\label{subsec:correlation}

To compare the spatial dependence of two configurations independently of their relative position, we use the cross-correlation function. The original K-Shape method assumes sequences with endpoints and pads them with zeros when they are shifted against each other. On a lattice with periodic boundary conditions such padding breaks the translational symmetry, and we replace it by a circular shift.

We write the Hermitian inner product of two configurations and the lattice translation by $\boldsymbol{\tau} \in \Lambda$ as
\begin{equation}
    \langle \Phi, \Psi \rangle = \sum_{\alpha} \sum_{\mathbf{n} \in \Lambda} \Phi_{\alpha, \mathbf{n}}^{*} \, \Psi_{\alpha, \mathbf{n}} \,, \qquad
    (\mathcal{T}_{\boldsymbol{\tau}} \Psi)_{\alpha, \mathbf{n}} = \Psi_{\alpha, (\mathbf{n}+\boldsymbol{\tau}) \bmod \mathbf{N}} \,,
    \label{eq:inner_product}
\end{equation}
where $\bmod\ \mathbf{N}$ acts element-wise with the lattice extent in each direction, and $\| \Phi \| = \langle \Phi, \Phi \rangle^{1/2}$. The translations $\mathcal{T}_{\boldsymbol{\tau}}$ are unitary and form the abelian group $G_{\text{trans}}$ of order $N_{\text{lat}}$. The multi-dimensional circular cross-correlation at the lattice shift $\boldsymbol{\tau}$ is
\begin{equation}
    (\Phi \star \Psi)_{\boldsymbol{\tau}} = \langle \Phi, \mathcal{T}_{\boldsymbol{\tau}} \Psi \rangle = \sum_{\alpha} \sum_{\mathbf{n} \in \Lambda} \Phi_{\alpha, \mathbf{n}}^* \, \Psi_{\alpha, (\mathbf{n}+\boldsymbol{\tau}) \bmod \mathbf{N}} \,,
    \label{eq:circular_cc}
\end{equation}
that is, the overlap of $\Phi$ with $\Psi$ translated by $\boldsymbol{\tau}$; the shift runs over all $D$ directions, including Euclidean time when it is not averaged over. By the correlation theorem, eq.~\eqref{eq:cc_fourier} of appendix~\ref{app:fourier}, all $N_{\text{lat}}$ shifts are obtained at once with fast Fourier transforms, at a cost $\mathcal{O}(N_v \log N_{\text{lat}})$ instead of $\mathcal{O}(N_v N_{\text{lat}})$ for a direct evaluation.

The shift that maximizes the real part of the correlation defines the alignment of the second configuration to the first:
\begin{equation}
    \boldsymbol{\tau}^\star = \underset{\boldsymbol{\tau} \in \Lambda}{\mathrm{argmax}} \; \mathrm{Re} \, (\Phi \star \Psi)_{\boldsymbol{\tau}} \,, \qquad
    \Psi^{\text{al}} = \mathcal{T}_{\boldsymbol{\tau}^\star} \Psi \,, \quad \text{i.e.} \quad
    \Psi^{\text{al}}_{\alpha, \mathbf{n}} = \Psi_{\alpha, (\mathbf{n}+\boldsymbol{\tau}^\star) \bmod \mathbf{N}} \,,
    \label{eq:alignment}
\end{equation}
so that $\mathrm{Re} \, \langle \Phi, \Psi^{\text{al}} \rangle = \max_{\boldsymbol{\tau}} \mathrm{Re} \, (\Phi \star \Psi)_{\boldsymbol{\tau}}$. If several shifts attain the maximum, the first one in the storage order of the lattice array is taken. Periodic boundary conditions are the appropriate ones for the field analysed in this paper: the clustered field is the bosonic auxiliary field $\sigma$, which as a bosonic field is periodic in Euclidean time and, on the spatial torus used here, also in the spatial direction.

\subsection{Extended Shape-Based Distance (exSBD)}
\label{subsec:exsbd}

To compare shapes independently of the overall level and amplitude of a configuration, both of which fluctuate from configuration to configuration, each configuration $\Phi$ is first Z-normalized,
\begin{equation}
    \tilde{\Phi}_{\alpha, \mathbf{n}} = \frac{\Phi_{\alpha, \mathbf{n}} - \bar{\Phi}_{\alpha}}{s_\Phi} \,,
    \label{eq:znorm}
\end{equation}
with the component-wise mean and a single scale common to all components,
\begin{equation}
    \bar{\Phi}_{\alpha} = \frac{1}{N_{\text{lat}}} \sum_{\mathbf{n} \in \Lambda} \Phi_{\alpha, \mathbf{n}} \in \mathbb{C} \,, \qquad
    s_\Phi^2 = \frac{1}{N_v} \sum_{\alpha} \sum_{\mathbf{n} \in \Lambda} \left| \Phi_{\alpha, \mathbf{n}} - \bar{\Phi}_{\alpha} \right|^2 \,,
    \label{eq:znorm_moments}
\end{equation}
where $s_\Phi > 0$ is assumed. Because all components are divided by the same real number, the relative amplitudes and the relative complex phases of the fluctuations are left unchanged. By construction,
\begin{equation}
    \sum_{\mathbf{n} \in \Lambda} \tilde{\Phi}_{\alpha, \mathbf{n}} = 0 \quad \text{for every } \alpha \,, \qquad
    \| \tilde{\Phi} \|^2 = N_v \,.
    \label{eq:znorm_properties}
\end{equation}
Without the normalization, configurations with large amplitudes would carry a disproportionate weight in the centroid of section~\ref{subsec:centroid}. The normalization does not declare $\bar{\Phi}_\alpha$ and $s_\Phi$ physically irrelevant: they are kept for every configuration and are used to express the centroid in the units of the field in eq.~\eqref{eq:restoration}, in line with the separation of spatial mean, fluctuation amplitude and normalized profile discussed in section~\ref{subsec:biased_states}.

For two Z-normalized configurations we define the normalized cross-correlation (NCC)
\begin{equation}
    \mathrm{NCC}(\tilde{\Phi}, \tilde{\Psi})_{\boldsymbol{\tau}} = \frac{(\tilde{\Phi} \star \tilde{\Psi})_{\boldsymbol{\tau}}}{\| \tilde{\Phi} \| \, \| \tilde{\Psi} \|} = \frac{1}{N_v} \, \langle \tilde{\Phi}, \mathcal{T}_{\boldsymbol{\tau}} \tilde{\Psi} \rangle \,.
    \label{eq:ncc}
\end{equation}
The denominator equals $N_v$ by eq.~\eqref{eq:znorm_properties}. Since $\mathcal{T}_{\boldsymbol{\tau}}$ is unitary, $\mathrm{Re} \, \mathrm{NCC}_{\boldsymbol{\tau}}$ is, at every shift, the cosine of the angle between $\tilde{\Phi}$ and $\mathcal{T}_{\boldsymbol{\tau}} \tilde{\Psi}$ regarded as real vectors, so that $| \mathrm{NCC}_{\boldsymbol{\tau}} | \le 1$ by the Cauchy--Schwarz inequality and the bound can be reached at any $\boldsymbol{\tau}$. The coefficient normalization of the original K-Shape method also divides by the norms of the unshifted sequences, but there the zero padding discards the entries shifted past the endpoints, so that the overlap is systematically reduced at large shifts; on the periodic lattice every shift compares all $N_v$ entries. The NCC is in general complex, and the extended Shape-Based Distance (exSBD) is the dissimilarity
\begin{equation}
    \mathrm{exSBD}(\tilde{\Phi}, \tilde{\Psi}) = 1 - \max_{\boldsymbol{\tau} \in \Lambda} \mathrm{Re} \left[ \mathrm{NCC}(\tilde{\Phi}, \tilde{\Psi})_{\boldsymbol{\tau}} \right] \,.
    \label{eq:exsbd}
\end{equation}
The maximizing shift is the $\boldsymbol{\tau}^\star$ of eq.~\eqref{eq:alignment}, so that evaluating eq.~\eqref{eq:exsbd} also aligns $\tilde{\Psi}$ to $\tilde{\Phi}$. The exSBD is symmetric in its arguments, because $\langle \tilde{\Phi}, \mathcal{T}_{\boldsymbol{\tau}} \tilde{\Psi} \rangle = \langle \tilde{\Psi}, \mathcal{T}_{-\boldsymbol{\tau}} \tilde{\Phi} \rangle^{*}$.

\paragraph{Range.}
Because every component of a Z-normalized configuration sums to zero, eq.~\eqref{eq:znorm_properties}, the $N_{\text{lat}}$ numbers $\mathrm{Re} \, \mathrm{NCC}_{\boldsymbol{\tau}}$ sum to zero over all shifts, so that their maximum is non-negative, and together with $| \mathrm{NCC}_{\boldsymbol{\tau}} | \le 1$,
\begin{equation}
    0 \le \mathrm{exSBD}(\tilde{\Phi}, \tilde{\Psi}) \le 1 \,.
    \label{eq:exsbd_range}
\end{equation}
The proof, the configurations that reach the upper bound, and the reason why the mean in eq.~\eqref{eq:znorm_moments} has to be subtracted component by component are given in appendix~\ref{app:range}.

\paragraph{Translations and the quotient.}
Since the maximum in eq.~\eqref{eq:exsbd} runs over all lattice translations, $\mathrm{exSBD}(\tilde{\Phi}, \mathcal{T}_{\boldsymbol{\tau}} \tilde{\Phi}) = 0$ for every $\boldsymbol{\tau}$. Conversely, $\mathrm{exSBD}(\tilde{\Phi}, \tilde{\Psi}) = 0$ requires equality in the Cauchy--Schwarz inequality, which for two vectors of the same norm $\sqrt{N_v}$ means $\mathcal{T}_{\boldsymbol{\tau}^\star} \tilde{\Psi} = \tilde{\Phi}$. The exSBD therefore depends on its two arguments only through their orbits under $G_{\text{trans}}$, i.e.\ it is a function on $(\mathcal{Z} / G_{\text{trans}}) \times (\mathcal{Z} / G_{\text{trans}})$, where $\mathcal{Z}$ is the set of Z-normalized configurations, and it vanishes only for identical orbits. Because $G_{\text{trans}}$ is a finite group, passing to this quotient removes the position of a structure but no continuous shape information, in accordance with the counting of appendix~\ref{app:phase_freedom}. Displacements by a fraction of a lattice spacing are not part of $G_{\text{trans}}$; they survive as a small residual dissimilarity, which vanishes with the displacement, and tend to smooth the centroid of section~\ref{subsec:centroid} slightly, except for a single harmonic, whose shape they leave unchanged (section~\ref{subsec:sinusoidal}).

A direct consequence is that a configuration and its sign-reversed counterpart have vanishing dissimilarity whenever the two are related by a translation. Examples are the idealized profiles discussed in section~\ref{subsec:kink_difficulty}. A sine wave, and the idealized kink--antikink lattice (square wave) of section~\ref{subsec:kink_difficulty}, are mapped onto minus themselves by a translation by half a period; when half a period is an integer number of lattice sites, $\mathrm{exSBD}(\tilde{\Phi}, -\tilde{\Phi}) = 0$ for both. This identification is intended: exchanging the roles of kinks and antikinks in a periodic kink--antikink lattice with domains of equal length amounts to such a translation, and the method is designed to disregard the position of a structure. A sign reversal that is not equivalent to a translation, for example that of a single localized bump, retains a non-zero dissimilarity.

\paragraph{Role of the real part.}
For a real field the use of $\mathrm{Re} \, \mathrm{NCC}$ rather than $| \mathrm{NCC} |$ only means that a sign reversal not equivalent to a translation is not quotiented out. For a complex field, $\mathrm{Re} \, \langle \Phi, \Psi \rangle$ is the Euclidean inner product of $\Phi$ and $\Psi$ regarded as real vectors in $\mathbb{R}^{2 N_v}$, with the real and imaginary parts of every component as separate coordinates. Equation~\eqref{eq:exsbd} therefore treats a complex component as two real components that are translated jointly, so that the local complex phase of the field, and the relative phases between components, enter the comparison. In particular, a global phase rotation $\tilde{\Psi} \to e^{i\theta} \tilde{\Psi}$ changes the exSBD. Whether it should is a physical choice, which the following generalization makes explicit.

\paragraph{Symmetry group.}
Let $G \supseteq G_{\text{trans}}$ be a group of unitary transformations of the configurations that preserve eq.~\eqref{eq:znorm_properties}, and whose elements relate configurations to be regarded as the same shape. We define
\begin{equation}
    d_G(\tilde{\Phi}, \tilde{\Psi}) = 1 - \max_{g \in G} \frac{\mathrm{Re} \, \langle \tilde{\Phi}, g \, \tilde{\Psi} \rangle}{\| \tilde{\Phi} \| \, \| \tilde{\Psi} \|} \,,
    \label{eq:dG}
\end{equation}
so that $\mathrm{exSBD} = d_{G_{\text{trans}}}$. The arguments above carry over: $d_G$ is symmetric, takes values in $[0, 1]$ because the maximum over $G$ is not smaller than that over $G_{\text{trans}}$, and depends on its arguments only through their $G$-orbits, vanishing only for identical orbits. For $G = G_{\text{trans}} \times U(1)$, with $U(1)$ acting as a common phase $e^{i\theta}$ on all components, the maximization over $\theta$ is carried out analytically, $\max_{\theta} \mathrm{Re} \, [ e^{i\theta} z ] = |z|$, and
\begin{equation}
    d_{G_{\text{trans}} \times U(1)}(\tilde{\Phi}, \tilde{\Psi}) = 1 - \max_{\boldsymbol{\tau} \in \Lambda} \left| \mathrm{NCC}(\tilde{\Phi}, \tilde{\Psi})_{\boldsymbol{\tau}} \right| \,.
    \label{eq:dG_phase}
\end{equation}
The aligned configuration is then $e^{-i \arg \mathrm{NCC}_{\boldsymbol{\tau}^\star}} \, \mathcal{T}_{\boldsymbol{\tau}^\star} \tilde{\Psi}$, with $\boldsymbol{\tau}^\star$ now maximizing $| \mathrm{NCC}_{\boldsymbol{\tau}} |$ instead of the real part in eq.~\eqref{eq:alignment}. For a real field the $U(1)$ factor is replaced by $\mathbb{Z}_2 = \{ +1, -1 \}$, and eq.~\eqref{eq:dG_phase} identifies $\tilde{\Phi}$ with $-\tilde{\Phi}$ also when the two are not related by a translation. A global internal transformation should be included in $G$ only if it is a symmetry of the theory; if it is explicitly broken, for instance by a bare fermion mass, the global phase carries physical information. Since $d_G$ can only decrease when $G$ grows, a larger group also makes pure noise appear more strongly aligned, and values of $d_G$ obtained with different groups are not directly comparable. Including a global sign in the alignment can fold symmetry-related variations onto a common orientation rather than average them out. The centroid may therefore reflect the typical magnitude of such variations even when their signed ensemble average vanishes.

The symmetry groups used in the GN$_2$ application are specified in section~\ref{subsec:setup}. Their effects on the extracted centroids and cluster assignments are examined in section~\ref{subsec:kink_extraction} and appendix~\ref{app:gn2_group}.

Although it carries the name Shape-Based Distance, $d_G$ is not a metric on $\mathcal{Z} / G$, since it violates the triangle inequality; the size of the violation, and the fact that $\sqrt{d_G}$ is a metric on $\mathcal{Z} / G$, are shown in appendix~\ref{app:triangle}.

\subsection{Vectorization and Centroid Extraction via Rayleigh Quotient}
\label{subsec:centroid}

Given a cluster $\mathcal{C}$, a set of $|\mathcal{C}|$ configuration indices, and a current estimate $\tilde{\Phi}_c$ of its centroid, we seek the shape that is, on average, most strongly correlated with the members of the cluster. Every member is first aligned to $\tilde{\Phi}_c$ by eq.~\eqref{eq:alignment}, $\tilde{\Phi}_i^{\text{al}} = \mathcal{T}_{\boldsymbol{\tau}_i^\star} \tilde{\Phi}_i$, with the additional phase factor (or sign) of eq.~\eqref{eq:dG_phase} if $G$ contains $U(1)$ (or $\mathbb{Z}_2$). It is then vectorized: its $N_v$ values, eq.~\eqref{eq:site_counting}, are arranged in a fixed order of $(\alpha, \mathbf{n})$ into a column vector $\boldsymbol{\tilde{\phi}}_i \in \mathbb{C}^{N_v}$. Since the alignment has already been carried out in the lattice geometry, this vectorization, used only in the present step, reduces the centroid computation to a matrix eigenvalue problem.

For fixed alignments, the direction of the centroid is defined as a maximizer of the sum of squared correlations, $J$ being invariant under a rescaling of its argument,
\begin{equation}
    \boldsymbol{v} \in \underset{\boldsymbol{v}' \in \mathbb{C}^{N_v},\, \| \boldsymbol{v}' \| = 1}{\mathrm{argmax}} \; J(\boldsymbol{v}') \,, \qquad
    J(\boldsymbol{v}) = \sum_{i \in \mathcal{C}} \frac{\left| \boldsymbol{v}^\dagger \boldsymbol{\tilde{\phi}}_i \right|^2}{\| \boldsymbol{v} \|^2 \, \| \boldsymbol{\tilde{\phi}}_i \|^2} \,,
    \label{eq:centroid_objective}
\end{equation}
where each term is the squared modulus of the normalized correlation, the first expression in eq.~\eqref{eq:ncc}, between $\boldsymbol{v}$ and the aligned member at zero shift. Using $| \boldsymbol{v}^\dagger \boldsymbol{\tilde{\phi}}_i |^2 = \boldsymbol{v}^\dagger \boldsymbol{\tilde{\phi}}_i \boldsymbol{\tilde{\phi}}_i^\dagger \boldsymbol{v}$ and $\| \boldsymbol{\tilde{\phi}}_i \|^2 = N_v$, one finds
\begin{equation}
    J(\boldsymbol{v}) = \frac{1}{N_v} \, \frac{\boldsymbol{v}^\dagger \hat{S} \, \boldsymbol{v}}{\boldsymbol{v}^\dagger \boldsymbol{v}} \,, \qquad
    \hat{S} = \sum_{i \in \mathcal{C}} \boldsymbol{\tilde{\phi}}_i \, \boldsymbol{\tilde{\phi}}_i^\dagger \,,
    \label{eq:rayleigh}
\end{equation}
a Rayleigh quotient of the Hermitian, positive semi-definite $N_v \times N_v$ matrix $\hat{S}$. Its maximum is $\lambda_{\max} / N_v$, where $\lambda_{\max}$ is the largest eigenvalue of $\hat{S}$, and it is attained by the corresponding eigenvector. This follows the shape-extraction step of the original K-Shape method~\cite{paparrizos2015k}. It is not the same as minimizing the sum of squared Euclidean norms $\sum_{i} \| \boldsymbol{\tilde{\phi}}_i - \boldsymbol{v} \|^2$ over the aligned members: at fixed $\| \boldsymbol{v} \|^2 = N_v$ this sum equals $2 |\mathcal{C}| N_v - 2 \, \mathrm{Re} \, \boldsymbol{v}^\dagger \sum_i \boldsymbol{\tilde{\phi}}_i$ and is minimized by the arithmetic mean, $\boldsymbol{v} \propto \sum_i \boldsymbol{\tilde{\phi}}_i$, rather than by an eigenvector.

The objective $J$ is invariant under $\boldsymbol{\tilde{\phi}}_i \to e^{i\theta_i} \boldsymbol{\tilde{\phi}}_i$ with an independent phase for every member (an independent sign for a real field), whereas the alignment and the assignment, which are based on $\mathrm{Re} \, \mathrm{NCC}$, are not; this mismatch is absent for $G \supseteq G_{\text{trans}} \times U(1)$, or $G \supseteq G_{\text{trans}} \times \mathbb{Z}_2$ for a real field. For any $G$, the centroid update maximizes a quadratic function of the correlations, while the assignment step of section~\ref{subsec:algorithm} is based on a linear one; the consequences for the iteration are discussed there.

\paragraph{Why the leading eigenvector.}
Three considerations support eq.~\eqref{eq:rayleigh} as the definition of the centroid, beyond its inheritance from the original K-Shape method. First, the centroid objective and the dissimilarity are both built from the normalized correlations of eq.~\eqref{eq:ncc}. For fixed alignments, $J$ sums their squared moduli, whereas $d_G$ is defined through the maximum of their real part over $G$. They therefore use the same correlations but optimize different functions of them. Second, $J$ does not require a sign, or a phase, to be fixed for each member. Where $G$ contains $\mathbb{Z}_2$ or $U(1)$, a configuration is only defined modulo that factor, and the arithmetic mean of the aligned members depends on the representative chosen for each of them: changing the signs of individual members changes the arithmetic mean and can cause cancellations, whereas each member contributes the same rank-one term $\boldsymbol{\tilde{\phi}}_i \boldsymbol{\tilde{\phi}}_i^\dagger$ to $\hat{S}$ whichever representative is taken, so that the leading eigenspace is unchanged. Third, the same construction is established in the statistical analysis of shapes: for planar shapes represented as centred complex vectors of fixed norm, on which rotations act as a global phase, the maximum-likelihood estimate of the modal shape under the complex Bingham distribution is the leading eigenvector of $\sum_i \boldsymbol{z}_i \boldsymbol{z}_i^\dagger$~\cite{Kent:1994}, which is eq.~\eqref{eq:rayleigh} for fixed alignments. The price of the quadratic objective is that the iteration is not a descent method for eq.~\eqref{eq:objective}, as discussed in section~\ref{subsec:algorithm}, and that the folding described in section~\ref{subsec:exsbd} is the counterpart of the invariance just used.

No explicit projection onto zero-mean vectors is needed: an eigenvector of $\hat{S}$ with a positive eigenvalue has zero mean in every component, so that the projector used for this purpose in ref.~\cite{paparrizos2015k} acts trivially and is omitted (appendix~\ref{app:projection}).

The eigenvector is determined only up to a factor $e^{i\theta}$, which is a sign for a real field. We fix this factor by requiring the summed overlap with the aligned members to be real and non-negative,
\begin{equation}
    \theta^\star = - \arg \Bigl[ \sum_{i \in \mathcal{C}} \boldsymbol{\tilde{\phi}}_i^\dagger \, \boldsymbol{v} \Bigr] \,, \qquad
    \boldsymbol{v}_\star = e^{i\theta^\star} \boldsymbol{v} \,,
    \label{eq:phase_fix}
\end{equation}
where $\boldsymbol{v}$ is the leading eigenvector normalized to $\| \boldsymbol{v} \| = 1$, and the factor is left unchanged if the sum vanishes. For a real field this selects the sign with $\sum_i \boldsymbol{\tilde{\phi}}_i^{\mathsf{T}} \boldsymbol{v}_\star \ge 0$, or equivalently $\sum_i \| \boldsymbol{\tilde{\phi}}_i - \boldsymbol{v}_\star \|^2 \le \sum_i \| \boldsymbol{\tilde{\phi}}_i + \boldsymbol{v}_\star \|^2$. Equation~\eqref{eq:phase_fix} selects one representative of the eigenvector. It is a convention that makes the output unique, not a quotient by a symmetry in the sense of eq.~\eqref{eq:dG}, and it does not identify configurations with each other. The Z-normalized centroid is finally
\begin{equation}
    \boldsymbol{\tilde{\phi}}_c = \sqrt{N_v} \; \boldsymbol{v}_\star \,,
    \label{eq:centroid_znorm}
\end{equation}
which has zero mean in every component and $\| \boldsymbol{\tilde{\phi}}_c \|^2 = N_v$; reshaping $\boldsymbol{\tilde{\phi}}_c$ back into the lattice geometry gives $\tilde{\Phi}_c$.

\paragraph{Direct and Gram routes.}
We diagonalize $\hat{S}$ when $N_v \le |\mathcal{C}|$ and otherwise the $|\mathcal{C}| \times |\mathcal{C}|$ Gram matrix $\hat{W}$ of the aligned members, eq.~\eqref{eq:gram}, whose leading eigenvector is mapped to that of $\hat{S}$ by eq.~\eqref{eq:gram_pullback}; in both cases a Hermitian eigensolver is used. The Gram route keeps the centroid computable when $N_v \gg |\mathcal{C}|$, as can happen on $(3+1)$-dimensional lattices. Its derivation, the position of the complex conjugation that it requires, and the costs of the two routes are given in appendix~\ref{app:gram}. In all global fits of section~\ref{sec:application}, with $N_v = 64$ after the Euclidean-time average or $N_v = 1536$ without it, only the direct route was used.

\paragraph{Restoration of physical units.}
The centroid $\tilde{\Phi}_c$ is a normalized shape. To express it in the units of the field, we apply a single affine map, common to the whole cluster and built from the moments~\eqref{eq:znorm_moments} of its members:
\begin{equation}
    \Phi_{c, \alpha, \mathbf{n}} = \bar{s} \, \tilde{\Phi}_{c, \alpha, \mathbf{n}} + \bar{m}_\alpha \,,
    \label{eq:restoration}
\end{equation}
\begin{equation}
    \bar{s} = \frac{1}{|\mathcal{C}|} \sum_{i \in \mathcal{C}} s_{\Phi_i} \,, \qquad
    \bar{m}_\alpha = \frac{1}{|\mathcal{C}|} \sum_{i \in \mathcal{C}} \bar{\Phi}_{i, \alpha} \,,
    \label{eq:restoration_constants}
\end{equation}
where $\bar{\Phi}_{i, \alpha}$ and $s_{\Phi_i}$ are the mean and scale of the unnormalized configuration $\Phi_i$. Translations leave these moments unchanged, so eq.~\eqref{eq:restoration_constants} does not depend on the alignment. The centroid profiles shown in section~\ref{sec:application} are $\Phi_c$ divided by $\sigma_0$. The two constants $\bar{s}$ and $\bar{m}_\alpha$ carry no information on the shape and can be reported separately from it. Conversely, $\Phi_c$ is not the arithmetic mean of the aligned unnormalized configurations; it is the normalized shape of eq.~\eqref{eq:centroid_znorm}, expressed with the average scale and offset of the cluster.

\subsection{The Spatiotemporal K-Shape Algorithm}
\label{subsec:algorithm}

The dissimilarity of section~\ref{subsec:exsbd} and the centroid of section~\ref{subsec:centroid} are combined into a clustering algorithm with alternating assignment and refinement steps, as in $k$-means clustering. For a given number of clusters $K$, the $N_{\text{conf}}$ Z-normalized configurations are given labels $\ell_i \in \{1, \dots, K\}$ and partitioned into clusters $\mathcal{C}_k = \{ i : \ell_i = k \}$, $k = 1, \dots, K$, by alternating two steps: an \textit{assignment} step, which gives every configuration the label of the most similar centroid, and a \textit{refinement} step, which recomputes every centroid from the configurations that carry its label. K-Shape~\cite{paparrizos2015k} modifies both steps of $k$-means. In the assignment, the squared Euclidean norm of the difference is replaced by the exSBD of eq.~\eqref{eq:exsbd}, or more generally by $d_G$ of eq.~\eqref{eq:dG}, which compares two configurations after the optimal transformation in the symmetry group $G$. In the refinement, the arithmetic mean of the members is replaced by the leading eigenvector of eq.~\eqref{eq:rayleigh}, built from the members aligned to the current centroid. For Z-normalized configurations and unitary $g$, eqs.~\eqref{eq:ncc} and \eqref{eq:znorm_properties} give
\begin{equation}
    d_G(\tilde{\Phi}, \tilde{\Psi}) = \min_{g \in G} \frac{\| \tilde{\Phi} - g \, \tilde{\Psi} \|^2}{2 N_v}
    \label{eq:dG_msd}
\end{equation}
(cf.\ appendix~\ref{app:triangle}), so that the assignment is the nearest-centroid rule of $k$-means in the squared Euclidean norm after optimal alignment. The refinement, however, maximizes $J$ of eq.~\eqref{eq:centroid_objective} and is not the corresponding $k$-means update, the normalized arithmetic mean of the aligned members (section~\ref{subsec:centroid}).

\paragraph{Objective, refinement and assignment.}
The quantity used to compare clusterings is the clustering objective
\begin{equation}
    \mathcal{L}\bigl( \{ \ell_i \}, \{ \tilde{\Phi}_c^{(k)} \} \bigr) = \sum_{i=1}^{N_{\text{conf}}} \mathrm{exSBD}\bigl( \tilde{\Phi}_c^{(\ell_i)}, \tilde{\Phi}_i \bigr) \,, \qquad \ell_i \in \{ 1, \dots, K \} \,,
    \label{eq:objective}
\end{equation}
in which the centroid is the reference, i.e.\ the first argument, and the configuration is aligned to it. If a group $G \supsetneq G_{\text{trans}}$ is used, the exSBD is replaced by $d_G$ in eqs.~\eqref{eq:objective} and \eqref{eq:assignment} and in algorithm~\ref{alg:sks}. We label the iterations by $r = 1, 2, \dots$ and write $\ell_i^{[r]}$, $\tilde{\Phi}_c^{(k)[r]}$ and $\mathcal{C}_k^{[r]} = \{ i : \ell_i^{[r]} = k \}$ for the state after iteration $r$. Iteration $r$ first refines the centroids with the labels of the previous iteration,
\begin{equation}
    \tilde{\Phi}_c^{(k)[r]} =
    \begin{cases}
        \mathcal{R}\bigl( \mathcal{C}_k^{[r-1]} ; \tilde{\Phi}_c^{(k)[r-1]} \bigr) & \text{if } \mathcal{C}_k^{[r-1]} \neq \emptyset \,, \\
        \tilde{\Phi}_c^{(k)[r-1]} & \text{otherwise} \,,
    \end{cases}
    \qquad k = 1, \dots, K \,,
    \label{eq:refinement}
\end{equation}
where $\mathcal{R}(\mathcal{C} ; \tilde{\Phi}_{\text{ref}})$ denotes the centroid extraction of section~\ref{subsec:centroid}: every member of $\mathcal{C}$ is aligned to the reference $\tilde{\Phi}_{\text{ref}}$ by eq.~\eqref{eq:alignment} (with the phase factor or sign of eq.~\eqref{eq:dG_phase} if $G$ contains $U(1)$ or $\mathbb{Z}_2$), the leading eigenvector of $\hat{S}$, eq.~\eqref{eq:rayleigh}, is obtained on the direct or the Gram route of section~\ref{subsec:centroid}, its phase is fixed by eq.~\eqref{eq:phase_fix}, and the result is normalized by eq.~\eqref{eq:centroid_znorm}. If the leading eigenvalue vanishes or the extracted centroid is numerically constant (in our implementation, if the standard deviation of all entries of $\boldsymbol{v}_\star$ does not exceed $10^{-8}$ times their root mean square), the extraction fails and the fit is aborted. The labels are then updated with the new centroids,
\begin{equation}
    \ell_i^{[r]} = \min \Bigl[ \, \underset{k \in \{ 1, \dots, K \}}{\mathrm{argmin}} \; \mathrm{exSBD}\bigl( \tilde{\Phi}_c^{(k)[r]}, \tilde{\Phi}_i \bigr) \Bigr] \,, \qquad i = 1, \dots, N_{\text{conf}} \,,
    \label{eq:assignment}
\end{equation}
where the outer minimum selects the smallest label in the case of ties. Let $\mathcal{E}^{[r]} = \{ k : \ell_i^{[r]} \neq k \ \text{for all } i \}$ be the set of clusters left empty by eq.~\eqref{eq:assignment}, evaluated before any replacement. For every $k \in \mathcal{E}^{[r]}$, in increasing order, one configuration is drawn uniformly at random from all $N_{\text{conf}}$ configurations and its label is set to $k$; from here on, $\ell_i^{[r]}$ and $\mathcal{C}_k^{[r]}$ denote the labels and clusters after these replacements, and $\mathcal{L}^{[r]} = \mathcal{L}(\{ \ell_i^{[r]} \}, \{ \tilde{\Phi}_c^{(k)[r]} \})$ is evaluated with them. A replacement can take the only member of another cluster, and two empty clusters can draw the same configuration. A cluster can therefore still be empty at the next refinement; this case, and empty clusters in the initial labels, are covered by the second line of eq.~\eqref{eq:refinement}.

\paragraph{The objective is not monotone.}
For fixed centroids, eq.~\eqref{eq:assignment} minimizes $\mathcal{L}$ over the labels, so the assignment step never increases $\mathcal{L}$, apart from the replacement of empty clusters. The refinement step, however, does not minimize $\mathcal{L}$ over the centroids. At fixed labels and alignments, the minimizer would be the normalized arithmetic mean mentioned above, whereas eq.~\eqref{eq:refinement} maximizes the quadratic function $J$, and it aligns the members to the previous centroid $\tilde{\Phi}_c^{(k)[r-1]}$ rather than to the new one. As a result, $\mathcal{L}^{[r]}$ can increase from one iteration to the next, and it does so in practice; examples are given in appendix~\ref{app:tolerance}. The iteration is therefore not a descent method for $\mathcal{L}$, nor an ascent method for $J$, whose sum over clusters depends on the alignments that change in every iteration. A converged state is an approximate fixed point of the alternating map, which in general is not a local minimum of $\mathcal{L}$; $\mathcal{L}$ is used only to rank independent restarts.

\paragraph{Stopping rule.}
Since the objective is not monotone, it cannot serve as a stopping criterion: stopping as soon as $\mathcal{L}$ fails to decrease can end the iteration in a transient state. Unchanged labels alone are not an adequate criterion either, because the refinement depends on the reference centroid, so that the centroids can keep moving while the labels stay fixed; for $K = 1$ the labels never change. We therefore stop a restart after iteration $r$ if
\begin{equation}
    \ell_i^{[r]} = \ell_i^{[r-1]} \ \ \text{for all } i \,, \qquad
    \mathcal{E}^{[r]} = \emptyset \,, \qquad
    \max_{k} \, d_G\bigl( \tilde{\Phi}_c^{(k)[r-1]}, \tilde{\Phi}_c^{(k)[r]} \bigr) \le \varepsilon \,,
    \label{eq:stopping_rule}
\end{equation}
with $d_G$ the dissimilarity used in the assignment, reducing to the exSBD for $G=G_{\text{trans}}$, and $\varepsilon = 10^{-6}$; otherwise the iteration continues up to $r_{\max}$ iterations. The requirement $\mathcal{E}^{[r]} = \emptyset$ ensures that the labels returned by a stopped restart are the assignment~\eqref{eq:assignment} to the returned centroids. By eq.~\eqref{eq:dG_msd}, the third condition states that the centroid of every cluster moved, after optimal alignment, by a root-mean-square amount of at most $\sqrt{2 \varepsilon}$ per entry, i.e.\ by at most $1.4 \times 10^{-3}$ in units of its own standard deviation.

Since eq.~\eqref{eq:stopping_rule} tests the change over a single iteration, a slowly drifting trajectory can stop before it reaches a fixed point. The tests that led to the choice of $\varepsilon$, and the size of this effect for the ensembles of section~\ref{sec:application}, are summarized in appendix~\ref{app:tolerance}. The stopping rule and the value of $\varepsilon$ have been tested for real fields with $G = G_{\text{trans}}$ and $G = G_{\text{trans}} \times \mathbb{Z}_2$, the settings used in this paper, and for complex fields with $G = G_{\text{trans}} \times U(1)$ on synthetic configurations.

\paragraph{Initialization, restarts and the adopted solution.}
The iteration is started $N_{\text{trial}}$ times independently. In each restart the initial labels $\ell_i^{[0]}$ are drawn independently and uniformly from $\{ 1, \dots, K \}$, and each initial centroid $\tilde{\Phi}_c^{(k)[0]}$ is obtained by drawing its $N_v$ entries independently and uniformly from $[0, 1)$ (the real and imaginary parts separately for a complex field) and Z-normalizing the result by eq.~\eqref{eq:znorm}. The first refinement thus aligns the members of a random partition to random references. All restarts draw from one pseudo-random sequence in a fixed order, so that a fit is fully determined by the state of the generator at its start. Among the $N_{\text{trial}}$ restarts, the one with the smallest final value of $\mathcal{L}$ is adopted, irrespective of whether it was stopped by eq.~\eqref{eq:stopping_rule} or reached $r_{\max}$; two values that agree within a relative difference of $10^{-12}$ are treated as equal, and the earlier restart is then kept. The relation between the returned labels and centroids, which differs between restarts stopped by eq.~\eqref{eq:stopping_rule} and restarts that reached $r_{\max}$, is described in appendix~\ref{app:returned_state}.

The dependence of the result on the initialization is an \textit{algorithmic} uncertainty, distinct from the statistical uncertainty of the Monte Carlo ensemble. Different restarts can end in different final states, which may differ in their partition as well as in their shapes. Even for $K = 1$, where the labels are fixed, restarts end at slightly different centroids, because the alignment in the refinement depends on the starting centroid. The restarts reduce the risk of adopting a solution with a larger final value of $\mathcal{L}$, but they do not remove the algorithmic uncertainty, which is not estimated separately. Part of it enters the error band of section~\ref{subsec:error_estimation} through the independent fits of the jackknife subsamples.

\paragraph{Computational cost.}
With fast Fourier transforms, one evaluation of the exSBD costs $\mathcal{O}(N_v \log N_{\text{lat}})$, section~\ref{subsec:correlation}, so that the dissimilarities evaluated in one iteration cost $\mathcal{O}(N_{\text{conf}} K N_v \log N_{\text{lat}})$, to which one eigenvalue problem per cluster is added. The full cost of an iteration, eq.~\eqref{eq:cost}, the cost of a global fit and of the jackknife, and the memory requirements are given in appendix~\ref{app:cost}.

\begin{figure}[htbp]
  \centering
  \includegraphics[width=0.75\textwidth]{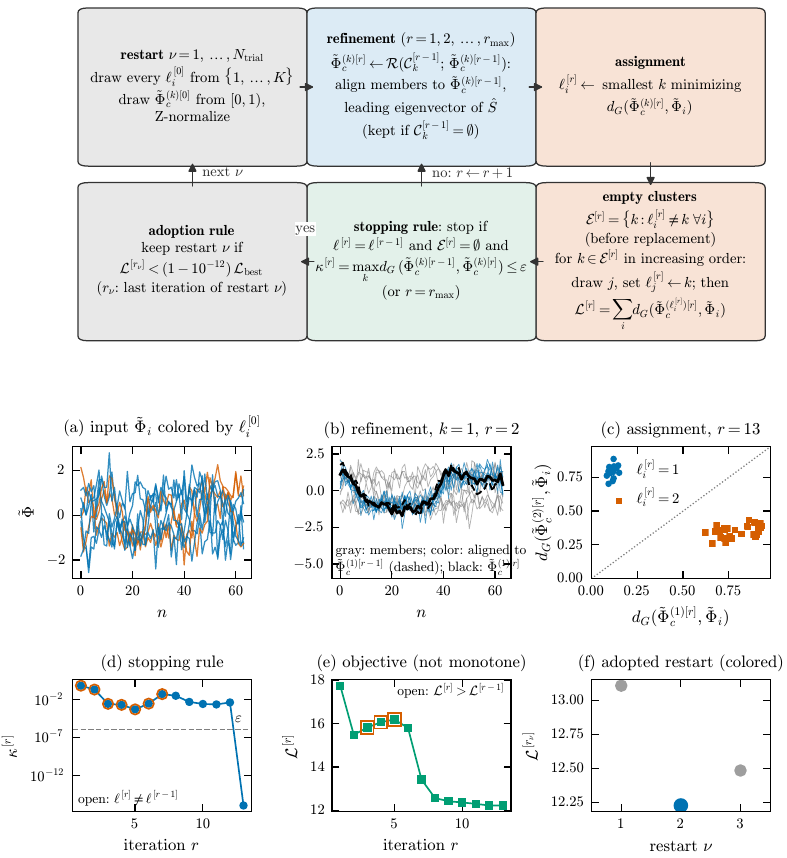}
  \caption{Schematic of one fit of the Spatiotemporal K-Shape algorithm (section~\ref{subsec:algorithm}). Top: the steps of algorithm~\ref{alg:sks} in the order in which they are executed, with $d_G$ the exSBD; the initialization of $\mathcal{L}_{\text{best}}$ and the abort on a failed extraction are omitted. Bottom: one fit to a synthetic ensemble of $45$ Z-normalized one-dimensional configurations on a periodic lattice of $64$ sites, fifteen from each of three shape families (a kink--antikink profile and sinusoids with wave numbers $q = 2$ and $q = 3$), each randomly translated and with Gaussian noise of standard deviation $0.5$ added before the Z-normalization to profiles of unit amplitude, fitted with $K = 2$, $N_{\text{trial}} = 3$, $r_{\max} = 50$ and $\varepsilon = 10^{-6}$. Panels (a)--(e) show the adopted restart; colours in (a)--(c) denote clusters. (a) Every fifth configuration, coloured by its initial label $\ell_i^{[0]}$. (b) Refinement of cluster $k = 1$ at iteration $r = 2$: eight members before (grey) and after (coloured) alignment to the previous centroid $\tilde{\Phi}_c^{(1)[1]}$ (dashed), and the new centroid $\tilde{\Phi}_c^{(1)[2]}$ (solid black). (c) Assignment at the last iteration: dissimilarities of every configuration to the two centroids, with the resulting labels $\ell_i^{[r]}$; the dotted line is the diagonal. (d) Centroid change $\kappa^{[r]}$ of the stopping rule, eq.~\eqref{eq:stopping_rule}, and the tolerance $\varepsilon$ (dashed); open circles mark iterations in which labels changed, and the value $\kappa^{[r]} = 0$ at the last iteration is drawn at the lower edge. (e) Objective $\mathcal{L}^{[r]}$, eq.~\eqref{eq:objective}; open squares mark increases. The objective increases at $r = 3$--$5$ because the refinement does not minimize $\mathcal{L}$ (section~\ref{subsec:algorithm}); in this example one cluster contains two of the three families, which makes the effect large. Examples for the GN$_2$ ensembles are given in appendix~\ref{app:tolerance}. (f) Final objective $\mathcal{L}^{[r_\nu]}$ of the three restarts; the adopted restart (coloured) has the smallest value.}
  \label{fig:sks_schematic}
\end{figure}

\paragraph{Hyperparameters and pseudocode.}
The parameters of the algorithm and of the error estimate are listed in table~\ref{tab:hyperparams}, together with the values used in section~\ref{sec:application}, and the complete procedure for one fit is summarized in algorithm~\ref{alg:sks} and illustrated in figure~\ref{fig:sks_schematic}.

\begin{table}[htbp]
  \centering
  \caption{Parameters of the Spatiotemporal K-Shape algorithm and of the jackknife error estimate, with the names of the corresponding parameters in our implementation, which is available on request (see Code availability), and the values used for the GN$_2$ ensembles of section~\ref{sec:application}; the second value of $G$ applies to the centroids for $K = 1$ of section~\ref{subsec:kink_extraction} and their bands. The value of $r_{\max}$ in parentheses applies to the global fits for $K = 1$ at $\mu/\sigma_0 = 0.388$ and for $K = 3$ at $\mu/\sigma_0 = 0.630$, and to the jackknife of the latter (table~\ref{tab:GN}); all other fits of this paper, including the null ensembles and the positive controls of appendix~\ref{app:null}, use $r_{\max} = 200$. The jackknife parameters apply to the ensembles for which an error band is quoted.}
  \label{tab:hyperparams}
  \vspace{0.2cm}
  \begin{tabular}{llp{0.32\textwidth}l}
    \hline
    Symbol & Implementation & Meaning & Value \\
    \hline \hline
    $K$ & \texttt{num\_clus} & Number of clusters & $1, 2, 3$ \\
    $N_{\text{trial}}$ & \texttt{num\_trial} & Independent restarts of a global fit & $20$ \\
    $r_{\max}$ & \texttt{max\_iter} & Maximum number of iterations per restart & $200$ ($2000$) \\
    $\varepsilon$ & \texttt{centroid\_tol} & Tolerance in eq.~\eqref{eq:stopping_rule} & $10^{-6}$ \\
    $G$ & \texttt{symmetry\_group} & Group in the dissimilarity, eq.~\eqref{eq:dG} & $G_{\text{trans}}$, $G_{\text{trans}} \times \mathbb{Z}_2$ \\
    $N_B$ & \texttt{num\_blocks} & Number of jackknife blocks, eq.~\eqref{eq:jk_blocks} & $6$ \\
    $N_{\text{trial}}^{\text{jk}}$ & \texttt{num\_trial} (jackknife) & Restarts of each subsample fit & $3$ \\
    $G_{\text{al}}$ & \texttt{align\_symmetry\_group} & Group used to match and align subsample centroids, eq.~\eqref{eq:jk_matching}; \texttt{auto} always includes the global sign (phase) & \texttt{auto}: $G_{\text{trans}} \times \mathbb{Z}_2$ \\
    \hline
  \end{tabular}
\end{table}

\begin{algorithm}[htbp]
  \caption{One fit of the Spatiotemporal K-Shape algorithm (section~\ref{subsec:algorithm}).}
  \label{alg:sks}
  \begin{algorithmic}[1]
    \Require Z-normalized configurations $\tilde{\Phi}_1, \dots, \tilde{\Phi}_{N_{\text{conf}}}$, eq.~\eqref{eq:znorm}; $K$, $N_{\text{trial}}$, $r_{\max}$, $\varepsilon$, $G$
    \Ensure labels $\{ \ell_i \}$, centroids $\{ \tilde{\Phi}_c^{(k)} \}$ and objective $\mathcal{L}$ of the adopted restart
    \State $\mathcal{L}_{\text{best}} \gets \infty$
    \For{$\nu = 1, \dots, N_{\text{trial}}$} \Comment{restarts}
      \State draw every $\ell_i^{[0]}$ uniformly from $\{ 1, \dots, K \}$
      \State draw the entries of every $\tilde{\Phi}_c^{(k)[0]}$ uniformly from $[0, 1)$ and Z-normalize
      \For{$r = 1, \dots, r_{\max}$}
        \For{$k = 1, \dots, K$} \Comment{refinement, eq.~\eqref{eq:refinement}}
          \If{$\mathcal{C}_k^{[r-1]} \neq \emptyset$}
            \State $\tilde{\Phi}_c^{(k)[r]} \gets \mathcal{R}( \mathcal{C}_k^{[r-1]} ; \tilde{\Phi}_c^{(k)[r-1]} )$ \Comment{abort if the extraction fails}
          \Else
            \State $\tilde{\Phi}_c^{(k)[r]} \gets \tilde{\Phi}_c^{(k)[r-1]}$
          \EndIf
        \EndFor
        \For{$i = 1, \dots, N_{\text{conf}}$} \Comment{assignment, eq.~\eqref{eq:assignment}}
          \State $\ell_i^{[r]} \gets$ smallest $k$ minimizing $d_G( \tilde{\Phi}_c^{(k)[r]}, \tilde{\Phi}_i )$
        \EndFor
        \State $\mathcal{E}^{[r]} \gets \{ k : \ell_i^{[r]} \neq k \text{ for all } i \}$ \Comment{before replacement}
        \For{$k \in \mathcal{E}^{[r]}$ in increasing order} \Comment{empty clusters}
          \State draw $j$ uniformly from $\{ 1, \dots, N_{\text{conf}} \}$ and set $\ell_j^{[r]} \gets k$
        \EndFor
        \State $\mathcal{L}^{[r]} \gets \sum_{i} d_G( \tilde{\Phi}_c^{(\ell_i^{[r]})[r]}, \tilde{\Phi}_i )$
        \State $\kappa^{[r]} \gets \max_k d_G( \tilde{\Phi}_c^{(k)[r-1]}, \tilde{\Phi}_c^{(k)[r]} )$
        \If{$\ell^{[r]} = \ell^{[r-1]}$ \textbf{and} $\mathcal{E}^{[r]} = \emptyset$ \textbf{and} $\kappa^{[r]} \le \varepsilon$} \Comment{eq.~\eqref{eq:stopping_rule}}
          \State \textbf{break}
        \EndIf
      \EndFor
      \State $r_\nu \gets$ the last iteration performed
      \If{$\mathcal{L}^{[r_\nu]} < (1 - 10^{-12}) \, \mathcal{L}_{\text{best}}$} \Comment{adoption rule}
        \State $\mathcal{L}_{\text{best}} \gets \mathcal{L}^{[r_\nu]}$; keep $\{ \ell_i^{[r_\nu]} \}$ and $\{ \tilde{\Phi}_c^{(k)[r_\nu]} \}$
      \EndIf
    \EndFor
    \State \Return the kept labels and centroids, and $\mathcal{L}_{\text{best}}$
  \end{algorithmic}
\end{algorithm}

\paragraph{Choice of the number of clusters.}
The number of clusters is an input of the algorithm. Rather than selecting a single $K$ by a pass/fail test, we run the algorithm for $K = 1$ and for several $K \ge 2$ and ask two separate questions about the resulting centroids. Both are formulated with the shape correlation
\begin{equation}
    \rho(\tilde{\Phi}, \tilde{\Psi}) = \max_{\boldsymbol{\tau} \in \Lambda} \bigl| \mathrm{NCC}(\tilde{\Phi}, \tilde{\Psi})_{\boldsymbol{\tau}} \bigr| = 1 - d_{G_{\text{trans}} \times U(1)}(\tilde{\Phi}, \tilde{\Psi}) \,,
    \label{eq:shape_correlation}
\end{equation}
cf.\ eq.~\eqref{eq:dG_phase}, which for a real field identifies two centroids that differ by a translation and a sign; for the GN$_2$ model the sign is the discrete chiral transformation $\sigma \to -\sigma$ (see section~\ref{subsec:setup}). Let $\tilde{\Phi}_c^{K=1}$ be the centroid for $K = 1$, $\tilde{\Phi}_c^{(k), K}$ the centroids for $K \ge 2$, and $\mathcal{K}$ the set of values $K \ge 2$ considered. Question~A asks whether the centroid for $K = 1$ coincides with one of the shapes found for $K \ge 2$, i.e.\ whether it represents the principal shape of the ensemble rather than a blend of several shapes:
\begin{equation}
    \rho_{\text{A}} = \min_{K \in \mathcal{K}} \; \max_{1 \le k \le K} \; \rho\bigl( \tilde{\Phi}_c^{K=1}, \tilde{\Phi}_c^{(k), K} \bigr) \,.
    \label{eq:criterion_A}
\end{equation}
Question~B asks whether the clusters obtained for $K \ge 2$ have different shapes:
\begin{equation}
    \rho_{\text{B}} = \min_{K \in \mathcal{K}} \; \min_{1 \le k < l \le K} \; \rho\bigl( \tilde{\Phi}_c^{(k), K}, \tilde{\Phi}_c^{(l), K} \bigr) \,.
    \label{eq:criterion_B}
\end{equation}
The two questions are not redundant. The centroid for $K = 1$ can coincide with one cluster found for $K \ge 2$ (large $\rho_{\text{A}}$) while another, smaller cluster has a different shape (small $\rho_{\text{B}}$).

Neither quantity can be interpreted by comparison with a fixed threshold, because the algorithm splits an ensemble into $K$ clusters whether or not it contains $K$ distinct shapes, and the resulting clusters differ slightly in shape even when all configurations share one underlying profile. We therefore calibrate $\rho_{\text{B}}$ with synthetic ensembles built from the data themselves. For the null hypothesis, one representative inhomogeneous ensemble of section~\ref{sec:application} is selected, and its centroid $\tilde{\Phi}_c^{K=1}$ is taken as the only profile. Every synthetic configuration is this profile at a lattice translation drawn uniformly at random, multiplied by an amplitude, plus noise. The amplitudes and the power spectrum of the coloured noise are taken from the selected ensemble, and the noise strength is tuned to reproduce the median of $1 - \rho(\tilde{\Phi}_c^{K=1}, \cdot)$ over its configurations; the construction of the amplitudes and the three noise models are specified in appendix~\ref{app:null}. The synthetic ensembles have the same number of configurations as the selected ensemble. They are Z-normalized and fitted with the same parameters and the same values of $K$ as the data, which yields the distribution of $\rho_{\text{B}}$ (and of $\rho_{\text{A}}$) when the splitting is forced; the only parameter that differs is $r_{\max}$, which is $200$ for these fits and for all but two of the fits of the data (table~\ref{tab:hyperparams}). The construction is repeated for a small number of independent realizations, with new random numbers for both the synthetic configurations and the fits. The null distribution obtained in this way is not rebuilt for every ensemble; it is used in common for all ensembles to which the two criteria are applied. As a positive control, synthetic ensembles with the same number of configurations as the null ensembles are built in the same way from two distinct profiles, the centroids of a $K = 2$ fit of an ensemble that contains them, mixed with the populations of that fit (appendix~\ref{app:null}). The value of $\rho_{\text{B}}$ for the data is then located relative to both distributions: a value within the null distribution is consistent with a forced splitting of one shape, whereas a value outside it indicates additional structure, whose strength is judged against the positive control. The selected ensemble, the number of realizations, the null ensembles and positive controls, and the resulting values of $\rho_{\text{A}}$ and $\rho_{\text{B}}$, are given in section~\ref{sec:application}.

\subsection{Statistical Uncertainty}
\label{subsec:error_estimation}

The centroid is a non-linear function of the whole ensemble: it depends on the data through the assignment, the alignments (maximizations over the lattice translations) and an eigenvalue problem, and in addition the configurations of a Markov chain are autocorrelated. We therefore estimate its statistical uncertainty with a delete-block jackknife~\cite{Efron:1979bxm, Kunsch:1989} in which the complete analysis is repeated on every subsample. The thermalized configurations are kept in the order of the Monte Carlo history and divided into $N_B$ contiguous blocks,
\begin{equation}
    \{ 1, \dots, N_{\text{conf}} \} = \mathcal{B}_1 \cup \dots \cup \mathcal{B}_{N_B} \,, \qquad
    \bigl| \, |\mathcal{B}_b| - |\mathcal{B}_{b'}| \, \bigr| \le 1 \,,
    \label{eq:jk_blocks}
\end{equation}
where each $\mathcal{B}_b$ is an interval of consecutive configurations and the $N_{\text{conf}} \bmod N_B$ remaining configurations are given to the first blocks. The subsample $(-b)$ consists of all configurations not in $\mathcal{B}_b$. The Euclidean-time average and the Z-normalization act on each configuration separately and are not affected by deleting a block. On every subsample, algorithm~\ref{alg:sks} is run anew with the same $K$, $r_{\max}$, $\varepsilon$ and $G$ as the global fit and $N_{\text{trial}}^{\text{jk}}$ restarts, with a random seed that depends only on $b$, so that the result does not depend on the order in which the subsamples are processed. It yields labels, clusters $\mathcal{C}_{k'}^{(-b)}$ and centroids $\tilde{\Phi}_c^{(k')(-b)}$, $k' = 1, \dots, K$.

\paragraph{Matching and alignment.}
The labels of an independent fit carry no information about the labels of the global fit, and a subsample centroid can be any representative of its equivalence class. Two ambiguities must therefore be removed before the subsample centroids can be compared with the global ones: a permutation of the $K$ labels, and a symmetry transformation of each centroid. The permutation is determined by solving the linear assignment problem
\begin{equation}
    \varpi_b = \underset{\varpi \in S_K}{\mathrm{argmin}} \; \sum_{k=1}^{K} d_{G_{\text{al}}}\bigl( \tilde{\Phi}_c^{(k)}, \tilde{\Phi}_c^{(\varpi(k))(-b)} \bigr) \,,
    \label{eq:jk_matching}
\end{equation}
with the Hungarian algorithm, where $S_K$ is the group of permutations of $\{ 1, \dots, K \}$ and $\tilde{\Phi}_c^{(k)}$ are the centroids of the global fit. Matching every global centroid to its most similar subsample centroid independently could assign the same subsample centroid twice. Each matched centroid is then aligned to its global counterpart,
\begin{equation}
    \tilde{\Phi}_c^{(k)(-b), \text{al}} = g_b^{(k)} \, \tilde{\Phi}_c^{(\varpi_b(k))(-b)} \,, \qquad
    g_b^{(k)} = \underset{g \in G_{\text{al}}}{\mathrm{argmax}} \; \mathrm{Re} \, \bigl\langle \tilde{\Phi}_c^{(k)}, g \, \tilde{\Phi}_c^{(\varpi_b(k))(-b)} \bigr\rangle \,,
    \label{eq:jk_alignment}
\end{equation}
which for $G_{\text{al}} = G_{\text{trans}} \times U(1)$ is the translation and phase factor given below eq.~\eqref{eq:dG_phase}. The operations are thus carried out in the order (i) matching of labels, (ii) alignment, (iii) evaluation of the variance; without step (ii), a mere displacement of a subsample centroid by a few lattice sites would be counted as a large statistical fluctuation of its shape. The matching is well defined only if the global centroids are pairwise distinct modulo $G_{\text{al}}$: two global clusters related by $\sigma \to -\sigma$, which the clustering with $G = G_{\text{trans}}$ may produce (appendix~\ref{app:gn2_group}), or clusters of nearly equal shape make eq.~\eqref{eq:jk_matching} degenerate or nearly so, and the assignment of subsample clusters to them is then arbitrary.

The group $G_{\text{al}}$ used here is at least as large as the group $G$ used in the clustering. It specifies which symmetry-related representatives are identified when comparing subsample centroids. As for $G$ (section~\ref{subsec:exsbd}), a global sign, or a global phase for a complex field, belongs in $G_{\text{al}}$ only if it is a symmetry of the theory. This choice is distinct from fixing the sign or phase of an eigenvector by eq.~\eqref{eq:phase_fix}. The groups used in the GN$_2$ application are given in section~\ref{subsec:setup}.

\paragraph{Jackknife error of the normalized centroid.}
A subsample is discarded if its fit fails (for instance because an extracted centroid is numerically constant) or if a matched cluster is empty. Let $\mathcal{V}$ be the set of the remaining subsamples and $N_B' = |\mathcal{V}| \le N_B$. The jackknife error of the normalized centroid at every entry is
\begin{equation}
    \delta \tilde{\Phi}_{c, \alpha, \mathbf{n}}^{(k)} = \Biggl[ \frac{(N_B - 1)^2}{N_B \, (N_B' - 1)} \sum_{b \in \mathcal{V}} \Bigl| \tilde{\Phi}_{c, \alpha, \mathbf{n}}^{(k)(-b), \text{al}} - \overline{\tilde{\Phi}}{}_{c, \alpha, \mathbf{n}}^{(k)} \Bigr|^2 \Biggr]^{1/2} \,, \qquad
    \overline{\tilde{\Phi}}{}_{c}^{(k)} = \frac{1}{N_B'} \sum_{b \in \mathcal{V}} \tilde{\Phi}_{c}^{(k)(-b), \text{al}} \,.
    \label{eq:jk_error}
\end{equation}
For $N_B' = N_B$ the prefactor is the familiar $(N_B - 1)/N_B$ of the delete-block jackknife; the choice of the prefactor for $N_B' < N_B$, and the condition under which it is appropriate, are explained in appendix~\ref{app:jk_prefactor}. Fewer than two valid subsamples leave the error undefined. The jackknife mean $\overline{\tilde{\Phi}}{}_{c}^{(k)}$ enters only through eq.~\eqref{eq:jk_error}; the central value is the centroid $\tilde{\Phi}_c^{(k)}$ of the global fit.

\paragraph{Error band.}
The centroid is shown in the units of the field, $\Phi_c^{(k)} = \bar{s}_k \tilde{\Phi}_c^{(k)} + \bar{m}_{k, \alpha}$, eqs.~\eqref{eq:restoration} and \eqref{eq:restoration_constants}, with $\bar{s}_k$ and $\bar{m}_{k, \alpha}$ computed from the members of $\mathcal{C}_k$ in the global fit. The same affine map, common to all subsamples, converts the jackknife error into the error band
\begin{equation}
    e_{\alpha, \mathbf{n}}^{(k)} = \bar{s}_k \, \delta \tilde{\Phi}_{c, \alpha, \mathbf{n}}^{(k)} \,,
    \label{eq:error_band}
\end{equation}
and the band shown is $\Phi_{c, \alpha, \mathbf{n}}^{(k)} \pm e_{\alpha, \mathbf{n}}^{(k)}$. The band is the uncertainty of the normalized shape, taken modulo $G_{\text{al}}$, expressed in the units of the field. It contains neither the fluctuation of the scale $\bar{s}_k$ nor that of the offset $\bar{m}_{k, \alpha}$, which are held at their values in the global fit. We treat them as separate observables, whose jackknife errors are obtained from the subsample values
\begin{equation}
    \bar{s}_k^{(-b)} = \frac{1}{| \mathcal{C}_{\varpi_b(k)}^{(-b)} |} \sum_{i \in \mathcal{C}_{\varpi_b(k)}^{(-b)}} s_{\Phi_i} \,, \qquad
    \bar{m}_{k, \alpha}^{(-b)} = \frac{1}{| \mathcal{C}_{\varpi_b(k)}^{(-b)} |} \sum_{i \in \mathcal{C}_{\varpi_b(k)}^{(-b)}} \bar{\Phi}_{i, \alpha} \,,
    \label{eq:restoration_jk}
\end{equation}
where the sums run over the members of the matched cluster of the subsample fit. The errors are computed with the same prefactor as in eq.~\eqref{eq:jk_error} and are reported separately as $\delta \bar{s}_k$ and $\delta \bar{m}_{k, \alpha}$. They are not added to the band. Adding them would require their covariance with the shape, which we do not estimate, and it would mix the uncertainty of the shape with that of two scalars. How much a band that included them would differ from eq.~\eqref{eq:error_band} depends strongly on the ensemble and on $K$ (appendix~\ref{app:gn2_band}). Nor does the band contain a fluctuation of the position of the structure, which the alignment~\eqref{eq:jk_alignment} removes by construction.

\paragraph{Interpretation.}
Six properties of the band should be kept in mind. (i) The subsample centroids are aligned to the global centroid, which is built from data that include most of every subsample. The alignment selects the transformation that brings a subsample centroid closest to this reference, which reduces the scatter of the aligned centroids compared with an alignment to an independent reference. Taken alone, this mechanism acts towards a smaller band; the net bias of the band, which also contains the effects of the discontinuous alignments and of the optimization in the subsample fits, has not been quantified. The reference itself is not resampled, so the band is an uncertainty conditional on it. (ii) The translations in the alignment are restricted to integer lattice shifts. A displacement by a fraction of a lattice spacing between subsamples cannot be removed and is counted in the band, mostly where the profile is steep. (iii) $e_{\alpha, \mathbf{n}}^{(k)}$ is the square root of the diagonal of the jackknife covariance matrix. Different sites, and without the Euclidean-time average different times, are strongly correlated, and with $N_B' \le N_B$ subsamples the jackknife estimate of the $N_v \times N_v$ covariance matrix has rank at most $N_B' - 1 < N_v$ and is therefore singular. A $\chi^2$ built from the band is therefore not meaningful, and statements about many sites at once cannot be turned into probabilities by treating the sites as independent. (iv) The band is the uncertainty of an ensemble-averaged shape. It is not a bound on the deviation of individual configurations, or of individual Euclidean-time slices of a centroid computed without the time average, from the centroid; such deviations are in general larger than the band. (v) The subsample fits start from random initializations with $N_{\text{trial}}^{\text{jk}}$ restarts, so the band also contains the variation of the local optimum reached in different subsamples, i.e.\ part of the algorithmic uncertainty of section~\ref{subsec:algorithm}. The subsample fits use the same stopping rule as the global fit. Their recorded stopping status in the GN$_2$ analysis is given in section~\ref{subsec:setup}. (vi) The band is built from subsamples of one ensemble analysed in one way, so it does not contain a bias that all subsamples share. In particular, if $G$ contains a global sign, the folding described in section~\ref{subsec:exsbd} acts in the same way in every subsample, and the band neither removes it nor signals its presence.

\paragraph{Number of blocks.}
The block length should be sufficiently large compared with the autocorrelation times relevant to the estimator~\cite{Wolff:2003sm}. The dependence of the jackknife error on the number of blocks should also be checked. A genuine statistical error is expected not to depend on $N_B$ once the blocks are longer than the autocorrelation time. In contrast, if the subsample estimates scatter by an amount that does not decrease as the deleted block becomes shorter, as is the case for optimization noise, the prefactor in eq.~\eqref{eq:jk_error} makes the band grow with $N_B$. The choice of the number of blocks and the associated limitations for the GN$_2$ ensembles are discussed in appendix~\ref{app:autocorr}.

\paragraph{When no band is quoted.}
If the centroid of an ensemble is not a stable shape, the subsample fits reach different local optima, the variation between subsamples consists mainly of this optimization noise at any affordable $N_{\text{trial}}^{\text{jk}}$, and the band grows with $N_B$ as described above. A band obtained in this situation would describe the optimizer rather than the ensemble. The corresponding treatment of the homogeneous GN$_2$ ensembles is described in section~\ref{subsec:homo_inhomo}.

\paragraph{The band is not used to compare values of $K$.}
Whether the centroid for $K = 1$ and a centroid for $K \ge 2$ describe the same shape is decided by the shape correlation~\eqref{eq:shape_correlation}, section~\ref{subsec:algorithm}, and not by asking whether the two centroids agree within their bands. The band is the estimation error of a centroid: it shrinks as $N_{\text{conf}}$ grows and, even for similar $N_{\text{conf}}$, varies by almost an order of magnitude between clusters with the spread of their members, so that agreement within the bands measures this spread and the size of the ensemble, not the similarity of the shapes. In practice, the pointwise difference of two aligned centroids in units of their combined band can be smaller for two clearly different shapes with wide bands than for two nearly identical shapes with narrow bands (appendix~\ref{app:gn2_band}). This comparison neglects the correlation between centroids obtained from the same configurations, which is expected to be positive; taking it into account would reduce the combined band.

\section{Benchmark Tests on Controlled Configurations}
\label{sec:benchmarks}

The tests collected in this section are controlled counterparts of the three aspects of section~\ref{sec:limitations}, taken one at a time and with the answer known by construction. Section~\ref{subsec:sinusoidal} isolates the translational collective coordinate of section~\ref{subsec:ensemble_loss}; section~\ref{subsec:kink_test} exhibits the loss of relative phase information of section~\ref{subsec:kink_difficulty}; section~\ref{subsec:biased_test} realizes the coexistence of a uniform and a modulated component of section~\ref{subsec:biased_states}; and section~\ref{subsec:superposed} addresses the second difficulty of section~\ref{subsec:ensemble_loss}, an ensemble whose members do not all carry the same shape. Section~\ref{subsec:2d_test} repeats the first of these on a two-dimensional lattice. Since the shape, the position and the noise of every configuration are put in by hand, what these tests can establish is that the method returns what was put in; none of the shapes below is a result.

The mock ensembles are summarized in table~\ref{tab:benchmarks}. Each consists of $N_{\text{conf}} = 1000$ configurations with $N_t = 24$ Euclidean times, an independent random translation, and Gaussian noise of width $\sigma_\eta = 0.1$ at every site. The one-dimensional tests of sections~\ref{subsec:sinusoidal}--\ref{subsec:superposed} use $N_s = 64$ spatial sites, matching the GN$_2$ lattices of table~\ref{tab:GN}; the statement applies to those tests alone, the two-dimensional ensemble of section~\ref{subsec:2d_test} having $32 \times 32$ sites. As in section~\ref{subsec:setup}, the configurations are averaged over Euclidean time before the analysis, so that the noise seen by the clustering has width $\sigma_\eta / \sqrt{N_t} = 0.0204$. The signal-to-noise ratio per site, the root-mean-square modulation divided by $\sigma_\eta / \sqrt{N_t}$, is then $36.7$ for the shapes of variance $0.560$, and $27.7$, $17.3$ and $14.7$ for the smaller amplitudes of table~\ref{tab:benchmarks}. These are favourable conditions, deliberately so: each test is built to isolate one mechanism, not to locate the noise level at which the method ceases to work, the single exception being the noise scan at the end of section~\ref{subsec:superposed}. The values of $C(n)$ quoted below are those of the input profiles. Since the $C(n)$ measured on the ensembles is computed on every time slice without the Euclidean-time average, as in eq.~\eqref{eq:correlator_def}, white noise contributes $\sigma_\eta^2 = 0.010$ to $C(0)$ alone in expectation and does not change the expected correlator at non-zero separation. The ratios of the expected correlators are therefore smaller in magnitude than the input-profile values: $-0.982$ for the shapes of variance $0.560$ and $-0.899$ for the smaller amplitude of section~\ref{subsec:sinusoidal}; the ratios measured on the ensembles, $-0.982$ and $-0.900$, agree with these expectations.

The ensembles of section~\ref{subsec:biased_test} carry a uniform component comparable to the modulation. They are controlled illustrations of the effect discussed in section~\ref{subsec:biased_states}, rather than models of the GN$_2$ ensembles. The size of the uniform contribution in the GN$_2$ ensembles is discussed in section~\ref{subsec:homo_inhomo}.

\begin{table}[htbp]
  \centering
  \caption{Mock ensembles used in the benchmark tests. All ensembles have $N_t = 24$, $N_{\text{conf}} = 1000$ and Gaussian noise of width $\sigma_\eta = 0.1$ per site; the last column is the signal-to-noise ratio per site after the average over Euclidean time, given for the mixed ensemble as sinusoidal / kink-like component. The clustering uses $G = G_{\text{trans}}$, $N_{\text{trial}} = 20$ restarts and at most $200$ iterations throughout. The one-dimensional sinusoidal ensembles are generated with a continuous random phase $\varphi$ uniform in $[0, 2\pi)$, the kink-like ones with a random integer shift, and the two-dimensional stripe with a random integer shift $(\tau_x, \tau_y) \in \mathbb{Z}_{32}^2$.}
  \label{tab:benchmarks}
  \vspace{0.2cm}
  \small
  \begin{tabular}{llccp{0.30\textwidth}c}
    \hline
    Section & Shape & $N_s$ & $K$ & Parameters & S/N \\
    \hline \hline
    \ref{subsec:sinusoidal} & sine & $64$ & $1$ & $A = 0.423$, $q = 2$ & $14.7$ \\
    \ref{subsec:kink_test} & zero-mean kink & $64$ & $1$ & $A = 0.8$, $w = 2$, $x_{1,2} = 16, 48$ & $36.7$ \\
    \ref{subsec:kink_test} & phase partner & $64$ & $1$ & moduli of the kink, phases redrawn (seed $20260920$, the $902$nd of $2000$ draws) & $36.7$ \\
    \ref{subsec:kink_test} & variance-matched sine & $64$ & $1$ & $A = 1.0583$, $q = 1$ & $36.7$ \\
    \ref{subsec:biased_test} & kink with background & $64$ & $1$ & $\sigma_{\text{const}} = 1.0$, $A = 0.8$, $w = 2$ & $36.7$ \\
    \ref{subsec:biased_test} & sine with background & $64$ & $1$ & $\sigma_{\text{const}} = 1.5$, $A = 0.5$, $q = 2$ & $17.3$ \\
    \ref{subsec:superposed} & $50\,\%$ sine, $50\,\%$ kink & $64$ & $2$ & $A = 0.8$, $q = 2$; $\sigma_{\text{const}} = 1.0$, $A = 0.8$, $w = 2$ & $27.7$ / $36.7$ \\
    \ref{subsec:2d_test} & two-dimensional stripe & $32 \times 32$ & $1$ & $A = 0.423$, $q = 2$ & $14.7$ \\
    \hline
  \end{tabular}
\end{table}

\subsection{Sinusoidal Modulation: Isolating the Translational Collective Coordinate}
\label{subsec:sinusoidal}

The first ensemble is a single harmonic at a random position,
\begin{equation}
    \sigma(t, m) = A \cos\left( \frac{2\pi q m}{N_s} + \varphi \right) + \eta(t, m) \,,
    \label{eq:mock_sine}
\end{equation}
with $A = 0.423$, $q = 2$, and $\varphi$ drawn independently for each configuration. This is the case in which the correlator is sufficient. For the noise-free input profile, eq.~\eqref{eq:sine_correlator} gives a cosine of the same wave number with $C_{\text{min}} = -A^2/2 = -0.0895$ and $C_{\text{min}}/C(0) = -1.000$, and, a single wave number being populated, the argument of section~\ref{subsec:kink_difficulty} shows that the shape itself is then fixed up to a translation. Nothing is gained here over the correlator, and nothing is claimed; the ensemble is used to display the alignment mechanism in isolation.

\begin{figure}[htbp]
    \centering
    \includegraphics[width=\textwidth]{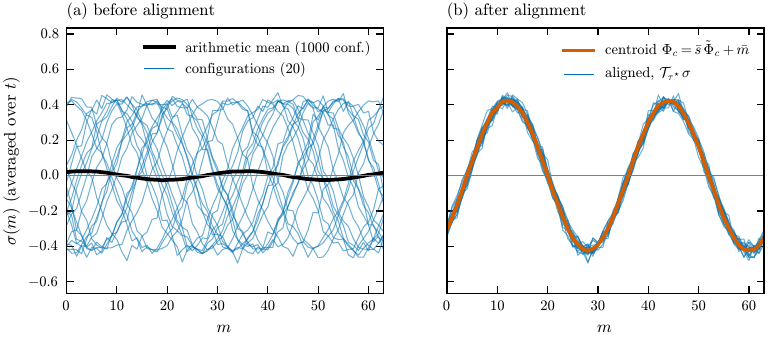}
    \caption{
        The alignment mechanism, for the sinusoidal ensemble of table~\ref{tab:benchmarks} at a signal-to-noise ratio of $14.7$.
        (a) Twenty of the $1000$ configurations, averaged over Euclidean time, at their random positions, and the arithmetic mean of all $1000$ (bold), flat up to a finite-sample residual of amplitude $0.025$.
        (b) The same twenty configurations after the alignment of eq.~\eqref{eq:alignment}, with the centroid restored to the scale of the field, $\Phi_c = \bar{s} \, \tilde{\Phi}_c + \bar{m}$, eqs.~\eqref{eq:restoration} and~\eqref{eq:restoration_constants}.
    }
    \label{fig:sinusoidal_alignment}
\end{figure}

Panel (a) of figure~\ref{fig:sinusoidal_alignment} shows $20$ of the configurations, averaged over Euclidean time, and the arithmetic mean of all $1000$, which is flat, as eq.~\eqref{eq:onepoint_average} implies, up to a finite-sample residual of amplitude $0.025$ at $q = 2$, accounted for by $A \, | \langle e^{i\varphi} \rangle |$ over the phases drawn, the noise leaving $6 \times 10^{-4}$ per site. Once aligned by eq.~\eqref{eq:alignment}, the members superpose and the centroid of eq.~\eqref{eq:centroid_znorm} reproduces the input cosine to $\mathrm{exSBD} = 3 \times 10^{-5}$ (panel (b)). The misalignment of up to half a lattice spacing left by the integer shifts (section~\ref{subsec:exsbd}) does not change this shape: a copy displaced by any amount lies in the span of the cosine and the sine at wave number $q$, so that without noise the leading eigenvector of $\hat{S}$ is again a single harmonic; the $\mathrm{exSBD}$ is set instead by the noise, which leaves a fraction $6 \times 10^{-5}$ of the power of the centroid at other wave numbers. The smoothing of section~\ref{subsec:exsbd} arises for profiles with several harmonics, the higher of which a fractional misalignment attenuates more strongly.

\subsection{Kink--Antikink Profiles: Recovering Information beyond a Single Fourier Mode}
\label{subsec:kink_test}

The second family of ensembles is built on a kink--antikink pair,
\begin{equation}
    \sigma(t, m) = \sigma_{\text{const}} + A \left[ \tanh\left( \frac{m - x_1}{w} \right) - \tanh\left( \frac{m - x_2}{w} \right) - 1 \right] + \eta(t, m) \,,
    \label{eq:mock_kink}
\end{equation}
translated by a random integer shift, with $A = 0.8$, $w = 2$ and walls at $x_{1,2} = 16, 48$, so that the two domains have equal length. In this section $\sigma_{\text{const}} = 0$; the same profile with $\sigma_{\text{const}} = 1.0$ is used in section~\ref{subsec:biased_test}. Equation~\eqref{eq:mock_kink} is a square wave rounded by walls of finite thickness rather than the idealized lattice of section~\ref{subsec:kink_difficulty}: with $w = 2$ the rounding lowers the variance to $C(0) = 0.560$, by $12.5\,\%$ with respect to $A^2 = 0.640$, and, once that difference in normalization is divided out, the shape of $C(n)$ still departs from the triangular wave of eq.~\eqref{eq:square_correlator} by up to $9.6\,\%$. What the rounding leaves intact is the antisymmetry under a shift by half a period, which holds up to the exponentially small overlap of the two walls, so that $C_{\text{min}}/C(0) = -1.000$ to the precision quoted, as it does for the idealized profile.

\paragraph{A single number is degenerate.}
Compare this ensemble with a sine wave of wave number $q = 1$ whose amplitude $A = 1.0583$ is chosen to match its variance. The two give
\begin{equation}
    C(0) = 0.560 \,, \qquad C_{\text{min}} = -0.560 \,, \qquad C_{\text{min}}/C(0) = -1.000 \,,
    \label{eq:degenerate_cmin}
\end{equation}
alike, with the minimum at the same separation $n = 32$. The criterion~\eqref{eq:conventional_criterion} therefore reads identically on the two, and it reads correctly on both, both being inhomogeneous; the statement is not that sharp walls escape $C_{\text{min}}$, but that they are not distinguished by it. The full correlator distinguishes the two profiles: the two curves differ by at most $0.0762$, which is $13.6\,\%$ of $C(0)$, and by $9.2\,\%$ in root mean square. The spectrum explains the proximity. The half-period antisymmetry suppresses the even harmonics, and the fundamental alone carries $89.8\,\%$ of the variance, followed by $7.8\,\%$ at $q = 3$ and $1.8\,\%$ at $q = 5$. The two shapes are nevertheless distinct, $\mathrm{exSBD} = 0.0526$.

\paragraph{The full correlator is degenerate as well.}
The residual family of appendix~\ref{app:phase_freedom} can be entered explicitly. Retaining the moduli of the kink and substituting the phases,
\begin{equation}
    \hat{\sigma}_{\text{p}}(q) = \left| \hat{\sigma}_{\text{k}}(q) \right| e^{i \theta_q} \,, \qquad
    \theta_{N_s - q} = -\theta_q \,, \qquad \theta_0 = \theta_{N_s/2} = 0 \,,
    \label{eq:phase_partner}
\end{equation}
produces a real field whose correlator is, by eq.~\eqref{eq:power_spectrum}, the same function; numerically
\begin{equation}
    \max_n \left| C_{\text{k}}(n) - C_{\text{p}}(n) \right| = 2.2 \times 10^{-16} \,, \qquad \text{that is} \quad 4 \times 10^{-16} \, C(0) \,,
    \label{eq:partner_degeneracy}
\end{equation}
which is machine precision. The phases $\theta_q$ of eq.~\eqref{eq:phase_partner} are drawn at random, and of $2000$ draws we retain the one farthest from the kink, at $\mathrm{exSBD} = 0.193$. That this is a different shape and not a displaced copy follows from that dissimilarity itself: the maximum over all $N_s$ shifts has already been taken in eq.~\eqref{eq:exsbd}, and by section~\ref{subsec:exsbd} a vanishing exSBD is what characterizes two configurations lying on the same translational orbit. Equation~\eqref{eq:phase_translation} says the same thing on the phases, a translation moving them by an amount linear in $q$, which a generic draw is not; that argument alone would not suffice here, the moduli at even $q$ being almost empty and the corresponding phases therefore almost free. The partner is the more strongly peaked of the two, $\max_m | \sigma(m) | = 1.197$ against $0.800$, while sharing every $| \hat{\sigma}(q) |$ with the kink. Its dissimilarity $0.193$ is $3.7$ times the $0.0526$ that separates the kink from the variance-matched sine. Thus profiles with identical full correlators can differ more strongly in shape than the kink and sine compared above. This is a comparison of these particular profiles, not a claim that fixing the full correlator leaves a larger ambiguity than fixing $C_{\text{min}}$ alone.

\begin{figure}[htbp]
    \centering
    \includegraphics[width=\textwidth]{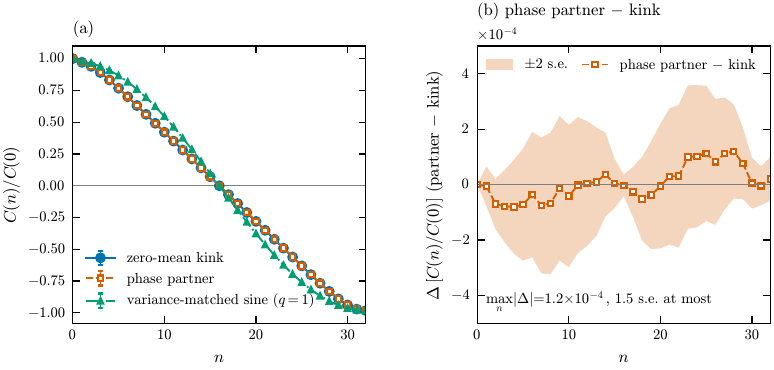}
    \caption{
        (a) The normalized correlator $C(n)/C(0)$ of the three ensembles of section~\ref{subsec:kink_test}: the zero-mean kink, its phase partner and the variance-matched sine. The curves are measured on the ensembles, with delete-1 jackknife errors over the $1000$ configurations. For the underlying profiles the first two are the same function, eq.~\eqref{eq:partner_degeneracy}, and here they agree within the statistical errors; the sine agrees with them at $n = 0$ and at the minimum and differs by at most $13.4\,\%$ in between. (b) The difference of $C(n)/C(0)$ between the partner and the kink, with a band of $\pm 2$ standard errors of the difference; it is at most $1.2 \times 10^{-4}$ and nowhere exceeds $1.5$ standard errors. The ratio $C_{\text{min}}/C(0)$ is $-1.000$ for the three profiles and $-0.982$ as measured, the difference being the noise contribution to $C(0)$. The signal-to-noise ratio is $36.7$.
    }
    \label{fig:degenerate_correlators}
\end{figure}

\begin{figure}[htbp]
    \centering
    \includegraphics[width=\textwidth]{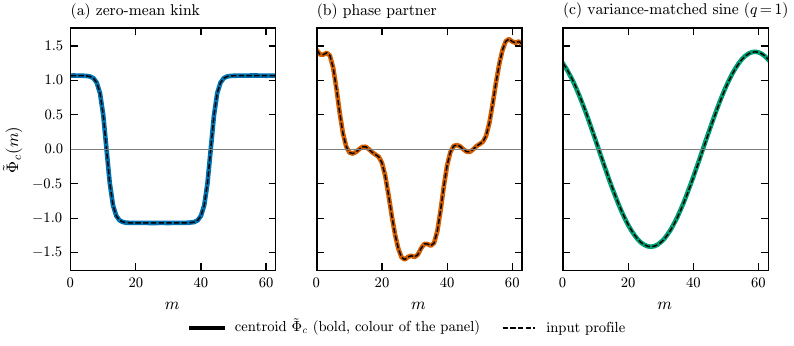}
    \caption{
        The extracted centroids (bold lines) and the Z-normalized input profiles (dashed) of the same three ensembles as in figure~\ref{fig:degenerate_correlators}: (a) the zero-mean kink, (b) its phase partner and (c) the variance-matched sine. For display, the centroids in (b) and (c) are translated to their best overlap with the one in (a), and each input profile to its best overlap with the centroid; since translations are quotiented out, the positions have no meaning. The shape dissimilarities are $\mathrm{exSBD} = 0.193$ between the kink and its phase partner and $\mathrm{exSBD} = 0.0526$ between the kink and the variance-matched sine, at a signal-to-noise ratio of $36.7$.
    }
    \label{fig:degenerate_shapes}
\end{figure}

Applied to the three ensembles separately with $K = 1$, the method returns three different centroids, each reproducing the profile it was given (figure~\ref{fig:degenerate_shapes}), whereas their correlators are those of figure~\ref{fig:degenerate_correlators}. The kink and its partner are thus an explicit pair of points of the residual family of appendix~\ref{app:phase_freedom}: no function of $\langle | \hat{\sigma}(q) |^2 \rangle$ can separate the kink from its partner, while the centroid, which is built from the configurations themselves, does.

\subsection{Coexisting Uniform and Modulated Components}
\label{subsec:biased_test}

Equation~\eqref{eq:cmin_offset} states that a uniform component raises $C(n)$ by $\sigma_{\text{const}}^2$ and, for a sinusoidal modulation, reverses the sign of $C_{\text{min}}$ once $| \sigma_{\text{const}} | > A/\sqrt{2}$. Two ensembles realize this with two different shapes: the sine of eq.~\eqref{eq:mock_sine} with a constant $\sigma_{\text{const}} = 1.5$ added to it, at $A = 0.5$ and $q = 2$, and the kink of eq.~\eqref{eq:mock_kink} with $\sigma_{\text{const}} = 1.0$. For the input profiles,
\begin{equation}
    C_{\text{min}} = +2.125 \qquad \text{and} \qquad C_{\text{min}} = +0.440 \,,
    \label{eq:biased_cmin}
\end{equation}
with $C_{\text{min}}/C(0) = +0.895$ and $+0.282$ respectively. The value $C_{\text{min}} = +2.125$ reproduces $\sigma_{\text{const}}^2 - A^2/2 = 2.25 - 0.125$ exactly. Measured on the ensembles, $C_{\text{min}}$ is the same to the digits quoted and the ratios are $+0.891$ and $+0.280$, the noise adding to $C(0)$ alone. By contrast, the same two shapes at zero background both have $C_{\text{min}}/C(0) = -1.000$, the kink at the $C_{\text{min}} = -0.560$ of eq.~\eqref{eq:degenerate_cmin}. The two examples quantify the point of section~\ref{subsec:biased_states}: a uniform component can reverse the sign of $C_{\text{min}}$ without changing the modulation shape, both for a smooth harmonic and for a pair of sharp walls.

\begin{figure}[htbp]
    \centering
    \includegraphics[width=\textwidth]{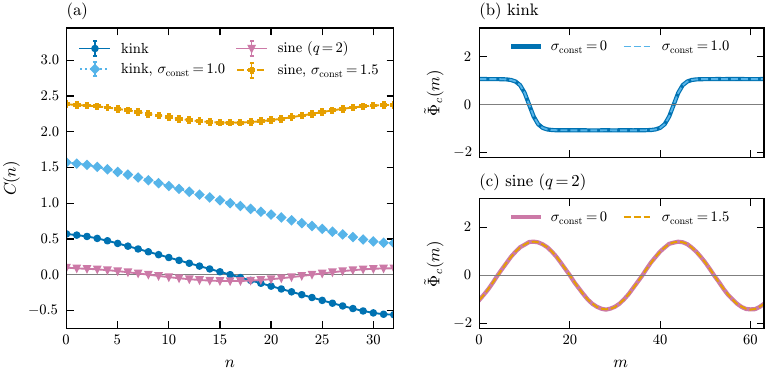}
    \caption{
        Effect of a uniform component.
        (a) $C(n)$ measured for the zero-mean kink, the kink with $\sigma_{\text{const}} = 1.0$, the sine of section~\ref{subsec:sinusoidal} ($q = 2$, $A = 0.423$) and the sine with $\sigma_{\text{const}} = 1.5$ ($q = 2$, $A = 0.5$), with $C_{\text{min}} = -0.560$, $+0.440$, $-0.0895$ and $+2.125$ respectively (eqs.~\eqref{eq:degenerate_cmin} and~\eqref{eq:biased_cmin}, section~\ref{subsec:sinusoidal}); the expected correlator at non-zero separation is unaffected by the noise, while the finite-sample minimum is subject to statistical fluctuations.
        (b) The centroids of the two kinks and (c) those of the two sines, after Z-normalization. The centroids of each pair coincide, to $\mathrm{exSBD} < 10^{-4}$, the underlying profiles being at $\mathrm{exSBD} = 0.000$; for display, the member with a background is translated to its best overlap with the other.
    }
    \label{fig:biased_centroid}
\end{figure}

The Z-normalization of eq.~\eqref{eq:znorm} subtracts the uniform component configuration by configuration, and does so exactly here: the biased and unbiased sine profiles are at $\mathrm{exSBD} = 0.000$ from each other, as are the biased and unbiased kink profiles, so the centroids are the shapes already obtained in sections~\ref{subsec:sinusoidal} and~\ref{subsec:kink_test} (figure~\ref{fig:biased_centroid}). What the subtraction discards is the ratio of the zero-momentum to the finite-momentum component. As described in section~\ref{subsec:exsbd}, that ratio is not lost but reported separately, through the constants $\bar{s}$ and $\bar{m}_\alpha$ of eq.~\eqref{eq:restoration_constants}.

\subsection{Mixed Ensembles: Separating Competing Spatial Structures}
\label{subsec:superposed}

\begin{figure}[htbp]
    \centering
    \includegraphics[width=\textwidth]{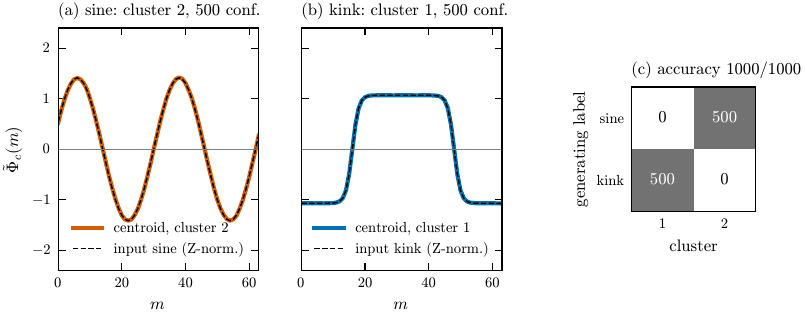}
    \caption{
        The two centroids extracted with $K = 2$ from the mixed ensemble of $50\,\%$ sinusoidal and $50\,\%$ kink-like configurations: (a) that of cluster 2, which contains the $500$ sinusoidal configurations, and (b) that of cluster 1, which contains the $500$ kink-like ones, each with the Z-normalized input shape (dashed) translated to its best overlap with the centroid. The colours denote the cluster index. (c) The confusion matrix against the generating labels, with the clusters in the order of their index: all $1000$ configurations are assigned according to their label. The two shapes are orthogonal after Z-normalization, $\mathrm{exSBD} = 1.000$.
    }
    \label{fig:superposed_centroids}
\end{figure}

A Markov chain may visit configurations of genuinely different shape, near a first-order boundary or otherwise, in which case, as noted in section~\ref{subsec:ensemble_loss}, the averaged correlator does not uniquely determine the constituent shapes or their relative populations. Whether such a mixture occurs in a given ensemble, and whether it signals a coexistence of phases rather than large fluctuations within one, depends on the dynamics and is not decided by the test below. The corresponding mock ensemble contains $1000$ configurations of which the first $500$ are sinusoidal, with $A = 0.8$ and $q = 2$, and the last $500$ are kinks with $\sigma_{\text{const}} = 1.0$; the labels are recorded at generation and are not used by the algorithm. With $K = 2$, the method returns two centroids, one of each shape, and an assignment that can be compared with the labels (figure~\ref{fig:superposed_centroids}). All $1000$ configurations are assigned according to their generating label, the $500$ sinusoidal ones to one cluster and the $500$ kinks to the other, so that the confusion matrix is diagonal once each cluster is labelled by the shape it contains.

After Z-normalization the two shapes are orthogonal, $\mathrm{exSBD} = 1.000$, with $\mathrm{Re} \, \mathrm{NCC}_{\boldsymbol{\tau}}$ vanishing at every shift, because the kink carries essentially no power at even wave numbers, as noted in section~\ref{subsec:kink_test}, while the sine sits at $q = 2$. A partition that respects the generating labels is therefore what the assignment step of section~\ref{subsec:algorithm} is expected to produce, and the test verifies that it does at the noise level of table~\ref{tab:benchmarks}; it is not evidence that shapes at a small mutual exSBD, such as the kink and the variance-matched sine of section~\ref{subsec:kink_test}, would be separated as cleanly. Whether a partition found in data reflects distinct shapes or a single one is decided by the criteria~\eqref{eq:criterion_A} and~\eqref{eq:criterion_B} rather than by the fit alone.

\begin{figure}[htbp]
    \centering
    \includegraphics[width=0.6\textwidth]{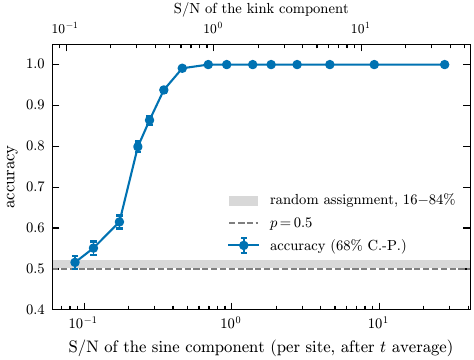}
    \caption{
        Accuracy of the assignment against the generating labels, the better of the two relabellings of the clusters being taken, for the mixed ensemble of figure~\ref{fig:superposed_centroids} regenerated at fifteen noise levels, $\sigma_\eta = 0.1$ to $32$, with the shapes, the positions and the labels unchanged. The lower abscissa is the signal-to-noise ratio per site of the sinusoidal component after the average over Euclidean time, from $27.7$ to $0.087$; the upper abscissa gives that of the kink component, larger by a factor $1.32$. Error bars are $68\,\%$ Clopper--Pearson intervals over the $1000$ configurations. The dashed line marks $0.5$, and the shaded band lies between the 16th and 84th percentiles, $0.503$ and $0.522$, of the accuracy of a random assignment, whose mean the choice of the better relabelling raises to $0.513$. Misassignments are first seen at the point $\sigma_\eta = 6$, a signal-to-noise ratio of $0.46$, and at $0.087$ ($\sigma_\eta = 32$) the accuracy cannot be distinguished from that of a random assignment.
    }
    \label{fig:accuracy_snr}
\end{figure}

How much of the outcome is owed to the noise being small is answered by repeating the test with the shapes, the positions and the labels held fixed and $\sigma_\eta$ alone varied, at fifteen values from $0.1$ to $32$; the signal-to-noise ratio per site of the sinusoidal component then runs from $27.7$ down to $0.087$. This is the one test of this section that is meant to locate where the method stops working, and it does so for the pair of orthogonal shapes above only. Figure~\ref{fig:accuracy_snr} shows the resulting accuracy, measured against the generating labels up to the relabelling of the two clusters, with $68\,\%$ Clopper--Pearson intervals. The ensemble above, with an accuracy of $1.000$, is the rightmost point of that curve, and the assignment remains without error down to a signal-to-noise ratio of $0.69$ ($\sigma_\eta = 4$). Misassignments are first seen at the point $\sigma_\eta = 6$, a ratio of $0.46$, where the accuracy is $0.991$; it falls to $0.938$ at $0.35$ ($\sigma_\eta = 8$) and to $0.799$ at $0.23$ ($\sigma_\eta = 12$). At $0.087$ ($\sigma_\eta = 32$) it is $0.516$, which cannot be distinguished from a random assignment: since the better of the two relabellings is taken, a partition of the $1000$ configurations unrelated to the labels has a mean accuracy of $0.513$ and reaches $0.516$ or more with probability $0.33$, whereas an accuracy of $0.551$ or more, as at $\sigma_\eta = 24$, would occur with probability $1.4 \times 10^{-3}$.

That the assignment survives per-site ratios well below unity is expected. A configuration is assigned through its projection onto a whole shape, so the relevant signal-to-noise ratio is that of the whole profile, the norm of its modulation divided by $\sigma_\eta / \sqrt{N_t}$, larger than the per-site one by $\sqrt{N_s} = 8$. For $\sigma_\eta \le 10$ the accuracy of the method agrees to within $0.004$, on the same configurations, with that of a matched filter that is given the two true shapes and assigns each configuration to the one with the larger $\max_{\boldsymbol{\tau}} \mathrm{Re} \, \mathrm{NCC}_{\boldsymbol{\tau}}$; at larger $\sigma_\eta$ it falls behind, presumably because the centroids, which have to be estimated from the data, are themselves degraded by the noise.

\subsection{A Two-Dimensional Test}
\label{subsec:2d_test}

\begin{figure}[htbp]
    \centering
    \includegraphics[width=\textwidth]{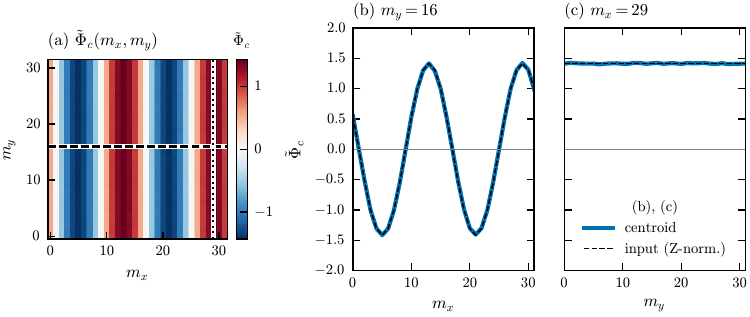}
    \caption{
        The centroid extracted with $K = 1$ from $1000$ two-dimensional stripe configurations, eq.~\eqref{eq:mock_stripe}, on a $32 \times 32$ spatial lattice, after Z-normalization: (a) colour map; (b) the row $m_y = 16$ and (c) the column $m_x = 29$, cut out of the centroid along the dashed and dotted lines in (a), each showing the centroid (solid) together with the Z-normalized input stripe (dashed) translated to its best overlap with the centroid; (b) and (c) share the line styles and the legend. The alignment uses the two-dimensional translation group $\mathbb{Z}_{N_x} \times \mathbb{Z}_{N_y}$.
    }
    \label{fig:stripe_2d}
\end{figure}

Nothing in section~\ref{sec:formulation} is specific to one spatial dimension, and a last ensemble checks this in practice. It consists of $1000$ stripe configurations on a $32 \times 32$ spatial lattice,
\begin{equation}
    \sigma(t, \mathbf{m}) = A \cos\left( \frac{2\pi q \, m_x}{N_x} + \varphi \right) + \eta(t, \mathbf{m}) \,,
    \label{eq:mock_stripe}
\end{equation}
with $A = 0.423$ and $q = 2$, each configuration carrying an independent integer translation $(\tau_x, \tau_y)$ drawn uniformly from $\mathbb{Z}_{N_x} \times \mathbb{Z}_{N_y}$, so that, unlike in eq.~\eqref{eq:mock_sine}, $\varphi$ takes only the discrete values $2\pi q \tau_x / N_x$. The alignment of eq.~\eqref{eq:alignment} is performed over that full group, all $N_x N_y$ shifts being obtained at once with fast Fourier transforms by eq.~\eqref{eq:cc_fourier}, and the centroid of eq.~\eqref{eq:centroid_znorm} reproduces the stripe, to $\mathrm{exSBD} = 2 \times 10^{-5}$, with no modulation along $y$ in the output any more than in the input (figure~\ref{fig:stripe_2d}). The noise-free input stripe is invariant under shifts along $y$, so its alignment is degenerate in that direction. Exact ties are broken by the convention stated below eq.~\eqref{eq:alignment}; for noisy configurations the fluctuations can select a shift along $y$, without changing the $y$-independent input signal. Modest as it is, this test is the practical support for the extensibility argument of section~\ref{subsec:extensibility}: the alignment is carried out in the lattice geometry, before the vectorization of section~\ref{subsec:centroid}, so that the only change with respect to the one-dimensional tests is the range of $\boldsymbol{\tau}$.

\section{Application to the $(1+1)$-Dimensional Gross--Neveu Model}
\label{sec:application}

We now apply the method to Monte Carlo ensembles of the GN$_2$ model~\cite{Lenz:2020bxk}: section~\ref{subsec:homo_inhomo} shows that $C(n)$ detects the inhomogeneous phase, section~\ref{subsec:kink_extraction} shows the shapes that it leaves undetermined, and section~\ref{subsec:multimodality} asks whether one shape suffices for each ensemble (supplementary material in appendix~\ref{app:gn2_supplement}).

\subsection{Lattice Simulation Setup and Preprocessing}
\label{subsec:setup}

The configurations were generated by our group with the lattice setup of ref.~\cite{Lenz:2020bxk} for naive fermions, with the parameters $N_f = 8$, $N_s = 64$ and $g^2 = 1.8132$ of table~1 of that reference, using a standard Hybrid Monte Carlo algorithm. Configurations separated by $10$ trajectories were stored, and the analysis uses the thermalized ones in the order of the Monte Carlo history. The scale is set by the vacuum expectation value $\sigma_0$ of the condensate. All six ensembles have $N_t = 24$, i.e.\ $T/\sigma_0 = 0.0985$, and differ in the chemical potential (table~\ref{tab:GN}); we refer to them by $\mu/\sigma_0$.

\begin{table}[htbp]
  \centering
  \caption{GN$_2$ ensembles analysed in section~\ref{sec:application}. Top: parameters common to all ensembles; $a$ is the lattice spacing, $N_f$, $N_s$ and $g^2$ are those of the naive-fermion setup of table~1 of ref.~\cite{Lenz:2020bxk}, and $T/\sigma_0 = 1/(N_t \, a\sigma_0)$. Bottom: chemical potential, phase as identified in section~\ref{subsec:homo_inhomo}, number $N_{\text{conf}}$ of thermalized configurations used, and number $N_B$ of jackknife blocks (no band is quoted for the homogeneous ensembles). All adopted restarts of the global fits of these ensembles were stopped by eq.~\eqref{eq:stopping_rule}. For $K = 1$ at $\mu/\sigma_0 = 0.388$ and $K = 3$ at $\mu/\sigma_0 = 0.630$ this required $r_{\max} = 2000$ instead of $200$ (table~\ref{tab:hyperparams}), with which all their restarts were stopped after at most $297$ and $231$ iterations; the three restarts of the other fits of these ensembles that reached $r_{\max} = 200$ were not adopted, and continuing them until they were stopped does not change the adopted restart.}
  \label{tab:GN}
  \vspace{0.2cm}
  \begin{tabular}{ccccccc}
    \hline
    fermion & $N_f$ & $N_s = L/a$ & $N_t = 1/(Ta)$ & $g^2$ & $a\sigma_0$ & $T/\sigma_0$ \\
    \hline \hline
    naive & $8$ & $64$ & $24$ & $1.8132$ & $0.4230367$ & $0.0985$ \\
    \hline
  \end{tabular}
  \\[0.4cm]
  \begin{tabular}{cclrc}
    \hline
    $a\mu$ & $\mu/\sigma_0$ & phase & $N_{\text{conf}}$ & $N_B$ \\
    \hline \hline
    $0.0000$ & $0.000$ & homogeneous & $4500$ & --- \\
    $0.1640$ & $0.388$ & homogeneous & $33679$ & --- \\
    $0.1845$ & $0.436$ & homogeneous & $26771$ & --- \\
    $0.2050$ & $0.485$ & inhomogeneous & $20001$ & $6$ \\
    $0.2460$ & $0.582$ & inhomogeneous & $14393$ & $6$ \\
    $0.2665$ & $0.630$ & inhomogeneous & $51000$ & $6$ \\
    \hline
  \end{tabular}
\end{table}

\paragraph{Euclidean-time average.}
As in section~\ref{sec:benchmarks}, the configurations are averaged over Euclidean time before the clustering ($N_v = 64$ instead of $1536$). Since a structure moving along $t$ within a configuration would be smeared by the average, we repeated the fit with $K = 1$ and $G = G_{\text{trans}}$ without it at $\mu/\sigma_0 = 0.582$ and $0.630$, aligning over the translations of the $N_t \times N_s$ lattice. The resulting centroid is nearly independent of $t$, and its time average agrees closely with the centroid from the time-averaged configurations (appendix~\ref{app:gn2_timeavg}), so that little of the normalized shape is lost there; $K \ge 2$ and the other ensembles have not been checked.

To characterize individual configurations we use $N_{\text{kink}}$, the number of sign changes along the periodic spatial direction of the time-averaged field of a configuration, counted after the Fourier modes with $|q| > 8$ have been removed and without subtracting the spatial mean. The configurations that enter the clustering are not filtered.

\paragraph{Choice of the group.}
Both groups of section~\ref{subsec:exsbd}, $G_{\text{trans}}$ and $G_{\text{trans}} \times \mathbb{Z}_2$, are used, each for a fixed set of quantities and for all ensembles alike. (i) The centroids for $K = 1$ of section~\ref{subsec:kink_extraction} and their bands are obtained with $G_{\text{trans}} \times \mathbb{Z}_2$, so that a shape is not combined with its sign reversal; the even harmonics of the centroid are then folded rather than averaged (section~\ref{subsec:exsbd}), which is taken up in section~\ref{subsec:kink_extraction}. (ii) The clustering into $K \ge 2$ clusters, the criteria~\eqref{eq:criterion_A} and~\eqref{eq:criterion_B}, including the centroid for $K = 1$ that enters $\rho_{\text{A}}$, and the null ensembles and positive controls that calibrate them use $G_{\text{trans}}$, so that the data and the synthetic ensembles are fitted with the same group; the calibration exists for this group only. (iii) The spectral concentration $R_1$ of section~\ref{subsec:homo_inhomo} is computed from the centroid for $K = 1$ with $G_{\text{trans}}$. (iv) Independently of $G$, the subsample centroids of the jackknife are matched and aligned with $G_{\text{al}} = G_{\text{trans}} \times \mathbb{Z}_2$ (section~\ref{subsec:error_estimation}). The choice in (iv) uses the discrete chiral symmetry of the GN$_2$ action: a shape and its sign reversal describe symmetry-equivalent states, and an independent subsample fit may select either. The comparison is therefore made modulo this symmetry, rather than because of an unfixed sign of the eigenvector. The setting \texttt{auto} of the implementation always includes the global sign, or the global phase for a complex field. The quantities of (i)--(iii) have also been computed with the other group and are quoted as cross-checks: the centroids for $K = 1$ with $G_{\text{trans}}$ in section~\ref{subsec:kink_extraction} and appendix~\ref{app:gn2_group}, $R_1$ with $G_{\text{trans}} \times \mathbb{Z}_2$ in the caption of table~\ref{tab:GN_indicators}, $\rho_{\text{A}}$ and $\rho_{\text{B}}$ with $G_{\text{trans}} \times \mathbb{Z}_2$ in section~\ref{subsec:multimodality}, and the partitions for $K \ge 2$ in appendix~\ref{app:gn2_group}. Values of the criteria obtained with $G_{\text{trans}} \times \mathbb{Z}_2$ are not strictly comparable with the null range.

\paragraph{Fit settings.}
All fits use the parameters of table~\ref{tab:hyperparams}, and the random seed is $20260907$ for the global fits of these ensembles and is derived from $20260906$ and $b$ for the jackknife subsample $(-b)$. There are three exceptions: $r_{\max} = 2000$ for the two global fits noted in tables~\ref{tab:hyperparams} and~\ref{tab:GN} and for the jackknife of the fit with $K = 3$ at $\mu/\sigma_0 = 0.630$; the stability test in $N_B$ of appendix~\ref{app:autocorr}, made with $K = 1$ and $G = G_{\text{trans}}$ at $\mu/\sigma_0 = 0.582$ from a global fit with a single restart and with different seeds; and the separate fits of the two orientations in appendix~\ref{app:gn2_group}, made with $G = G_{\text{trans}}$ and $5$ restarts. Error bands, from $N_B = 6$ blocks, are quoted for the inhomogeneous ensembles only. All restarts of the global fits with $G=G_{\text{trans}}\times\mathbb{Z}_2$ were stopped by eq.~\eqref{eq:stopping_rule}.

\paragraph{Jackknife checks and limitations.}
In all jackknife analyses reported here, no subsample was discarded. Whether every subsample fit stopped by eq.~\eqref{eq:stopping_rule} or reached $r_{\max}$ was recorded only for the fits with $G = G_{\text{trans}} \times \mathbb{Z}_2$ and for the fit with $K = 3$ at $\mu/\sigma_0 = 0.630$ with $G_{\text{trans}}$; all of these subsample fits were stopped by that rule. With $N_B=6$, the variance estimate has only five degrees of freedom, so that for Gaussian block averages the error itself has a relative statistical uncertainty of about $30\,\%$. The block-number test and autocorrelation measurements are given in appendix~\ref{app:autocorr}. The autocorrelation estimates do not reach a plateau for most observables, so convergence of the error estimates has not been established. Under a single-exponential model, errors of the scalar observables are underestimated by a factor of about $1.1$--$1.5$, with no established upper bound. This estimate has not been propagated into a correction of the centroid band, which is a non-linear function of the ensemble. All jackknife errors quoted below should be read with these limitations in mind.

\subsection{Detecting the Inhomogeneous Phase}
\label{subsec:homo_inhomo}

\begin{figure}[htbp]
    \centering
    \includegraphics[width=0.62\textwidth]{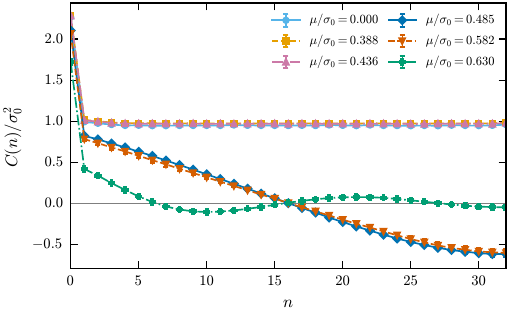}
    \caption{
        Spatial correlation $C(n)$ of eq.~\eqref{eq:correlator_def}, in units of $\sigma_0^2$, for the six ensembles of table~\ref{tab:GN}, computed on every time slice without the Euclidean-time average, for $n = 0, \dots, 32$. Colours and markers denote $\mu/\sigma_0$, as in all figures of this section in which ensembles are compared. The errors are delete-block jackknife errors with $N_B = 6$; with only six blocks, the errors themselves have a relative
statistical uncertainty of about $30\,\%$ under the Gaussian-block
approximation. Residual autocorrelation can additionally lead to
underestimated errors, as discussed in section~\ref{subsec:setup}. The values $C(0) = 1.72$--$2.29$ contain the site-local fluctuations.
    }
    \label{fig:gn2_correlator}
\end{figure}

Figure~\ref{fig:gn2_correlator} shows $C(n)$ of eq.~\eqref{eq:correlator_def}, in units of $\sigma_0^2$ and computed on every time slice without the Euclidean-time average; table~\ref{tab:GN_indicators} collects the numbers used below. The ensembles fall into the two classes of eq.~\eqref{eq:conventional_criterion}. At $\mu/\sigma_0 \le 0.436$, $C(n)$ decreases from $C(1) \approx 1.0$ to a positive plateau, $C(32) = 0.958$--$0.984$, and is almost entirely the pedestal $P = \frac{1}{N_s} \sum_n C(n) = 0.971$--$0.996$ that the spatial mean of a time slice contributes through eq.~\eqref{eq:offset_decomposition}. At $\mu/\sigma_0 \ge 0.485$, $C(n)$ has a negative minimum,
\begin{equation}
    C_{\text{min}} = -0.620 \pm 0.030 \,, \qquad -0.606 \pm 0.042 \,, \qquad -0.108 \pm 0.015 \,,
    \label{eq:gn2_cmin}
\end{equation}
at $n = 31$, $32$ and $10$ for $\mu/\sigma_0 = 0.485$, $0.582$ and $0.630$, while the pedestal $P$ is smaller than $|C_{\text{min}}|$ by factors of $12$, $13$ and $1.9$. The pedestal corresponds to a root mean square of $0.22$--$0.24\,\sigma_0$ for the spatial mean of a time slice. These ensembles are not in the regime illustrated by the mock backgrounds of section~\ref{subsec:biased_test}, in which the uniform contribution reverses the sign of $C_{\text{min}}$. For a zero-mean profile antisymmetric under a half-period translation, section~\ref{subsec:kink_difficulty} gives $C_{\text{min}}/C(0)=-1$, both for a sine wave and for the idealized kink--antikink profile. A deep negative minimum therefore does not by itself distinguish these shapes. The ratio $C_{\text{min}}/C(0) \approx -0.29$ of the first two ensembles is not comparable with the value $-1$ derived there, since $C(0)$ contains site-local fluctuations not averaged over time (table~\ref{tab:GN_indicators} gives $C_{\text{min}}/C(1)$). The position of the minimum indicates the wavelength but not the shape: at the first two ensembles the wavelength is the length of the lattice, the longest it admits, so that it may be set by the volume, and at the third it is much shorter. The shallow minimum at $\mu/\sigma_0 = 0.630$ could come from a modulation of small amplitude or from a mixture of shapes.

\begin{table}[htbp]
  \centering
  \caption{Phase indicators of the six ensembles. $C(1)$, the minimum $C_{\text{min}}$ with its separation $n$ in parentheses, and the pedestal $P = \frac{1}{N_s} \sum_n C(n)$ are in units of $\sigma_0^2$, with jackknife errors ($N_B = 6$) in the last digits; for the homogeneous ensembles $C(n)$ has no well-defined minimum and $C_{\text{min}}$ is not quoted. $\bar{m}/\sigma_0$, with its jackknife error, and $\bar{s}/\sigma_0$ are the restoration constants~\eqref{eq:restoration_constants} for $K = 1$. $f_0$ is the fraction of configurations with $N_{\text{kink}} = 0$; for the inhomogeneous ensembles the distribution of $N_{\text{kink}}$ is shown in figure~\ref{fig:gn2_kink_hist}. $R_1$ is the fraction of the power of the centroid for $K = 1$ with $G = G_{\text{trans}}$, after subtraction of its mean, that is carried by its strongest Fourier mode, the power being summed over the modes $q = 1, \dots, N_s/2$, each counted once; for the centroids with $G = G_{\text{trans}} \times \mathbb{Z}_2$ of section~\ref{subsec:kink_extraction} the corresponding fractions are $92.6$, $91.8$ and $62.3\,\%$. $K = 1$ fits of white noise and of Gaussian noise with the power spectrum of the homogeneous data give $4.6$--$5.6\,\%$ and $15.8$--$18.4\,\%$.}
  \label{tab:GN_indicators}
  \vspace{0.2cm}
  \small
  \setlength{\tabcolsep}{4pt}
  \begin{tabular}{ccccccccc}
    \hline
    $\mu/\sigma_0$ & $C(1)$ & $C_{\text{min}}$ ($n$) & $C_{\text{min}}/C(1)$ & $P$ & $\bar{m}/\sigma_0$ & $\bar{s}/\sigma_0$ & $f_0$ & $R_1$ (\%) \\
    \hline \hline
    $0.000$ & $0.991(17)$ & --- & --- & $0.971(17)$ & $0.9727(91)$ & $0.247$ & $1.000$ & $15.1$ \\
    $0.388$ & $1.015(4)$ & --- & --- & $0.996(4)$ & $0.9861(20)$ & $0.245$ & $1.000$ & $14.3$ \\
    $0.436$ & $1.010(6)$ & --- & --- & $0.991(6)$ & $0.9835(31)$ & $0.246$ & $1.000$ & $15.1$ \\
    $0.485$ & $0.821(8)$ & $-0.620(30)$ ($31$) & $-0.755(36)$ & $0.0520(63)$ & $+0.015(61)$ & $0.889$ & --- & $96.3$ \\
    $0.582$ & $0.781(24)$ & $-0.606(42)$ ($32$) & $-0.777(34)$ & $0.0473(56)$ & $-0.016(45)$ & $0.866$ & --- & $93.2$ \\
    $0.630$ & $0.418(14)$ & $-0.108(15)$ ($10$) & $-0.258(37)$ & $0.0577(51)$ & $+0.056(34)$ & $0.589$ & --- & $69.4$ \\
    \hline
  \end{tabular}
\end{table}

For the homogeneous ensembles, the subsample variation is expected to be dominated by optimization noise, so no error band is quoted (section~\ref{subsec:error_estimation}). We do not show their centroids, because they carry no information on a modulation: by eq.~\eqref{eq:znorm_properties} the restored centroid fluctuates about $\bar{m}$ with root mean square $\bar{s}$ whatever the data, in a normalized noise pattern. The homogeneous phase is identified instead by three numbers of table~\ref{tab:GN_indicators}: the offset $\bar{m}/\sigma_0 = 0.973$--$0.986$; the fraction $f_0 = 1$ of configurations with $N_{\text{kink}} = 0$; and the spectral concentration $R_1 = 14$--$15\,\%$ of the centroid, comparable with that of $K = 1$ fits of Gaussian noise with the power spectrum of these data and far below that of the inhomogeneous ensembles.

That $\bar{m}/\sigma_0$ is close to one reflects the absence, within these chains, of transitions between the sectors related by $\sigma \to -\sigma$, every configuration having a positive spatial mean; this does not contradict $\langle \sigma \rangle = 0$ in a finite volume (section~\ref{subsec:biased_states}). A connected correlator built with the sample average subtracts the square of that average, but retains a contribution from fluctuations of the spatial mean between configurations and time slices (section~\ref{subsec:biased_states}). Equation~\eqref{eq:conventional_criterion} uses the disconnected $C(n)$.

\subsection{Shapes of the Inhomogeneous Configurations}
\label{subsec:kink_extraction}

\begin{figure}[htbp]
    \centering
    \includegraphics[width=\textwidth]{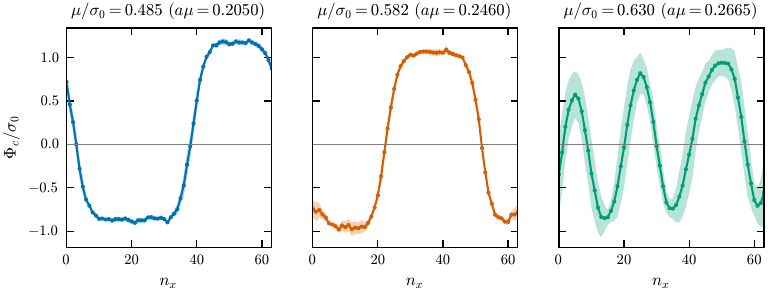}
    \caption{
        Centroids $\Phi_c/\sigma_0$ for $K = 1$ of the three inhomogeneous ensembles as functions of the spatial site $n_x$, obtained with $G = G_{\text{trans}} \times \mathbb{Z}_2$ from the time-averaged configurations, with the band $\pm e/\sigma_0$ of eq.~\eqref{eq:error_band} from the jackknife of the same fits ($N_B = 6$). The site-averaged band is $\langle e \rangle/\sigma_0 = 0.046$, $0.047$ and $0.285$, i.e.\ $0.052$, $0.055$ and $0.481$ of the root mean square of the centroid. The position of a centroid on the periodic lattice has no meaning, and its amplitude is fixed by $\bar{s}$ (section~\ref{subsec:homo_inhomo}). At $\mu/\sigma_0 = 0.485$ and $0.582$ the even Fourier components of these centroids, and with them the asymmetry of the profiles, contain the two orientations of the unequal domain lengths of the individual configurations folded onto each other by the alignment with a sign, and do not indicate an asymmetry of the ensemble (section~\ref{subsec:kink_extraction}). At $\mu/\sigma_0 = 0.630$ the centroid is a blend of several shape families (section~\ref{subsec:multimodality}). These centroids are not those from which $\rho_{\text{A}}$ in table~\ref{tab:GN_criteria} is computed.
    }
    \label{fig:gn2_centroids_K1}
\end{figure}

\begin{figure}[htbp]
    \centering
    \includegraphics[width=\textwidth]{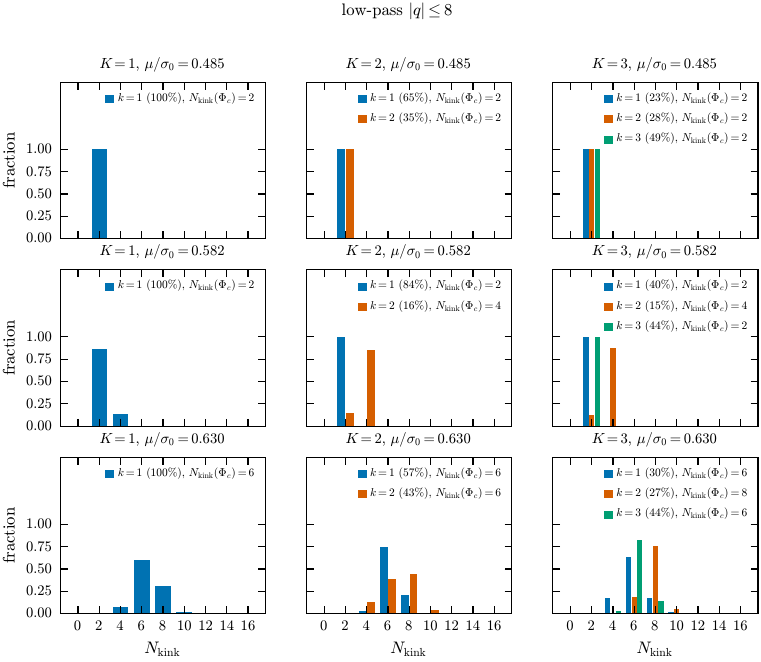}
    \caption{
        Distribution of $N_{\text{kink}}$, the number of sign changes of the time-averaged field of a configuration after the removal of the Fourier modes with $|q| > 8$, for the three inhomogeneous ensembles (rows) and the clusters of the fits with $K = 1$, $2$ and $3$ with $G = G_{\text{trans}}$ (columns). Bars denote clusters, coloured by the cluster index in a fixed order ($k = 1$, $2$, $3$: blue, orange, green), a colour code that differs from that of $\mu/\sigma_0$ in the other figures. The legend gives the population of each cluster and $N_{\text{kink}}(\Phi_c)$, the same count applied to the restored centroid of the cluster with the same cut-off, a single value rather than a distribution over configurations. In the three homogeneous ensembles all configurations have $N_{\text{kink}} = 0$. $N_{\text{kink}}$ describes smoothed configurations and is not a conserved or topological quantity; the cut-offs $|q| \le 4$ and $|q| \le 16$ are shown in appendix~\ref{app:gn2_kink_robust}.
    }
    \label{fig:gn2_kink_hist}
\end{figure}

Figure~\ref{fig:gn2_centroids_K1} shows the centroids for $K = 1$ of the inhomogeneous ensembles, obtained with $G = G_{\text{trans}} \times \mathbb{Z}_2$, with the band of eq.~\eqref{eq:error_band} from the jackknife of the same fits. Figure~\ref{fig:gn2_kink_hist} shows the distribution of $N_{\text{kink}}$.

At $\mu/\sigma_0 = 0.485$ and $0.582$ the centroid changes sign twice, and so do the configurations: $N_{\text{kink}} = 2$ for all configurations at $0.485$ and for $86.5\,\%$ at $0.582$, the others having $N_{\text{kink}} = 4$. Neither profile is a single harmonic. The strongest Fourier mode carries $92.6\,\%$ and $91.8\,\%$ of the power of these centroids ($96.3\,\%$ and $93.2\,\%$ for the centroids with $G_{\text{trans}}$, table~\ref{tab:GN_indicators}), and the odd ratios $h_j = |\hat{c}(jq)| / |\hat{c}(q)|$ of the Fourier moduli of the centroid at odd multiples of its strongest wave number $q$ (appendix~\ref{app:gn2_supplement}), $h_3 = 0.19$ and $0.23$, do not depend on the group and exceed the first-order uncertainty of $0.044$--$0.057$ that the band implies for them. The same holds for the power outside the strongest mode, for which the band corresponds to an uncertainty of about $0.4\,\%$ of the power. That power cannot be read off $C(n)$: it enters $C(n)$, eq.~\eqref{eq:power_spectrum}, only together with that of the fluctuations. We describe these profiles as non-sinusoidal and make no further identification. The values $\Phi_c/\sigma_0 \approx \pm 1$ away from the sign changes are fixed by $\bar{s}$ (section~\ref{subsec:homo_inhomo}) and are not a measurement of the condensate within a domain.

The centroids for $G_{\text{trans}} \times \mathbb{Z}_2$ are in addition asymmetric: the third moment $\langle c^3 \rangle = N_s^{-1} \sum_n c(n)^3$ over the sites $n$ of the Z-normalized centroid $c = \tilde{\Phi}_c$, which vanishes for every profile mapped onto minus itself by a translation, the sine wave and the idealized kink--antikink lattice of section~\ref{subsec:kink_difficulty} alike, has magnitude $0.263$ and $0.183$ (its sign depends on the representative under $\mathbb{Z}_2$); correspondingly, $\max_\tau \mathrm{NCC}_\tau(c, -c) = 0.919$ and $0.944$ rather than one. To first order, the band corresponds to an uncertainty of $\langle c^3 \rangle$ of at most $0.08$ and $0.06$, whatever the correlation between the sites.

This asymmetry does not indicate a preferred asymmetric shape, and the band does not bear on the question. In the individual configurations the positive and the negative domain have different lengths: for the configurations with $N_{\text{kink}} = 2$, the length $d_+$ of the positive domain, measured on the low-pass filtered field used for $N_{\text{kink}}$, deviates from $N_s/2$ by $5.2$ and $4.2$ lattice spacings (root mean square), and it is anticorrelated with the third moment of the configuration (correlation coefficients $-0.99$ and $-0.97$). The alignment with a sign folds the two orientations of this difference onto each other (section~\ref{subsec:exsbd}), so that the centroid reflects the typical size of $|d_+ - N_s/2|$ rather than an imbalance of the ensemble; with $G_{\text{trans}} \times \mathbb{Z}_2$ the centroid is by construction the same for the ensemble symmetrized under $\sigma \to -\sigma$. Like $\langle c^3 \rangle$, the unequal domain lengths of the individual configurations are a property of the relative phases of their Fourier modes, which $C(n)$ does not retain; what the alignment with a sign makes visible is their typical size rather than their sign.

Mock ensembles of kink--antikink profiles with $d_+$ drawn from a single Gaussian centred at $N_s/2$, with a width taken from the data, and with noise of the spectrum of the residuals, fitted in the same way, approximately reproduce these values: at $\mu/\sigma_0 = 0.485$ they give $\langle c^3 \rangle = 0.26$ and $h_2 = 0.17$, against $\langle c^3 \rangle = 0.263$ and $h_2 = 0.183$ in the data, and at $0.582$ they give $\langle c^3 \rangle = 0.22$ and $h_2 = 0.16$, against $\langle c^3 \rangle = 0.183$ and $h_2 = 0.153$ in the data. At $0.582$ the mock thus reproduces $h_2$ but exceeds the measured $\langle c^3 \rangle$ by about $20\,\%$, a difference the band does not resolve; we have not identified the cause, the candidates being the $13.5\,\%$ of configurations with $N_{\text{kink}} = 4$, which the mock does not contain, and the mean of the measured $d_+$ lying $2.0$ lattice spacings below $N_s/2$ ($P(d_+ > N_s/2) = 0.27$, the same phenomenon as the unbalanced orientations of appendix~\ref{app:gn2_group}), so that a mock with a Gaussian centred at $N_s/2$ and the measured width enters this shift as a fluctuation in both directions. The centroid for $G_{\text{trans}}$ at $\mu/\sigma_0 = 0.485$, where the two orientations occur in nearly equal proportions, is by contrast nearly symmetric, $\langle c^3 \rangle = -0.03$. Whether the distribution of $d_+$ has one maximum or more cannot be decided from the present histories (section~\ref{subsec:limitations_analysis}).

At $\mu/\sigma_0 = 0.630$ the centroid changes sign six times, with unequal extrema and spacings, carries $62.3\,\%$ of its power in the strongest Fourier mode, and carries a band of nearly half the signal. It is not a typical configuration: most configurations have $N_{\text{kink}} = 6$ or $8$ (appendix~\ref{app:gn2_kink_robust}), and they fall into families of different shape (section~\ref{subsec:multimodality}), of which this centroid is a blend; we describe it neither as sinusoidal nor as a profile with six walls.

\paragraph{Dependence on the group.}
The two groups differ in how they treat the two orientations of the unequal domain lengths discussed above, which occur in nearly equal proportions at $\mu/\sigma_0 = 0.485$: $G_{\text{trans}}$ averages them into a nearly symmetric profile, and $G_{\text{trans}} \times \mathbb{Z}_2$ folds them into an asymmetric one. With $G = G_{\text{trans}}$ the centroid can combine a pattern with its sign reversal (section~\ref{subsec:exsbd}). We compared the centroids of the two groups, aligned by the best translation and sign, through the pointwise difference $\Delta$ and the combined band $\sigma_{\text{comb}} = (e_{\text{T}}^2 + e_{\mathbb{Z}_2}^2)^{1/2}$ of the two fits, both in units of $\bar{s}$. The difference exceeds the band only at $\mu/\sigma_0 = 0.485$, where $\mathrm{rms}\, \Delta / \mathrm{rms}\, \sigma_{\text{comb}} = 1.63$, the largest $|\Delta|/\sigma_{\text{comb}}$ at a site is $3.08$, and $18$ of the $64$ sites exceed $2 \sigma_{\text{comb}}$; the values at the other two ensembles, and why the difference exceeds the combined band at $0.485$ only, are given in appendix~\ref{app:gn2_group}. Since $\sigma_{\text{comb}}$ neglects the expected positive correlation of two fits of the same configurations, the excess at $0.485$ is conservative, whereas the agreement at $0.582$ and $0.630$ is not a bound on the difference. In terms of the shape correlation~\eqref{eq:shape_correlation} (appendix~\ref{app:gn2_group}) the shapes differ most at $0.630$, and remain within the band there only because it is broad.

\subsection{Multiple Shapes within One Ensemble}
\label{subsec:multimodality}

\begin{figure}[htbp]
    \centering
    \includegraphics[width=\textwidth]{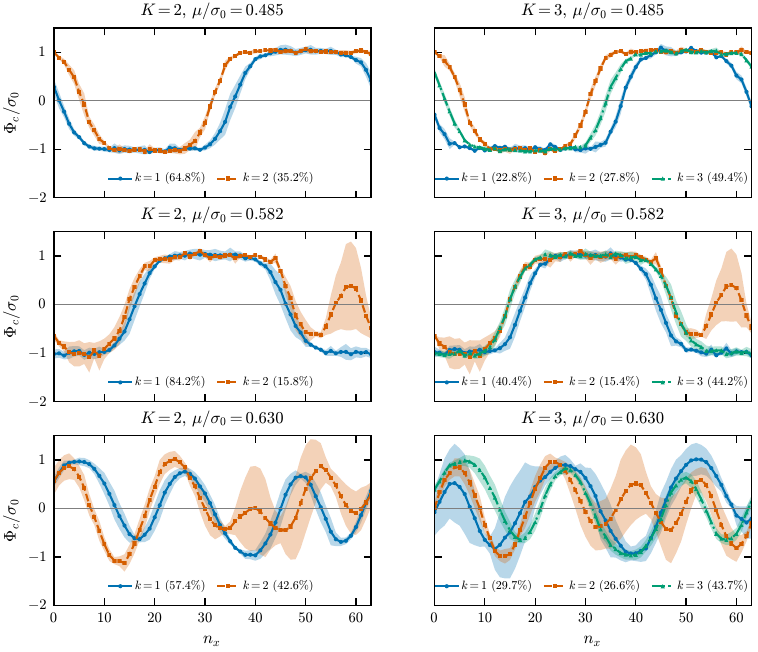}
    \caption{
        Centroids $\Phi_c^{(k)}/\sigma_0$ for $K = 2$ (left) and $K = 3$ (right) of the three inhomogeneous ensembles (rows), fitted with $G = G_{\text{trans}}$, with the band $\pm e^{(k)}/\sigma_0$ of eq.~\eqref{eq:error_band} from the jackknife of the same fits ($N_B = 6$) and the cluster populations in the legend; colours denote clusters as in figure~\ref{fig:gn2_kink_hist}. For display, every centroid and its band are translated to the best overlap, $\max_\tau \mathrm{Re}\, \mathrm{NCC}_\tau$, of the centroid with the centroid for $K = 1$ of the fit with $G_{\text{trans}}$ of the same ensemble, without sign reversal; since translations are quotiented out, the positions have no meaning. The subsample centroids are matched and aligned with $G_{\text{al}} = G_{\text{trans}} \times \mathbb{Z}_2$ (section~\ref{subsec:error_estimation}), so that a band is the uncertainty of the shape of a cluster irrespective of its orientation. For $K = 3$ at $\mu/\sigma_0 = 0.485$, two clusters are mapped onto each other by $\sigma \to -\sigma$ and a translation (section~\ref{subsec:multimodality}); modulo $G_{\text{al}}$ they are nearly identical, so that the matching~\eqref{eq:jk_matching} is nearly degenerate and their bands do not tell which of the two orientations a subsample cluster was assigned to. The bands are not used to compare values of $K$, nor clusters with each other (section~\ref{subsec:error_estimation}).
    }
    \label{fig:gn2_centroids_K23}
\end{figure}

Table~\ref{tab:GN_criteria} gives the criteria~\eqref{eq:criterion_A} and~\eqref{eq:criterion_B} with $\mathcal{K} = \{ 2, 3 \}$, and figure~\ref{fig:gn2_centroids_K23} the centroids for $K = 2$ and $3$ with their bands, all with $G = G_{\text{trans}}$. The null ensembles of section~\ref{subsec:algorithm} are built from the ensemble at $\mu/\sigma_0 = 0.582$, with the three noise models of appendix~\ref{app:null} and two realizations of each, and used for all three ensembles, conservatively at $0.485$, where a null built from that ensemble has a higher lower edge of $\rho_{\text{B}}$; the positive controls are built from the two centroids for $K = 2$ at $0.630$, with two noise models and two realizations of each. $\rho_{\text{A}}$ is computed from the centroid for $K = 1$ with $G_{\text{trans}}$, as the calibration requires.

Two properties of this calibration limit what the criteria can carry, and we state them before using them. First, each realization requires a complete set of fits with $K = 1$, $2$ and $3$ of an ensemble of the size of the selected one, so that the six null and four positive-control realizations already amount to thirty global fits; the number of realizations was set by the computational cost and not by a target precision, and the ranges they give are minima and maxima over a handful of realizations rather than probabilities. Second, a null ensemble contains one fixed profile at random translations, with an amplitude and noise taken from the data, and therefore no variants of that profile: an ensemble whose configurations differ from one another in ways that the noise model does not reproduce will give a $\rho_{\text{B}}$ below the null range even if it carries a single shape family. Both points bear on the ensembles at $\mu/\sigma_0 = 0.485$ and $0.582$ below.

\begin{table}[htbp]
  \centering
  \caption{Criteria~\eqref{eq:criterion_A} and~\eqref{eq:criterion_B} for the inhomogeneous ensembles, from fits with $G = G_{\text{trans}}$ and $K = 2, 3$; $\rho_{\text{B}}$ is the smaller of the two columns for $K = 2$ and $K = 3$. The column $\rho_{\text{A}}$ ($\times \mathbb{Z}_2$) uses fits with $G_{\text{trans}} \times \mathbb{Z}_2$ for all $K$ and is not strictly comparable with the null range, which was obtained with $G_{\text{trans}}$. Null: minimum and maximum over the six null realizations built from the ensemble at $\mu/\sigma_0 = 0.582$; positive control: over the four realizations built from the fit with $K = 2$ at $\mu/\sigma_0 = 0.630$. The value for $K = 2$ alone at $\mu/\sigma_0 = 0.485$ is not compared with the common null, which does not represent that ensemble for $K = 2$ alone. The populations (lower part) are given in per cent, in the order of the cluster index.}
  \label{tab:GN_criteria}
  \vspace{0.2cm}
  \small
  \begin{tabular}{lcccc}
    \hline
    $\mu/\sigma_0$ & $\rho_{\text{A}}$ & $\rho_{\text{A}}$ ($\times \mathbb{Z}_2$) & $\rho_{\text{B}}$, $K = 2$ & $\rho_{\text{B}}$, $K = 3$ \\
    \hline \hline
    $0.485$ & $0.9933$ & $0.9904$ & $0.9819$ & $0.9338$ \\
    $0.582$ & $0.9938$ & $0.9964$ & $0.9101$ & $0.8894$ \\
    $0.630$ & $0.8317$ & $0.9236$ & $0.5869$ & $0.4949$ \\
    \hline
    null ($6$) & $[0.9918, 0.9988]$ & --- & $[0.9762, 0.9968]$ & $[0.9722, 0.9958]$ \\
    positive control ($4$) & $[0.9050, 0.9273]$ & --- & $[0.5888, 0.6038]$ & $[0.5717, 0.6074]$ \\
    \hline
  \end{tabular}

  \vspace{0.3cm}
  \begin{tabular}{lcc}
    \hline
    $\mu/\sigma_0$ & populations, $K = 2$ & populations, $K = 3$ \\
    \hline \hline
    $0.485$ & $64.8$, $35.2$ & $22.8$, $27.8$, $49.4$ \\
    $0.582$ & $84.2$, $15.8$ & $40.4$, $15.4$, $44.2$ \\
    $0.630$ & $57.4$, $42.6$ & $29.7$, $26.6$, $43.7$ \\
    \hline
  \end{tabular}
\end{table}

At $\mu/\sigma_0 = 0.630$ the ensemble carries more than one shape family. The evidence we rely on is the distribution of $N_{\text{kink}}$, which is counted on the configurations themselves and does not involve the criteria. The configurations carry different numbers of sign changes, $N_{\text{kink}} = 6$, $8$, $4$ and $10$ for $59.6$, $31.2$, $6.9$ and $2.2\,\%$ of them (appendix~\ref{app:gn2_kink_robust}), and the clustering separates them: for $K = 2$ the larger cluster consists mainly of configurations with $N_{\text{kink}} = 6$ and the smaller one has a broader distribution peaked at $8$; for $K = 3$ a family dominated by $N_{\text{kink}} = 8$ separates, although the distributions of $N_{\text{kink}}$ in the clusters overlap (figure~\ref{fig:gn2_kink_hist}, appendix~\ref{app:gn2_kink_robust}). This multimodality does not depend on the group: with $G_{\text{trans}} \times \mathbb{Z}_2$ the partitions agree for about $85\,\%$ of the configurations, and the clusters are not sign-reversed partners and keep their composition in $N_{\text{kink}}$ (appendix~\ref{app:gn2_group}). The centroid for $K = 1$ is therefore a blend of more than one shape. The criteria agree with this reading: both lie far outside the null range, $\rho_{\text{A}} = 0.832$ and $\rho_{\text{B}} = 0.495$, so that the clusters differ in shape and not only in their composition in $N_{\text{kink}}$. Their quantitative use is more limited. The positive controls being built from this ensemble, the comparison with them checks the consistency of a two-family description rather than calibrating it; for $K = 3$, $\rho_{\text{B}} = 0.495$ lies below the $0.572$--$0.607$ of the controls, which two families built in this way do not reproduce. Since a control contains two fixed profiles and no variants of either, this is consistent with more than two families, but it does not establish their number.

At $\mu/\sigma_0 = 0.582$, $\rho_{\text{B}} = 0.889$ lies below the null range, by $0.083$, while $\rho_{\text{A}} = 0.9938$ lies inside it. Here too the statement we make rests on $N_{\text{kink}}$: one dominant shape, that of the centroid for $K = 1$, is accompanied by a minority of about $16\,\%$ in which most configurations have $N_{\text{kink}} = 4$ instead of $2$ (appendix~\ref{app:gn2_kink_robust}), a difference of the configurations themselves. The value of $\rho_{\text{B}}$ is consistent with it but does not by itself distinguish a second family from variants of one profile, which the null does not contain. The partition hardly depends on the group for $K = 2$, but noticeably for $K = 3$ (appendix~\ref{app:gn2_group}).

At $\mu/\sigma_0 = 0.485$, $\rho_{\text{B}} = 0.934$ is also below the null range, by $0.038$, but we do not read this as evidence of distinct shape families. All clusters for $K \ge 2$ have $N_{\text{kink}} = 2$, two of the clusters for $K = 3$ are mapped onto each other by $\sigma \to -\sigma$ and a translation up to a shape correlation of $0.997$, and the fits with $G_{\text{trans}} \times \mathbb{Z}_2$ reassign about a third of the configurations (appendix~\ref{app:gn2_group}). The clusters thus differ in orientation and in variants of one profile with two sign changes, which a null built from one orientation of one fixed profile does not contain. Orientation alone does not account for the value: with $G_{\text{trans}} \times \mathbb{Z}_2$, $\rho_{\text{B}} = 0.938$ remains below the null range (not strictly comparable, the null being obtained with $G_{\text{trans}}$), and which variation of the profile is responsible has not been determined; a candidate is the varying distance between the two sign changes (section~\ref{subsec:kink_extraction}). With $G_{\text{trans}}$, $\rho_{\text{A}} = 0.9933$ lies inside the null range, whereas with $G_{\text{trans}} \times \mathbb{Z}_2$ it is $0.9904$, below that range (again not strictly comparable, the null being calibrated with $G_{\text{trans}}$ only); we therefore do not use $\rho_{\text{A}}$ as evidence that the centroid for $K = 1$ captures the dominant shape, and interpret the criteria together with the orientation dependence and the configuration-level diagnostics described above.

The populations are not equilibrium mixing ratios. At $\mu/\sigma_0 = 0.630$ the fraction of the larger cluster for $K = 2$ varies between $0.33$ and $0.92$ over the six blocks of the history, and that of the family dominated by $N_{\text{kink}} = 8$ between $0.001$ and $0.482$; at $0.582$ the minority appears in two of the six blocks only. Each block contains only about one to seven effectively independent configurations, judged by the integrated autocorrelation time of the cluster membership (table~\ref{tab:tau_int}), so that this variation is not a measure of an equilibrium spread. Whether the families reflect coexisting structures or slow fluctuations of the Markov chain is not decided by these data.

\section{Discussion}
\label{sec:discussion}

\subsection{The Centroid as a Morphological Descriptor}
\label{subsec:centroid_descriptor}

The centroid complements $C(n)$ by describing the real-space shape left undetermined by the power spectrum and ensemble average (sections~\ref{subsec:kink_difficulty} and~\ref{subsec:ensemble_loss}). The GN$_2$ application illustrates both a departure from a single harmonic and the need for more than one representative shape (sections~\ref{subsec:kink_extraction} and~\ref{subsec:multimodality}). We therefore regard the centroid as a morphological descriptor of an ensemble, and not as an order parameter. It is neither the expectation value of a local operator nor the one-point function in a fixed frame, which is independent of the position by eq.~\eqref{eq:onepoint_average}; it is a representative normalized shape, obtained after the alignment has removed the position of every configuration, and it depends on choices that are part of the analysis: the group $G$, as shown in section~\ref{subsec:kink_extraction}, the number of clusters $K$, the Z-normalization and, here, the Euclidean-time average. The constants $\bar{s}$ and $\bar{m}$ carry information that the shape does not, such as the offset that identified the homogeneous phase in section~\ref{subsec:homo_inhomo}. Whether the centroid encodes a symmetry of the ground state cannot be decided from a normalized shape alone, and we make no such claim. The possible relation of the profiles with two sign changes to the baryons of the model is discussed in section~\ref{subsec:baryons}.

\subsection{Relation to the Baryons of the GN$_2$ Model}
\label{subsec:baryons}

At $\mu/\sigma_0 = 0.485$ and $0.582$ the dominant profiles have a single period in the box, with one kink and one antikink. With baryon number normalized as the total fermion number divided by $N_f$, one period of the zero-temperature, large-$N_f$ kink--antikink crystal carries baryon number one, with one half associated with each sign change~\cite{Thies:2003kk, Schnetz:2004vr, Lenz:2020cuv}; at non-zero temperature the density per period is not quantized. In infinite volume a single baryon is a static self-consistent fermion bag of kink--antikink shape~\cite{Feinberg:2003qz, Thies:2006ti}. On the lattice at $N_f = 8$, and at larger chemical potentials, the baryon number of a configuration was found to be almost identical to the number of periods of its modulation, in a box of nearly the same size as ours ($L \approx 25.8/\sigma_0$ there against $L \approx 27.1/\sigma_0$ here)~\cite{Lenz:2020cuv}. The configurations with $N_{\text{kink}} = 2$ are thus compatible with an interpretation as a single baryonic modulation in the box. The shape alone does not distinguish an isolated baryon from a periodic structure constrained to one period by the finite volume, or establish a crystal in infinite volume. We have not measured the baryon number, and the agreement just quoted was established away from the region studied here, so that this reading is an extrapolation. In a finite periodic box the wavelength of a modulation is restricted to $L/\nu$ with integer $\nu$~\cite{Lenz:2020cuv}, and at large $N$, if the condensate is assumed to carry a single momentum, the inhomogeneous region accordingly splits into sub-phases labelled by that integer~\cite{Narayanan:2020uqt}; the two ensembles here would fall in the first of them. Finite-size effects from the incommensurability of the box with the modulation are sizeable, and large volumes are needed~\cite{deForcrand:2006zz, Lenz:2020cuv}, so that we draw no conclusion about the infinite-volume state from a single volume. The centroid asymmetry must also be interpreted with the folding induced by the alignment in mind (section~\ref{subsec:kink_extraction}).

\subsection{Practical Considerations on Z-Normalization and Noise}
\label{subsec:practical_considerations}

Where the field has no macroscopic modulation, the Z-normalization of section~\ref{subsec:exsbd} scales the fluctuations about the spatial mean to unit variance, and the clustering operates on noise. The centroid is then a normalized fluctuation pattern whose restored amplitude is $\bar{s}$ by construction. Spatially correlated noise can itself produce a pattern under alignment, so the existence or apparent stability of a centroid alone does not establish a macroscopic modulation (sections~\ref{subsec:error_estimation} and~\ref{subsec:homo_inhomo}). In section~\ref{sec:application} the homogeneous ensembles were therefore recognized by quantities other than the shape (section~\ref{subsec:homo_inhomo}). This sufficed because they lie deep in the broken phase, with $\bar{m}/\sigma_0 = 0.97$--$0.99$; it would not in a symmetric phase or near a phase boundary, where both the offset and the modulation are small.

When the method is used to map a phase diagram, we therefore recommend a pre-screening step that decides, before the clustering, whether a configuration carries a modulation of significant amplitude, for instance through its spatial variance evaluated after a mild smoothing such as the low-pass filter used for $N_{\text{kink}}$. Configurations compatible with thermal noise would be set aside from shape clustering, and the clustering applied to those with a detectable modulation. Distinguishing a homogeneous broken regime from a symmetric one would require the uniform component and other ensemble observables in addition to this screening. The screening threshold requires a null model of the noise, like the one used for $R_1$ and for the criterion~\eqref{eq:criterion_B}. We have not implemented such a test: the procedure of section~\ref{subsec:homo_inhomo} is a check on ensembles whose phase was already clear from $C(n)$, not a validated screening.

\subsection{Extension to Higher Dimensions}
\label{subsec:extensibility}

The alignment and the dissimilarity of section~\ref{sec:formulation} are defined for any number of lattice directions, and the vectorization of section~\ref{subsec:centroid} turns the centroid into an ordinary matrix eigenvalue problem in any dimension. The method has been run beyond one dimension twice in this paper: on the two-dimensional mock ensemble of section~\ref{subsec:2d_test}, and on the $(1+1)$-dimensional space-time field of the GN$_2$ model without the Euclidean-time average, with $N_v = 1536$ (section~\ref{subsec:setup}). The alignment scales well, at a cost $\mathcal{O}(N_v \log N_{\text{lat}})$ per pair of configurations with fast Fourier transforms.

The difficulty in $3+1$ dimensions is the size of the eigenvalue problem. Already after the average over Euclidean time, a $32^3$ spatial lattice gives $N_v = 32768$, so that $\hat{S}$ has about $1.1 \times 10^9$ entries, or $8.6$~GB in double precision for a real field, and a full diagonalization costs $\mathcal{O}(N_v^3)$ operations. The Gram route of appendix~\ref{app:gram} helps only when $|\mathcal{C}| \ll N_v$, which need not hold for ensembles as large as those of section~\ref{sec:application}, with $N_{\text{conf}}$ up to $51000$ (table~\ref{tab:GN}). Only the leading eigenvector is needed, however, and $\hat{S}$ need not be formed at all: its action on a vector,
\begin{equation}
    \hat{S} \boldsymbol{v} = \sum_{i \in \mathcal{C}} \boldsymbol{\tilde{\phi}}_i \bigl( \boldsymbol{\tilde{\phi}}_i^\dagger \boldsymbol{v} \bigr) \,,
    \label{eq:matrix_free}
\end{equation}
costs $\mathcal{O}(|\mathcal{C}| N_v)$ operations, i.e.\ $\mathcal{O}(N_{\text{conf}} N_v)$ summed over the clusters, without explicitly storing $\hat{S}$. Power iteration or a Lanczos method could obtain the leading eigenvector at this cost per iteration, with a number of iterations set by the gap between the two largest eigenvalues, which can become small when $K$ is too small (appendix~\ref{app:tolerance}). This route has neither been implemented nor tested and remains future work, as do applications to $(3+1)$-dimensional models of dense matter. In higher-dimensional four-fermion models the existence of inhomogeneous phases is itself delicate: in the $(2+1)$-dimensional GN model at large $N_f$ an inhomogeneous phase was reported at finite lattice spacing~\cite{Winstel:2019zfn}, but the inhomogeneous region found there shrinks as the cutoff is increased and, at non-zero temperature, disappears when it is removed~\cite{Buballa:2020nsi}, and with an additional chiral chemical potential its absence in the continuum limit has been suggested~\cite{Pannullo:2021edr}. A method that extracts representative real-space profiles from clusters of aligned configurations could help to distinguish modulations tied to the lattice scale from physical ones, but we have not applied it there. The application to complex fields such as the chiral spiral $\Delta = \sigma + i\pi$ of the chiral GN model, which our implementation handles as a single complex component, remains to be carried out on Monte Carlo data (section~\ref{subsec:notations}).

\subsection{Limitations of the Present Analysis}
\label{subsec:limitations_analysis}

We collect here what the analysis of section~\ref{sec:application} does not establish. (i) Only real scalar fields have been analysed (section~\ref{subsec:notations}). (ii) No single $K$ is selected; the criteria are calibrated against forced splits built from one ensemble and used for all three, and the ranges from six null and four positive-control realizations are not probabilities. (iii) The band is conditional on the global centroid and taken modulo $G_{\text{al}}$; its pointwise half-width is the square root of the corresponding diagonal entry of a singular covariance matrix, and has a relative statistical uncertainty of about $30\,\%$ with $N_B = 6$, a choice of $N_B$ that rests on a stability test at one ensemble; it is not used to compare values of $K$ and not quoted for the homogeneous ensembles. The jackknife blocks are only $0.9$ to $14$ times as long as $2\tau_{\text{int}}$ of scalar observables related to the centroid, so that the band, like every other jackknife error quoted in section~\ref{sec:application}, is expected to be underestimated; the factor has been estimated only for scalar observables and not propagated to the band (section~\ref{subsec:setup}, appendix~\ref{app:autocorr}). The band also does not detect a bias common to all subsamples, such as the folding of the even harmonics with $G_{\text{trans}} \times \mathbb{Z}_2$. (iv) Populations are not equilibrium mixing ratios: the integrated autocorrelation time of quantities related to the shape is $10^2$ to $2 \times 10^3$ configurations and is itself only a lower bound (appendix~\ref{app:autocorr}), the number of effectively independent configurations that follows from it is between about $5$ and $70$ for these observables, and families appear in parts of the histories only. (v) $N_{\text{kink}}$ counts sign changes of a smoothed, time-averaged field and is neither conserved nor topological. (vi) The Euclidean-time average has been checked for $K = 1$ at two ensembles only. (vii) The even Fourier components of the centroids for $K = 1$, and their asymmetry, depend on whether $\sigma \to -\sigma$ is included in $G$: with it they reflect the typical difference between the lengths of the two domains of a configuration, folded by the alignment, and without it they reflect the imbalance of the two orientations in the sample. At one of the three inhomogeneous ensembles the difference exceeds the combined band. We fixed the group per quantity, and a different choice would change what is shown there. Whether the distribution of the domain lengths has more than one maximum cannot be decided from the present histories, whose autocorrelation time for this quantity is $10^3$ configurations or longer. (viii) All ensembles share one volume, one temperature and one number of flavours. The wavelength and the number of sign changes change with $\mu$, but the comparisons at the selected chemical potentials do not establish a systematic dependence of the normalized profile within one period on $\mu$. No infinite-volume or continuum limit is taken, and no quantitative comparison with the analytic kink--antikink crystal is made; the profiles are therefore described only as non-sinusoidal. At two of the ensembles the wavelength of the modulation is the length of the lattice, so that it may be fixed by the volume. (ix) The algorithmic uncertainty of section~\ref{subsec:algorithm} is not estimated separately. Two global fits required $r_{\max} = 2000$ rather than $200$ for their adopted restart to be stopped by eq.~\eqref{eq:stopping_rule} (tables~\ref{tab:hyperparams} and~\ref{tab:GN}), and for $K = 3$ at $\mu/\sigma_0 = 0.630$ this changed which restart is adopted (appendix~\ref{app:gn2_group}). (x) Every configuration is aligned by a single lattice translation, combined with a sign where $G$ contains it. Configurations in which several structures move independently, for example walls or defects whose relative positions fluctuate, cannot be brought into coincidence by one translation, so that the centroid averages over their relative positions; whether this contributes to the unequal spacings of the centroid and to the broad band at $\mu/\sigma_0 = 0.630$ has not been determined. At $\mu/\sigma_0 = 0.485$ and $0.582$ the distance between the two walls varies from configuration to configuration (section~\ref{subsec:kink_extraction}); with $G_{\text{trans}} \times \mathbb{Z}_2$ this variation is folded into the even harmonics of the centroid.

\section{Conclusion}
\label{sec:conclusion}

Spatial correlation functions are the standard tool for locating inhomogeneous phases in lattice simulations, and on the data analysed here they do so without difficulty. They do not in general uniquely determine the real-space shape of the configurations: up to a Fourier transform the correlator is the ensemble-averaged power spectrum and discards the relative phases of the Fourier modes. Profiles containing several modes can therefore have different shapes while sharing the same $C(n)$ (section~\ref{sec:limitations}). We have constructed an explicit pair of profiles whose correlators agree to machine precision.

To recover the shape, we proposed the Spatiotemporal K-Shape method, which extends K-Shape clustering to lattice fields with periodic boundary conditions. Configurations are Z-normalized and compared modulo lattice translations and, where it is a symmetry, a global sign or phase. The centroid of a cluster is the leading eigenvector of a Hermitian matrix built from its aligned members and is given with a delete-block jackknife band. Shape-correlation criteria, calibrated against forced splits of a single profile and against two-shape positive controls, compare the shapes obtained at different $K$ to assess whether one representative shape suffices. Mock ensembles showed that, at the noise level of table~\ref{tab:benchmarks}, the method returns the profiles put into it, including two profiles with identical correlators, and that a uniform component can reverse the sign of $C_{\text{min}}$ without changing the modulation shape.

Applied to GN$_2$ ensembles with $N_f = 8$ at $T/\sigma_0 = 0.0985$ and six chemical potentials, the correlator identified three homogeneous and three inhomogeneous ensembles, and the homogeneous ones were confirmed by their offset, by the absence of sign changes and by the spectral concentration of the centroid. For the inhomogeneous ensembles, the centroids added what the correlator leaves open. At $\mu/\sigma_0 = 0.485$ and $0.582$, where $C_{\text{min}} \approx -0.6$, the centroids show a non-sinusoidal profile with two sign changes. In the smoothed, time-averaged configurations, two sign changes are found in all configurations at $0.485$ and most at $0.582$; the latter ensemble also contains a minority cluster of about $16\,\%$, mostly with four sign changes. At $\mu/\sigma_0 = 0.630$, where $C_{\text{min}} = -0.108$, the ensemble mixes families dominated by six and eight sign changes, whose separation by clustering is observed with both symmetry groups considered, and the centroid for $K = 1$ is a blend of them. Including $\sigma \to -\sigma$ in the symmetry group changed the centroid by more than the combined band of the two fits at one ensemble, where the two orientations of unequal domain lengths occur in nearly equal proportions: the group decides whether this configuration-level asymmetry is averaged out or folded into the centroid, and it is a physical choice that has to be stated with the result.

These conclusions are limited to real fields, to one volume, temperature and number of flavours, and to shapes described without comparison with analytic solutions (section~\ref{subsec:limitations_analysis}). Long autocorrelations limit the statistical precision, and cluster populations are not interpreted as equilibrium mixing ratios. Natural next steps are a scan of the GN$_2$ phase diagram with the descriptors used here, a quantitative comparison of the extracted shapes with the kink--antikink crystal, the pre-screening of section~\ref{subsec:practical_considerations}, complex fields such as the chiral spiral and fields with several components, and an eigenvalue solver that does not form $\hat{S}$, which could lower the computational barriers to applications on $(3+1)$-dimensional lattices.

\section*{Code availability}
The implementation of the Spatiotemporal K-Shape method used in this work is available from the corresponding author upon reasonable request. The parameter names in table~\ref{tab:hyperparams} are those of this implementation.

\acknowledgments
Numerical computation in this work was carried out at the Yukawa Institute Computer Facility. This work was supported by JSPS KAKENHI Grant Numbers JP20H00156, JP20H11581, and JP25H00449 (C.N.) and the World Premier International Research Center Initiative (WPI) under MEXT, Japan (C.N.).

The Spatiotemporal K-Shape method, its implementation, the figures and the manuscript were originally developed and written by the authors. In revising this work, K.H. used Claude (Anthropic; Claude Opus 5 and Opus 5.5) as an assistant to review the manuscript and the code, to draft revisions of the text, and to update, test and run parts of the analysis code and of the figure-generation scripts; Claude Fable 5.1 and Fable 5.2 were used to review the manuscript. All changes were checked by the authors, who take full responsibility for the content of this article.

\appendix

\section{Mathematical Details of the exSBD and Centroid Extraction}
\label{app:math}

This appendix collects derivations that support section~\ref{sec:formulation} but are not needed to follow the definitions given there.

\subsection{Lattice Fourier Transform and the Correlation Theorem}
\label{app:fourier}

With the unitary lattice Fourier transform
\begin{equation}
    \hat{\Phi}_{\alpha}(\mathbf{q}) = \frac{1}{\sqrt{N_{\text{lat}}}} \sum_{\mathbf{n} \in \Lambda} \Phi_{\alpha, \mathbf{n}} \, e^{-2\pi i \, \mathbf{q} \cdot \mathbf{n} / \mathbf{N}} \,, \qquad
    \frac{\mathbf{q} \cdot \mathbf{n}}{\mathbf{N}} \equiv \sum_{j=1}^{D} \frac{q_j n_j}{N_j} \,,
    \label{eq:lattice_fourier}
\end{equation}
where $j$ runs over the $D$ lattice directions and $N_j$ is the extent in direction $j$, and which reduces to eq.~\eqref{eq:fourier_coefficients} in one dimension, the circular cross-correlation~\eqref{eq:circular_cc} obeys the correlation theorem
\begin{equation}
    (\Phi \star \Psi)_{\boldsymbol{\tau}} = \sum_{\alpha} \sum_{\mathbf{q}} \hat{\Phi}_{\alpha}(\mathbf{q})^{*} \, \hat{\Psi}_{\alpha}(\mathbf{q}) \, e^{2\pi i \, \mathbf{q} \cdot \boldsymbol{\tau} / \mathbf{N}} \,.
    \label{eq:cc_fourier}
\end{equation}
It allows the implementation to evaluate eq.~\eqref{eq:circular_cc} for all $N_{\text{lat}}$ shifts at once with fast Fourier transforms, as stated in section~\ref{subsec:correlation}.

\subsection{Phase Freedom at Fixed Power Spectrum}
\label{app:phase_freedom}

For a real field, $\hat{\sigma}(N_s-q)=\hat{\sigma}(q)^*$. For even $N_s$, the independent continuous phases are therefore those at $q=1,\dots,N_s/2-1$, whereas the zero and Nyquist coefficients are real and carry a sign but no continuous phase. For odd $N_s$, the Nyquist mode is absent and there are $(N_s-1)/2$ independent phases.

In the generic case in which every Fourier modulus is non-zero, fixing all moduli and the signs of the real coefficients leaves a torus of configurations with the same spatial correlator $C(n)$, which we call the residual family. Its dimension is $N_s/2-1$ for even $N_s$ and $(N_s-1)/2$ for odd $N_s$. Integer lattice translations act on these phases through eq.~\eqref{eq:phase_translation} and form a finite group. Taking the quotient by this group does not reduce the dimension. Thus fixing the power spectrum and removing lattice translations still leaves continuous freedom in the configurations. As discussed in section~\ref{subsec:kink_difficulty}, general changes of the relative phases alter the shape rather than merely its position.

\subsection{Range of the exSBD}
\label{app:range}

Summed over all shifts, the correlation factorizes component by component, and eq.~\eqref{eq:znorm_properties} gives
\begin{equation}
    \sum_{\boldsymbol{\tau} \in \Lambda} (\tilde{\Phi} \star \tilde{\Psi})_{\boldsymbol{\tau}} = \sum_{\alpha} \Bigl( \sum_{\mathbf{n} \in \Lambda} \tilde{\Phi}_{\alpha, \mathbf{n}} \Bigr)^{*} \Bigl( \sum_{\mathbf{n} \in \Lambda} \tilde{\Psi}_{\alpha, \mathbf{n}} \Bigr) = 0 \,.
    \label{eq:ncc_sum_rule}
\end{equation}
The $N_{\text{lat}}$ numbers $\mathrm{Re} \, \mathrm{NCC}_{\boldsymbol{\tau}}$ thus sum to zero, their maximum is non-negative, and together with $| \mathrm{NCC}_{\boldsymbol{\tau}} | \le 1$ the bounds~\eqref{eq:exsbd_range} follow. The upper bound is reached if and only if $\mathrm{Re} \, \mathrm{NCC}_{\boldsymbol{\tau}} = 0$ for every $\boldsymbol{\tau}$, e.g.\ for two real configurations with disjoint Fourier support, or for $\tilde{\Psi} = i \tilde{\Phi}$ with real $\tilde{\Phi}$. It is the component-wise mean in eq.~\eqref{eq:znorm_moments} that makes every factor in eq.~\eqref{eq:ncc_sum_rule} vanish; a single mean subtracted from all components would not, and for $N_{\text{comp}} > 1$ the upper bound in eq.~\eqref{eq:exsbd_range} can then fail.

\subsection{Triangle Inequality}
\label{app:triangle}

The dissimilarity $d_G$ of eq.~\eqref{eq:dG}, and in particular the exSBD, violates the triangle inequality. Write $1 - d_G = \cos\gamma_G$ with $\gamma_G \in [0, \pi/2]$. The angle $\gamma_G(\tilde{\Phi}, \tilde{\Psi}) = \min_{g \in G} \angle(\tilde{\Phi}, g \tilde{\Psi})$ between two $G$-orbits on the sphere $\| \tilde{\Phi} \|^2 = N_v$ of $\mathbb{R}^{2N_v}$ obeys the triangle inequality, because $G$ acts by isometries, but $1 - \cos\gamma_G$ does not. Maximizing $d_G(A, C) - d_G(A, B) - d_G(B, C)$ subject to these constraints gives the bound $\sqrt{2} - 1 \simeq 0.414$. It is attained for $G = G_{\text{trans}}$ by two Z-normalized cosines $A$ and $C$ with different wave numbers and the Z-normalized $B \propto A + C$, for which $d_G(A, C) = 1$ and $d_G(A, B) = d_G(B, C) = 1 - 1/\sqrt{2}$. Equivalently, $2 N_v \, d_G(\tilde{\Phi}, \tilde{\Psi}) = \min_{g \in G} \| \tilde{\Phi} - g \tilde{\Psi} \|^2$, eq.~\eqref{eq:dG_msd}, so that $\sqrt{d_G}$ is a metric on $\mathcal{Z} / G$.

\subsection{Zero Mean of the Centroid without Projection}
\label{app:projection}

No explicit projection onto zero-mean vectors is needed in the centroid extraction of section~\ref{subsec:centroid}. By eq.~\eqref{eq:znorm_properties}, which translations and global phases preserve, every $\boldsymbol{\tilde{\phi}}_i$ is orthogonal to each vector $\boldsymbol{e}_\alpha \in \mathbb{C}^{N_v}$ that equals $1$ on the entries of component $\alpha$ and $0$ elsewhere, so that $\hat{S} \boldsymbol{e}_\alpha = 0$. An eigenvector with a positive eigenvalue lies in the span of the $\boldsymbol{\tilde{\phi}}_i$ and hence has zero mean in every component. The projector $Q = I - N_v^{-1} \boldsymbol{1} \boldsymbol{1}^{\mathsf{T}}$, used in ref.~\cite{paparrizos2015k} to obtain a zero-mean centroid, therefore acts trivially, $Q \hat{S} Q = \hat{S}$, and is omitted; in ref.~\cite{paparrizos2015k} it is not trivial, because a zero-padded shift changes the mean of a sequence.

\subsection{Gram Route for the Centroid}
\label{app:gram}

Let $Y \in \mathbb{C}^{|\mathcal{C}| \times N_v}$ be the matrix whose rows are the transposed aligned members $\boldsymbol{\tilde{\phi}}_i^{\mathsf{T}}$. Then
\begin{equation}
    \hat{S} = Y^{\mathsf{T}} \, \bar{Y} \,, \qquad
    \hat{W} = \bar{Y} \, Y^{\mathsf{T}} \in \mathbb{C}^{|\mathcal{C}| \times |\mathcal{C}|} \,, \qquad
    \hat{W}_{ij} = \boldsymbol{\tilde{\phi}}_i^\dagger \, \boldsymbol{\tilde{\phi}}_j \,,
    \label{eq:gram}
\end{equation}
where the bar denotes complex conjugation and $\hat{W}$ is the Gram matrix of the aligned members. The position of the conjugation matters for complex fields: $Y^\dagger Y = \bar{\hat{S}}$ has the complex-conjugate eigenvectors, and using it would, for instance, reverse the winding direction of a spiral. The matrices $\hat{S}$ and $\hat{W}$ are both Hermitian and positive semi-definite and share their non-zero eigenvalues. If $\hat{W} \boldsymbol{u} = \lambda \boldsymbol{u}$ with $\lambda > 0$, then
\begin{equation}
    \hat{S} \, (Y^{\mathsf{T}} \boldsymbol{u}) = Y^{\mathsf{T}} \hat{W} \boldsymbol{u} = \lambda \, Y^{\mathsf{T}} \boldsymbol{u} \,, \qquad
    \| Y^{\mathsf{T}} \boldsymbol{u} \|^2 = \boldsymbol{u}^\dagger \hat{W} \boldsymbol{u} = \lambda \, \| \boldsymbol{u} \|^2 \,,
    \label{eq:gram_pullback}
\end{equation}
so that the leading eigenvector of $\hat{S}$ is $\boldsymbol{v} \propto Y^{\mathsf{T}} \boldsymbol{u}$ for the leading eigenvector $\boldsymbol{u}$ of $\hat{W}$, up to the factor fixed by eq.~\eqref{eq:phase_fix}. Building and diagonalizing the matrix costs $\mathcal{O}(|\mathcal{C}| N_v^2 + N_v^3)$ on the direct route and $\mathcal{O}(|\mathcal{C}|^2 N_v + |\mathcal{C}|^3)$ on the Gram route, which keeps the centroid computable when $N_v \gg |\mathcal{C}|$, as can happen on $(3+1)$-dimensional lattices, where $\hat{S}$ could not even be stored.

\section{Convergence and Implementation Details}
\label{app:impl}

This appendix collects details of the algorithm of section~\ref{subsec:algorithm} and of the error estimate of section~\ref{subsec:error_estimation} that matter for a reimplementation or for the assessment of the numerical results, but not for the definitions.

\subsection{Non-Monotone Objective and Choice of the Tolerance}
\label{app:tolerance}

The clustering objective~\eqref{eq:objective} can increase from one iteration to the next (section~\ref{subsec:algorithm}). For a synthetic test ensemble of $1000$ configurations with $K = 1$, a single iteration started from an intermediate state raised $\mathcal{L}$ from $293.08$ to $310.07$, and the iteration subsequently settled at a fixed point with $\mathcal{L} = 309.9995$. With an earlier version of the implementation, which stopped once the labels were unchanged and $\mathcal{L}$ had decreased by less than a tolerance, $\mathcal{L}$ had increased at the stopping iteration of the adopted restart, by relative amounts of at most $6.4 \times 10^{-6}$, in $12$ of the $14$ fits of the GN$_2$ ensembles of section~\ref{sec:application}.

The value of $\varepsilon$ was chosen in tests on four synthetic ensembles of $1000$ configurations each. With $\varepsilon = 10^{-5}$, a run with $K = 1$ stopped while its centroid was still changing, at $1 - \max_{\boldsymbol{\tau}} | \mathrm{NCC}_{\boldsymbol{\tau}} | \approx 1.6 \times 10^{-3}$ from the solution obtained with the earlier stopping criterion. With $\varepsilon = 10^{-6}$, the centroids of all tests ($K = 1, 2, 3$) agreed within $1 - \max_{\boldsymbol{\tau}} | \mathrm{NCC}_{\boldsymbol{\tau}} | \le 10^{-4}$ with those of reference runs iterated to a fixed point. Smaller values appear impractical: on the homogeneous GN$_2$ ensembles, whose centroids keep drifting slowly, a change below $10^{-7}$ was not reached within twenty further iterations, which suggests that many restarts would run to $r_{\max}$. For real fields with $G = G_{\text{trans}} \times \mathbb{Z}_2$, all restarts of the GN$_2$ fits to the inhomogeneous ensembles and of their jackknife subsamples were stopped by eq.~\eqref{eq:stopping_rule} (section~\ref{subsec:setup}). For $K = 1$ on two of the inhomogeneous GN$_2$ ensembles, the centroids at which eq.~\eqref{eq:stopping_rule} stopped the iteration differed from those of longer runs by $1 - \max_{\boldsymbol{\tau}} | \mathrm{NCC}_{\boldsymbol{\tau}} | \lesssim 3 \times 10^{-5}$; for the homogeneous ensembles no such bound applies, and no error band is quoted for them (section~\ref{subsec:error_estimation}). For complex fields with $G = G_{\text{trans}} \times U(1)$, tests on three synthetic ensembles of $1000$ configurations each with $N_v = 64$, i.e.\ on the direct route of section~\ref{subsec:centroid}, gave the following results: with $\varepsilon = 10^{-6}$, every restart was stopped by eq.~\eqref{eq:stopping_rule}, after $2$ to $14$ iterations, and for $K$ equal to the number of distinct shapes in the ensemble the centroids agreed within $1 - \max_{\boldsymbol{\tau}} | \mathrm{NCC}_{\boldsymbol{\tau}} | \le 2.4 \times 10^{-5}$ with those of reference runs iterated to a fixed point. If $K$ is too small, the leading eigenvalue of $\hat{S}$ can be nearly degenerate, and the centroid is then not fixed by the data: for an ensemble of two equally populated shapes fitted with $K = 1$, the iteration stopped up to $7 \times 10^{-4}$ from the fixed point, whereas with $G = G_{\text{trans}}$ almost all restarts on this ensemble reached $r_{\max}$.

\subsection{State Returned by a Restart}
\label{app:returned_state}

For a restart that reached $r_{\max}$, the returned centroids are those extracted from the labels at the beginning of the last iteration, the returned labels are the assignment to these centroids (unless an empty cluster was replaced in that iteration), and the returned value of $\mathcal{L}$ is that of the returned labels and centroids. A new extraction from the returned labels does not in general reproduce the returned centroids: the labels may have changed in the last iteration, and even if they did not, the reference of such an extraction, the returned centroid itself, differs by more than $\varepsilon$ from the reference that was used. For a restart stopped by eq.~\eqref{eq:stopping_rule}, the labels did not change, and the returned centroids are the refinement of the returned labels computed with a reference that differs from them by at most $\varepsilon$ in $d_G$.

\subsection{Computational Cost}
\label{app:cost}

With fast Fourier transforms, one evaluation of the exSBD costs $\mathcal{O}(N_v \log N_{\text{lat}})$ (section~\ref{subsec:correlation}). Per iteration, the assignment evaluates $N_{\text{conf}} K$ dissimilarities, the refinement aligns every configuration once to the centroid of its cluster and solves one eigenvalue problem per cluster, and the stopping rule costs $K$ further dissimilarities. The cost of one iteration is therefore
\begin{equation}
    \mathcal{O}\bigl( N_{\text{conf}} \, K \, N_v \log N_{\text{lat}} \bigr)
    + \sum_{k=1}^{K} \mathcal{O}\Bigl( \min \bigl[ \, |\mathcal{C}_k| N_v^2 + N_v^3 , \; |\mathcal{C}_k|^2 N_v + |\mathcal{C}_k|^3 \, \bigr] \Bigr) \,,
    \label{eq:cost}
\end{equation}
where the two entries of the minimum are the costs of the direct and the Gram route of section~\ref{subsec:centroid}, and the cheaper route is taken, i.e.\ the direct route for $N_v \le |\mathcal{C}_k|$ and the Gram route for $N_v > |\mathcal{C}_k|$. A global fit requires at most $N_{\text{trial}} \, r_{\max}$ iterations. The jackknife of section~\ref{subsec:error_estimation} repeats the fit $N_B$ times on $(1 - 1/N_B) N_{\text{conf}}$ configurations with $N_{\text{trial}}^{\text{jk}}$ restarts each. The memory required is $\mathcal{O}(N_{\text{conf}} N_v)$ for the data and, per cluster, $\mathcal{O}(|\mathcal{C}_k| N_v + N_v^2)$ or $\mathcal{O}(|\mathcal{C}_k| N_v + |\mathcal{C}_k|^2)$ for the eigenvalue problem.

\subsection{Null Ensembles and Positive Controls}
\label{app:null}

The synthetic ensembles used to calibrate $\rho_{\text{B}}$ in section~\ref{subsec:algorithm} are built from the selected ensemble and its centroid $\tilde{\Phi}_c^{K=1}$ as follows. Every synthetic configuration is this profile at a lattice translation drawn uniformly at random, multiplied by an amplitude drawn with replacement from the amplitudes $s_{\Phi_i} \beta_i$ of the configurations of the selected ensemble, where $\beta_i = \mathrm{Re} \, \langle \tilde{\Phi}_c^{K=1}, \tilde{\Phi}_i^{\text{al}} \rangle / N_v$ is the projection of the aligned configuration on the profile, plus noise. Three noise models are used: white Gaussian noise; coloured noise whose Fourier coefficients have moduli fixed to the square root of the mean power spectrum of the residuals $\tilde{\Phi}_i^{\text{al}} - \beta_i \tilde{\Phi}_c^{K=1}$ and uniformly random phases, which is not Gaussian; and a stationary Gaussian field with the same mean power spectrum. In all three models the noise is normalized to unit mean square per lattice site, exactly for fixed moduli and in expectation otherwise, multiplied by the noise strength and added, in the units of the field, to the amplitude times the translated profile before the synthetic configuration is Z-normalized. The two coloured models have no zero mode, and their Nyquist coefficient, which must be real, is given a random sign for fixed moduli and a real Gaussian value for the Gaussian field. For each model, the noise strength is tuned such that the median over the configurations of $1 - \rho(\tilde{\Phi}_c^{K=1}, \cdot)$ agrees with that of the selected ensemble.

The positive controls are built in the same way from the two centroids of a $K = 2$ fit of an ensemble that contains two distinct profiles, mixed with the populations of that fit and with the amplitudes and the residual spectrum obtained from it. Their noise, white or coloured with fixed moduli, is tuned to the median of $1 - \rho$ with respect to the nearer of the two profiles. The null and positive-control fits use $r_{\max} = 200$ (table~\ref{tab:hyperparams}); in one of them the adopted restart reached it.

\subsection{Jackknife Prefactor with Discarded Subsamples}
\label{app:jk_prefactor}

For $N_B' < N_B$ subsamples, the prefactor in eq.~\eqref{eq:jk_error} keeps the variance estimate unbiased for a linear estimator if the discarded subsamples are unrelated to the data: with blocks of equal length, the deviation of a delete-block estimate from the mean is $-1/(N_B - 1)$ times the deviation of the corresponding block average, so that the $N_B'$ available estimates determine the variance of the block averages, which the prefactor converts into the variance of the full-sample estimate. Since a fit can fail for reasons that depend on the data, this condition is not guaranteed.

\subsection{Autocorrelation and the Length of the Jackknife Blocks}
\label{app:autocorr}

The general error prescription is given in section~\ref{subsec:error_estimation}; this appendix collects the block-number test and autocorrelation measurements underlying the choices and qualifications in section~\ref{subsec:setup}. For the three inhomogeneous ensembles of section~\ref{sec:application}, the autocorrelation estimates from block averages keep growing with block length, so a rule of the form ``block length $\ge$ a fixed multiple of $\tau_{\text{int}}$'' does not fix $N_B$ reliably. We therefore also tested the dependence of the band on $N_B$ and adopted the common value $N_B=6$. For $K=1$ with $G=G_{\text{trans}}$ at $\mu/\sigma_0=0.582$, the site-averaged band varies non-monotonically, by at most a factor of $1.6$, over $N_B\in\{3,4,5,6,8,10\}$. Subsample estimates scattering by an amount independent of $N_B$ would instead make it grow in proportion to $(N_B-1)/\sqrt{N_B}$, by a factor of $2.5$ from $N_B=3$ to $10$. This test was made on one ensemble only and does not establish convergence of the error estimates.

On each of the three inhomogeneous ensembles we formed one number per configuration for five observables that are related to the quantities of section~\ref{sec:application}, of which three are listed in table~\ref{tab:tau_int}: the shape correlation $\rho(\tilde{\Phi}_c^{K=1}, \tilde{\Phi}_i)$ of eq.~\eqref{eq:shape_correlation} of a configuration with the centroid for $K = 1$ of the fit with $G_{\text{trans}} \times \mathbb{Z}_2$ (section~\ref{subsec:kink_extraction}); the value $C_i(n_{\min})$ of the correlator of a configuration at the separation $n_{\min} = 31$, $32$ and $10$ at which the ensemble average of eq.~\eqref{eq:gn2_cmin} is minimal; and the membership of a configuration in the larger cluster of the fit with $K = 2$ and $G_{\text{trans}}$ (section~\ref{subsec:multimodality}). The two remaining observables, the squared spatial mean of a configuration and the spatial standard deviation of its time average, give $\tau_{\text{W}}$ between $117$ and $642$ configurations, the smallest value of the whole set, so that $L_{\text{jk}} / 2\tau_{\text{W}}$ reaches $14$ there.

For the five observables, the automatic-window estimates range from approximately $10^2$ to $1.6 \times 10^3$ configurations, and the estimates for the two additional quantities discussed below reach about $2 \times 10^3$; in trajectories, these are $10^3$ to $2 \times 10^4$.

\begin{figure}[htbp]
    \centering
    \includegraphics[width=\textwidth]{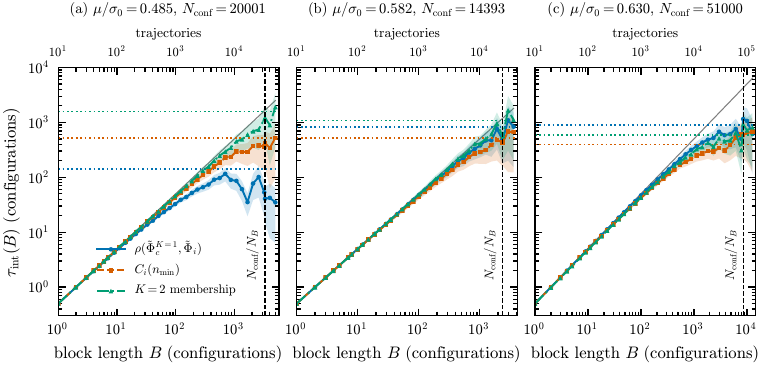}
    \caption{
        Integrated autocorrelation time $\tau_{\text{int}}(B) = B \, \mathrm{Var}(\text{block mean}) / (2 \, \mathrm{Var})$ from non-overlapping blocks of length $B$, for the three observables of table~\ref{tab:tau_int} and the three inhomogeneous ensembles: (a) $\mu/\sigma_0 = 0.485$, (b) $0.582$ and (c) $0.630$. The shaded regions are the errors of the block estimate. Dotted horizontal lines: the automatic-window estimates $\tau_{\text{W}}$ of the same colour; dashed vertical line: the jackknife block length $N_{\text{conf}}/N_B$; grey line: $\tau_{\text{int}} = B/2$, the value of a completely correlated series. The lower abscissa counts configurations and the upper one trajectories, ten per configuration (section~\ref{subsec:setup}). The centroid entering $\rho(\tilde{\Phi}_c^{K=1}, \tilde{\Phi}_i)$ is that of the fit with $G_{\text{trans}} \times \mathbb{Z}_2$ (figure~\ref{fig:gn2_centroids_K1}), and $n_{\min}$ is the separation at which the ensemble-averaged $C(n)$ is minimal, $31$, $32$ and $10$.
    }
    \label{fig:gn2_tau_int}
\end{figure}

The integrated autocorrelation time is estimated in two ways, with the convention $\tau_{\text{int}} = \frac{1}{2} + \sum_{t \ge 1} \rho(t)$: from the variance of the means of non-overlapping blocks of length $B$, $\tau_{\text{int}}(B) = B \, \mathrm{Var}(\text{block mean}) / (2 \, \mathrm{Var})$, and with the automatic windowing procedure of ref.~\cite{Wolff:2003sm} at $S = 1.5$. Figure~\ref{fig:gn2_tau_int} shows $\tau_{\text{int}}(B)$, and table~\ref{tab:tau_int} the values at the block length $L_{\text{jk}} = N_{\text{conf}}/N_B$ of the jackknife, the largest value reached over $B$, and the windowing estimate $\tau_{\text{W}}$. For most of these observables $\tau_{\text{int}}(B)$ is still growing at $B = L_{\text{jk}}$, so that both estimates are lower bounds; the values quoted are those of the observables shown. Two further quantities that enter section~\ref{subsec:kink_extraction} are of the same size or larger: the orientation of a configuration, i.e.\ whether a sign reversal is needed to align it to the centroid (appendix~\ref{app:gn2_group}), for which the block estimate reaches $2.0 \times 10^3$ and $1.3 \times 10^3$ configurations at $\mu/\sigma_0 = 0.485$ and $0.582$ without a plateau; and the bispectral combination $\mathrm{Re}\,[\hat{\sigma}(q)^2 \hat{\sigma}(2q)^*] / |\hat{\sigma}(q)|^3$ of a configuration, with $q$ its dominant wave number, which measures the asymmetry discussed there and gives $\tau_{\text{W}} = 1.9 \times 10^3$ configurations at $\mu/\sigma_0 = 0.485$, the one case in which $L_{\text{jk}}$ is shorter than $2\tau_{\text{int}}$, at $L_{\text{jk}}/2\tau_{\text{W}} = 0.9$.

\begin{table}[htbp]
  \centering
  \caption{Integrated autocorrelation time of three configuration-level observables of the inhomogeneous ensembles, in units of configurations; configurations are separated by $10$ trajectories (section~\ref{subsec:setup}). $\tau_{\text{int}}(L_{\text{jk}})$ is the block estimate at the jackknife block length $L_{\text{jk}} = N_{\text{conf}}/N_B$, which is $3333$, $2398$ and $8500$ configurations; $\max_B \tau_{\text{int}}(B)$ is the largest value reached over the block lengths scanned, with that block length in parentheses; $\tau_{\text{W}}$ is the estimate from the automatic windowing procedure of ref.~\cite{Wolff:2003sm}, with the window in parentheses. The errors are those of the respective procedures. Since $\tau_{\text{int}}(B)$ does not reach a plateau (figure~\ref{fig:gn2_tau_int}), the estimates of $\tau_{\text{int}}$ are lower bounds and the ratio in the last column is correspondingly an upper bound.}
  \label{tab:tau_int}
  \vspace{0.2cm}
  \small
  \begin{tabular}{llrrrr}
    \hline
    $\mu/\sigma_0$ & observable & $\tau_{\text{int}}(L_{\text{jk}})$ & $\max_B \tau_{\text{int}}(B)$ ($B$) & $\tau_{\text{W}}$ ($W$) & $L_{\text{jk}}/2\tau_{\text{W}}$ \\
    \hline \hline
    $0.485$ & $\rho(\tilde{\Phi}_c^{K=1}, \tilde{\Phi}_i)$ & $41(26)$ & $116$ ($700$) & $142(42)$ ($574$) & $11.7$ \\
    $0.485$ & $C_i(n_{\min})$ & $420(266)$ & $520$ ($5000$) & $515(221)$ ($1432$) & $3.2$ \\
    $0.485$ & cluster, $K = 2$ & $1217(770)$ & $1905$ ($5000$) & $1589(717)$ ($2608$) & $1.0$ \\
    $0.582$ & $\rho(\tilde{\Phi}_c^{K=1}, \tilde{\Phi}_i)$ & $489(309)$ & $1093$ ($2917$) & $825(392)$ ($1639$) & $1.5$ \\
    $0.582$ & $C_i(n_{\min})$ & $441(279)$ & $677$ ($2917$) & $521(238)$ ($1274$) & $2.3$ \\
    $0.582$ & cluster, $K = 2$ & $569(360)$ & $1661$ ($2917$) & $1076(496)$ ($1839$) & $1.1$ \\
    $0.630$ & $\rho(\tilde{\Phi}_c^{K=1}, \tilde{\Phi}_i)$ & $1153(729)$ & $1153$ ($8500$) & $899(351)$ ($2835$) & $4.7$ \\
    $0.630$ & $C_i(n_{\min})$ & $587(371)$ & $677$ ($12750$) & $398(121)$ ($1571$) & $10.7$ \\
    $0.630$ & cluster, $K = 2$ & $918(580)$ & $918$ ($8500$) & $597(207)$ ($2123$) & $7.1$ \\
    \hline
  \end{tabular}
\end{table}

Three consequences are used in sections~\ref{subsec:setup} and~\ref{subsec:limitations_analysis}. First, with $N_B = 6$ the jackknife blocks are $0.9$--$14$ times as long as $2\tau_{\text{int}}$, so that the blocks are not independent. Second, modelling the autocorrelation as a single exponential, $\tau_{\text{int}}(L)/\tau_{\text{int}} = 1 - (\tau_{\text{int}}/L)(1 - e^{-L/\tau_{\text{int}}})$, the error of such an observable is underestimated by a factor of about $1.1$--$1.5$ at these block lengths; because $\tau_{\text{int}}(B)$ does not stop growing, this is a lower estimate and no upper bound can be given. Third, and independently of the autocorrelation, an error obtained from $N_B = 6$ blocks has a relative statistical uncertainty of about $30\,\%$ for Gaussian block averages (section~\ref{subsec:setup}), which is the dominant source of the scatter between $\tau_{\text{int}}(L_{\text{jk}})$ and $\max_B \tau_{\text{int}}(B)$ in table~\ref{tab:tau_int} at the entries with $L_{\text{jk}} \gg 2\tau_{\text{int}}$. Neither factor has been propagated to the band of a centroid, which is not a linear function of the ensemble.

\section{Supplementary Figures for the GN$_2$ Analysis}
\label{app:gn2_supplement}

This appendix collects figures and numbers that support section~\ref{sec:application} but are not needed to follow it.

The departure of the centroids of section~\ref{subsec:kink_extraction} from a single harmonic is also seen in the ratios $h_j = |\hat{c}(jq)|/|\hat{c}(q)|$ of the Fourier moduli of the Z-normalized centroid $c$ for $K = 1$ (eq.~\eqref{eq:fourier_coefficients} applied to $c$), $q$ being its strongest wave number: at $\mu/\sigma_0 = 0.485$ and $0.582$, where $q = 1$, $h_3 = 0.19$ and $0.23$ and $h_5 = 0.04$ and $0.07$, against $0$ for a single harmonic and $0.29$ and $0.14$ for the kink of section~\ref{subsec:kink_test}. The even ratio $h_2$, which vanishes for a profile mapped onto minus itself by a translation, is $0.183$ and $0.153$ at the two chemical potentials for the centroids with $G_{\text{trans}} \times \mathbb{Z}_2$, and $0.023$ and $0.109$ for those with $G_{\text{trans}}$. The first-order uncertainties from the band are $0.044$--$0.057$ whatever the correlation between the sites, so that $h_3$ is resolved at both chemical potentials and $h_5$ only marginally at $0.582$. The value of $h_2$ also exceeds this uncertainty, but the band does not bound the folding described in sections~\ref{subsec:exsbd} and~\ref{subsec:kink_extraction}: mock ensembles with a profile mapped onto minus itself by a translation and Gaussian noise with the spectrum of the residuals already give $h_2 \approx 0.12$--$0.13$ when they are fitted with $G_{\text{trans}} \times \mathbb{Z}_2$, so that $h_2$ alone is not a measure of an asymmetry of the ensemble. Unlike $\langle c^3 \rangle$, the ratios $h_j$ do not involve the relative phases, but they enter $C(n)$ only together with the fluctuations (section~\ref{subsec:kink_extraction}).

\subsection{Centroids without the Euclidean-Time Average}
\label{app:gn2_timeavg}

\begin{figure}[htbp]
    \centering
    \includegraphics[width=\textwidth]{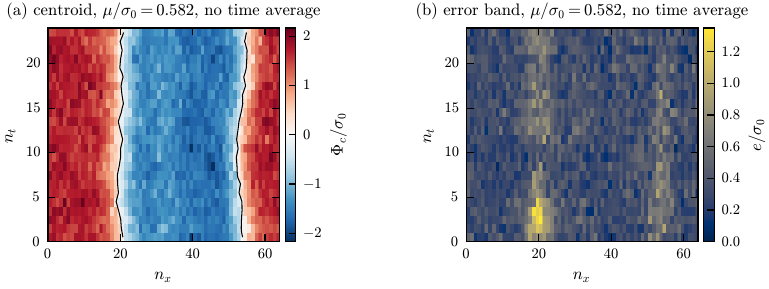}
    \\[0.2cm]
    \includegraphics[width=\textwidth]{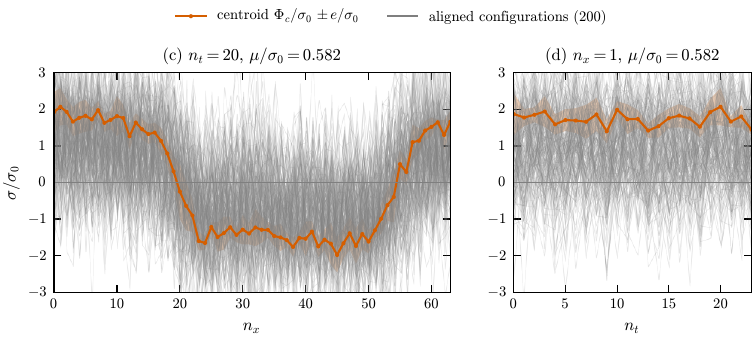}
    \caption{
        Centroid for $K = 1$ at $\mu/\sigma_0 = 0.582$ without the Euclidean-time average. Top: (a) $\Phi_c(n_t, n_x)/\sigma_0$ on the $N_t \times N_s$ lattice, with the zero contour, and (b) the band $e(n_t, n_x)/\sigma_0$; $n_t$ and $n_x$ are the Euclidean-time and spatial components of the lattice coordinate $\mathbf{n}$. Bottom: (c) the slice at $n_t = 20$ as a function of $n_x$ and (d) the slice at $n_x = 1$ as a function of $n_t$, with the band and with $200$ configurations (grey, in units of $\sigma/\sigma_0$) aligned to the centroid by two-dimensional translations. The slices are chosen from the time average $\bar{c}(n_x)$ of the normalized centroid: $n_x$ maximizes $|\bar{c}(n_x)|$, and $n_t$ is the time slice that agrees least with $\bar{c}$, at $\max_{\boldsymbol{\tau}} |\mathrm{NCC}_{\boldsymbol{\tau}}| = 0.9888$; the position of the centroid on the periodic lattice is arbitrary. At this extremum the centroid lies outside the bulk of the configurations, its amplitude being fixed by $\bar{s}$, which contains the site-local fluctuations (see the text above). The grey curves illustrate that the band does not bound the deviation of individual configurations.
    }
    \label{fig:gn2_D7_0582}
\end{figure}

\begin{figure}[htbp]
    \centering
    \includegraphics[width=\textwidth]{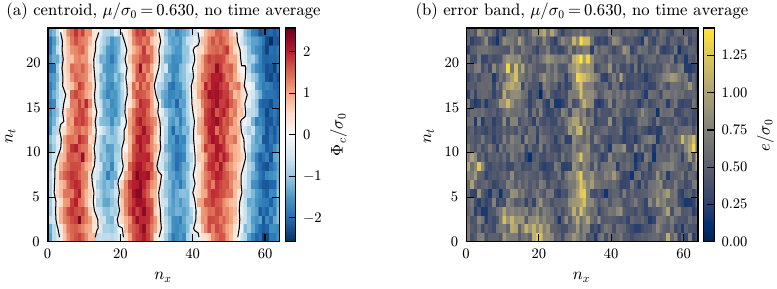}
    \\[0.2cm]
    \includegraphics[width=\textwidth]{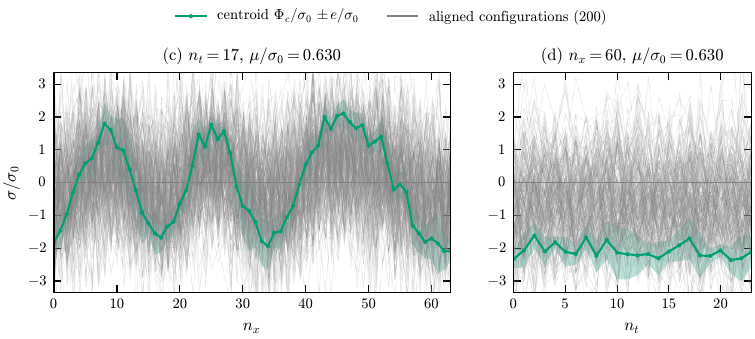}
    \caption{
        As figure~\ref{fig:gn2_D7_0582}, at $\mu/\sigma_0 = 0.630$, with the slices (c) at $n_t = 17$ ($\max_{\boldsymbol{\tau}} |\mathrm{NCC}_{\boldsymbol{\tau}}| = 0.9504$) and (d) at $n_x = 60$.
    }
    \label{fig:gn2_D7_0630}
\end{figure}

Figures~\ref{fig:gn2_D7_0582} and~\ref{fig:gn2_D7_0630} show the centroids for $K = 1$ obtained without the average over Euclidean time at $\mu/\sigma_0 = 0.582$ and $0.630$, fitted with $G = G_{\text{trans}}$ on the $24 \times 64$ lattice, $N_v = 1536$, on the direct route; the adopted restarts were stopped by eq.~\eqref{eq:stopping_rule} after $88$ and $116$ iterations. The mode $k_t = 0$ carries $97.9\,\%$ and $93.8\,\%$ of the power of these centroids, and their time average agrees with the centroid from the time-averaged configurations to $\max_{\boldsymbol{\tau}} |\mathrm{NCC}_{\boldsymbol{\tau}}| = 0.9993$ and $0.9975$. The individual time slices of the centroid agree with its time average to $\max_{\boldsymbol{\tau}} |\mathrm{NCC}_{\boldsymbol{\tau}}| = 0.9888$--$0.9954$ at $\mu/\sigma_0 = 0.582$ and $0.9504$--$0.9892$ at $0.630$, with medians $0.9921$ and $0.9746$. At every site, the standard deviation of the centroid along $t$ has been compared with the band at that site. The median ratio is $0.47$ and $0.56$, and the ratio exceeds one at a single site of $64$ at $\mu/\sigma_0 = 0.630$, where it is $1.23$ and where the band is locally smaller than at the neighbouring sites. The band is larger than for the time-averaged fits with the same group, $\langle e \rangle / \mathrm{rms}\, \Phi_c = 0.27$ and $0.43$ against $0.075$ and $0.325$, and $\bar{s}/\sigma_0 = 1.43$ and $1.30$ contains the site-local fluctuations; the amplitudes in these figures are therefore not comparable with those of figure~\ref{fig:gn2_centroids_K1}, and only normalized shapes are compared. As stated in property~(iv) of section~\ref{subsec:error_estimation}, individual time slices can deviate from the centroid by more than the band.

\subsection{Dependence of the Centroid on the Group}
\label{app:gn2_group}

With $G_{\text{trans}}$, sign-reversed patterns that are not related by a translation are not identified. Their relative populations can therefore affect the centroid, whereas the additional $\mathbb{Z}_2$ factor aligns their orientations. The clustering definitions do not use the Monte Carlo ordering of the configurations. Persistence of an orientation or shape over a long part of the history instead matters for autocorrelation and the block error estimate; it can also make the measured populations unrepresentative of equilibrium. With $G_{\text{trans}}$ and $K \ge 2$, the two orientations of one pattern can thus appear as separate clusters; clusters related by a sign and a translation can be recognized a posteriori by their shape correlation, $\max_{\boldsymbol{\tau}}|\mathrm{NCC}_{\boldsymbol{\tau}}|$.

\begin{figure}[htbp]
    \centering
    \includegraphics[width=0.62\textwidth]{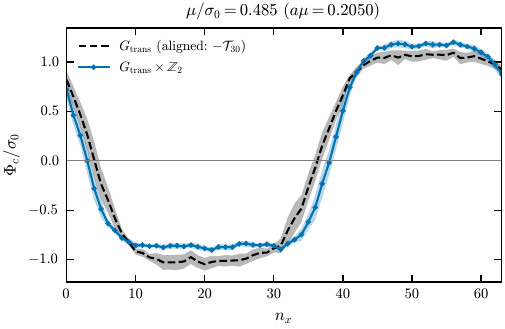}
    \caption{
        Centroids for $K = 1$ at $\mu/\sigma_0 = 0.485$ obtained with $G = G_{\text{trans}} \times \mathbb{Z}_2$ (colour, with band) and with $G = G_{\text{trans}}$ (black dashed, grey band). For display, the latter is aligned to the former by $-\mathcal{T}_\tau$ with $\tau = 30$ in the Z-normalized space, and both are restored with the common $\bar{s}$ and $\bar{m}$, which are the same for the two groups at $K = 1$; its band is translated by the same $\tau$. The difference exceeds twice the combined band at $18$ of the $64$ sites (section~\ref{subsec:kink_extraction}).
    }
    \label{fig:gn2_group_comparison}
\end{figure}

Figure~\ref{fig:gn2_group_comparison} overlays the centroids for $K = 1$ at $\mu/\sigma_0 = 0.485$ obtained with $G = G_{\text{trans}} \times \mathbb{Z}_2$ and with $G = G_{\text{trans}}$. The latter is aligned to the former by the element of $G_{\text{trans}} \times \mathbb{Z}_2$ that maximizes their overlap, the translation $\mathcal{T}_\tau$ of eq.~\eqref{eq:inner_product} with $\tau = 30$ combined with a sign reversal ($|\mathrm{NCC}| = 0.983$), applied in the Z-normalized space. At $0.485$ four conditions coincide: one dominant shape, whose individual configurations are not mapped onto themselves by $\sigma \to -\sigma$ combined with a translation, two orientations in nearly equal proportions ($52.6\,\%$ in one of them, as identified by whether a sign reversal is needed to align a configuration to the centroid for $G_{\text{trans}} \times \mathbb{Z}_2$), and a narrow band. At $0.582$ the orientations are unbalanced ($22.5\,\%$), and at $0.630$ no single shape dominates and no sector structure can be measured through the spatial mean of a configuration. The orientation of a configuration is defined relative to the centroid and not by the sign of its spatial mean, but the two are closely related: it follows the sign of $d_+ - N_s/2$ for $95\,\%$ and $93\,\%$ of the configurations with $N_{\text{kink}} = 2$ at $0.485$ and $0.582$, and, through the unequal domain lengths, the sign of the spatial mean, with which it agrees for $89\,\%$ of the configurations at $0.485$ and for $83\,\%$ of those with $N_{\text{kink}} = 2$ at $0.582$. Splitting the ensemble by the sign of the spatial mean therefore separates the two orientations rather than providing an independent test: at $0.485$ the centroid for $G_{\text{trans}} \times \mathbb{Z}_2$ agrees to a shape correlation of at least $0.9993$ with the centroids fitted separately to the two subsets defined by that sign, taken after the Euclidean-time average and before the Z-normalization, whereas the centroid for $G_{\text{trans}}$ agrees with them only to $0.982$--$0.986$. The orientation changes slowly along the Monte Carlo history, with an integrated autocorrelation time of at least about $2 \times 10^3$ and $10^3$ configurations at $0.485$ and $0.582$ (appendix~\ref{app:autocorr}); the proportion $22.5\,\%$ at $0.582$ deviates from one half by at most about $1.5$ times its expected fluctuation under $\sigma \to -\sigma$, the autocorrelation time being a lower bound, and we do not interpret it as a physical imbalance. At $0.485$ and $0.582$ the spatial means are smaller than the modulation (table~\ref{tab:GN_indicators}) and take both signs, and their average, $\bar{m}/\sigma_0 = -0.016(45)$ at $0.582$, has an uncertainty that does not exclude the offset that the unbalanced orientations would produce if the same correlation held there. The bands of the two groups differ as well: the root mean square over the sites of $e/\bar{s}$ changes by factors $0.71$, $0.71$ and $1.48$ at $\mu/\sigma_0 = 0.485$, $0.582$ and $0.630$ from $G_{\text{trans}}$ to $G_{\text{trans}} \times \mathbb{Z}_2$. This is not a fixed relation between the two groups, and why the first two factors coincide is not determined by the present calculation.

In the comparison of section~\ref{subsec:kink_extraction}, the band is the scale of a change of one shape under two analysis choices, not a test of whether two shapes are the same (section~\ref{subsec:error_estimation}). There $\mathrm{rms}\, \Delta / \mathrm{rms}\, \sigma_{\text{comb}} = 0.65$ and $0.51$ at $\mu/\sigma_0 = 0.582$ and $0.630$, the largest $|\Delta|/\sigma_{\text{comb}}$ at a site is $1.59$ and $0.79$, and no site exceeds $2 \sigma_{\text{comb}}$. One minus the shape correlation~\eqref{eq:shape_correlation} of the centroids of the two groups is $0.017$, $0.0024$ and $0.050$ at $\mu/\sigma_0 = 0.485$, $0.582$ and $0.630$.

For $K \ge 2$, the partitions obtained with $G_{\text{trans}} \times \mathbb{Z}_2$ agree with those for $G_{\text{trans}}$ (section~\ref{subsec:multimodality}) for $67.6\,\%$ and $71.5\,\%$ of the configurations ($K = 2$, $3$) at $\mu/\sigma_0 = 0.485$, for $99.4\,\%$ and $77.6\,\%$ at $0.582$, and for $85.4\,\%$ and $84.9\,\%$ at $0.630$. At $0.630$ the two partitions differ mostly at the cluster boundaries, the clusters obtained with $G_{\text{trans}} \times \mathbb{Z}_2$ are not sign-reversed partners (mutual shape correlations $0.52$--$0.64$), each contains both signs of the spatial mean, and the composition in $N_{\text{kink}}$ is kept. A fit with another random seed reproduces the partition for $K = 2$ at $0.630$ for $99.30\,\%$ of the configurations. For $K = 3$ at $0.630$ the restart that is adopted with $r_{\max} = 200$ is stopped by eq.~\eqref{eq:stopping_rule} after $231$ iterations, at a larger value of $\mathcal{L}$ than the one it had at iteration $200$, so that with $r_{\max} = 2000$ another restart, itself stopped by eq.~\eqref{eq:stopping_rule}, has the smallest $\mathcal{L}$ and is adopted instead; the two partitions agree for $98.7\,\%$ of the configurations and the matched centroids to shape correlations of at least $0.9985$. Fits with $r_{\max} = 2000$ and five random seeds, in which all restarts were stopped by eq.~\eqref{eq:stopping_rule}, agree for $99.8$--$99.97\,\%$ of the configurations and give $\rho_{\text{B}} = 0.4949$--$0.4954$.

\subsection{Cut-Off Dependence of the Kink Numbers}
\label{app:gn2_kink_robust}

With the cut-off $|q| \le 8$ of figure~\ref{fig:gn2_kink_hist}, the configurations at $\mu/\sigma_0 = 0.630$ have $N_{\text{kink}} = 6$ ($59.6\,\%$), $8$ ($31.2\,\%$), $4$ ($6.9\,\%$) and $10$ ($2.2\,\%$). For $K = 2$, $75.0\,\%$ of the members of the larger cluster ($57.4\,\%$ of the configurations) have $N_{\text{kink}} = 6$, and $8$ is the most frequent value in the smaller one ($42.6\,\%$), at $44.7\,\%$ of its members; for $K = 3$, $75.4\,\%$ of the members of the family that separates ($26.6\,\%$ of the configurations) have $N_{\text{kink}} = 8$. At $\mu/\sigma_0 = 0.582$, $85$--$87\,\%$ of the members of the minority cluster ($15.8\,\%$ of the configurations for $K = 2$, $15.4\,\%$ for $K = 3$) have $N_{\text{kink}} = 4$.

\begin{figure}[htbp]
    \centering
    \includegraphics[width=\textwidth]{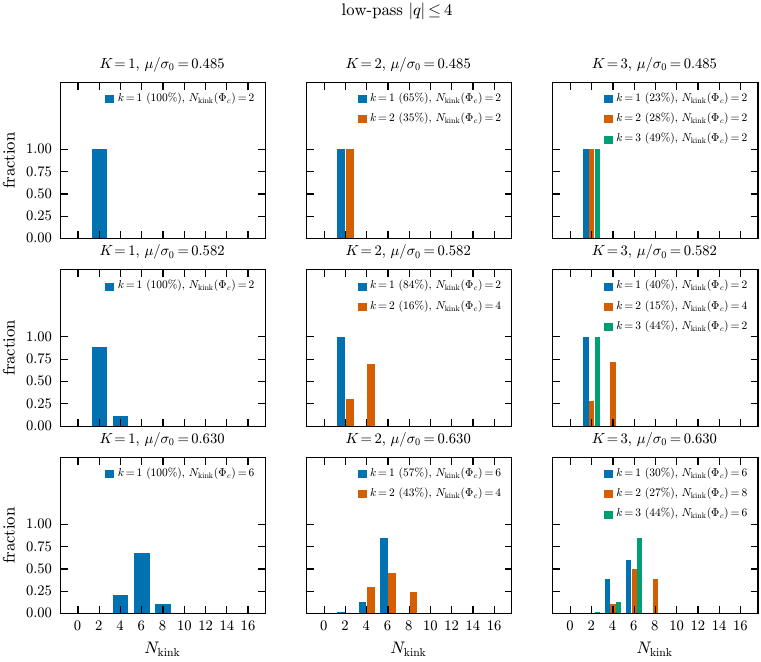}
    \caption{
        As figure~\ref{fig:gn2_kink_hist}, with the Fourier modes $|q| \le 4$ retained.
    }
    \label{fig:gn2_kink_hist_lp4}
\end{figure}

\begin{figure}[htbp]
    \centering
    \includegraphics[width=\textwidth]{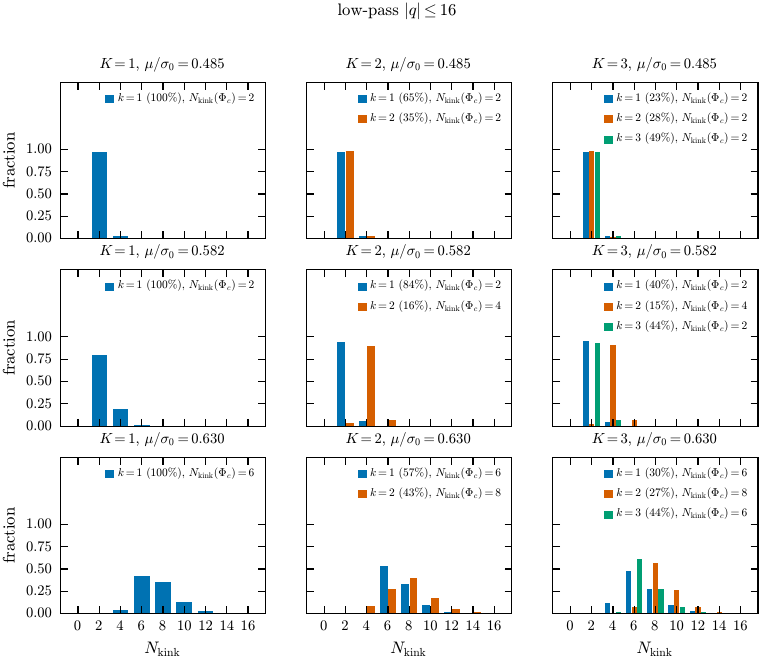}
    \caption{
        As figure~\ref{fig:gn2_kink_hist}, with the Fourier modes $|q| \le 16$ retained.
    }
    \label{fig:gn2_kink_hist_lp16}
\end{figure}

Figures~\ref{fig:gn2_kink_hist_lp4} and~\ref{fig:gn2_kink_hist_lp16} repeat figure~\ref{fig:gn2_kink_hist} with the Fourier modes $|q| \le 4$ and $|q| \le 16$ retained. The fractions, and in one cluster of each fit the most frequent value, depend on the cut-off, as does $N_{\text{kink}}(\Phi_c)$ of the smaller cluster for $K = 2$ at $\mu/\sigma_0 = 0.630$ ($4$, $6$ and $8$ for $|q| \le 4$, $8$ and $16$), whereas the qualitative statements of section~\ref{subsec:multimodality} do not.

\subsection{Discriminating Power of the Band}
\label{app:gn2_band}

\begin{figure}[htbp]
    \centering
    \includegraphics[width=0.62\textwidth]{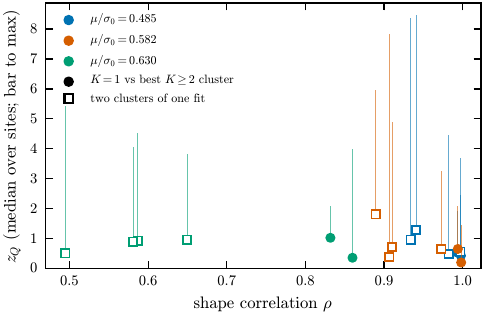}
    \caption{
        Discriminating power of the band. For pairs of aligned centroids of the inhomogeneous ensembles, the abscissa is the shape correlation $\rho$ of eq.~\eqref{eq:shape_correlation} and the ordinate is $z_Q$, the median over the sites of $|\Phi_c^{a} - \Phi_c^{b}| / [(e^{a})^2 + (e^{b})^2]^{1/2}$, with a bar extending to the maximum over the sites. Filled symbols: the centroid for $K = 1$ and the best-matching cluster for $K \ge 2$; open symbols: two clusters of the same fit. Colours denote $\mu/\sigma_0$. $z_Q$ is a descriptive number and not a probability.
    }
    \label{fig:gn2_band_discrimination}
\end{figure}

For two centroids $\Phi_c^{a}$ and $\Phi_c^{b}$ with bands $e^{a}$ and $e^{b}$, aligned such that a sign reversal, where needed, acts on the whole restored field, offset included ($\sigma \to -\sigma$), let $z_Q$ be the median over the sites of $|\Phi_c^{a} - \Phi_c^{b}| / [(e^{a})^2 + (e^{b})^2]^{1/2}$. Figure~\ref{fig:gn2_band_discrimination} shows $z_Q$ against the shape correlation~\eqref{eq:shape_correlation} for pairs of centroids of the three inhomogeneous ensembles. For the four pairs of nearly identical shapes formed by the centroid for $K = 1$ and the best-matching clusters for $K = 2$ and $3$ at $\mu/\sigma_0 = 0.485$ and $0.582$, $z_Q = 0.21$--$0.66$, while for the best-matching pair for $K = 3$ at $0.630$, whose shapes differ ($\rho = 0.860$), $z_Q = 0.36$ falls within that range. Pairs of nearly identical shapes and pairs of clearly different shapes are thus not separated by $z_Q$, which is the reason why the band is not used to compare values of $K$ (section~\ref{subsec:error_estimation}); $z_Q$ is a descriptive number, not a probability. The jackknife error $\delta \bar{m}_k$ of the offset, which is not part of the band, ranges from $6.1\,\%$ to $498\,\%$ of the site-averaged band over the $20$ clusters of the fits with $K = 1$, $2$ and $3$ and of the two fits without the Euclidean-time average, so that no single factor relates the two.

\clearpage
\bibliographystyle{JHEP}
\bibliography{refs}
\end{document}